\documentclass[12]{article}
\usepackage{times}
\usepackage{natbib}

\usepackage{color}
\usepackage{subfigure}

\usepackage{hyperref}

\usepackage{amsmath, amssymb, fullpage, amsthm, array, algorithm2e,graphicx}

\begin{document}

\title{Utilizing Distance Metrics on Lineups to Examine What People Read From Data Plots}
\author{Niladri Roy Chowdhury, Dianne Cook, Heike Hofmann, Mahbubul Majumder, Yifan Zhao}
\date{\today}  
\maketitle

\begin {abstract} 
Graphics play a crucial role in statistical analysis and data mining. This paper describes metrics developed to assist the use of lineups for making inferential statements. Lineups embed the plot of the data among a set of null plots, and engage a human observer to select the plot that is most different from the rest. If the data plot is selected it corresponds to the rejection of a null hypothesis. Metrics are calculated in association with lineups, to measure the quality of the lineup, and help to understand what people see in the data plots.  The null plots represent a finite sample from a null distribution, and the selected sample potentially affects the ease or difficulty of a lineup. Distance metrics are designed to describe how close the true data plot is to the null plots, and how close the null plots are to each other. The distribution of the distance metrics is studied to learn how well this matches to what people detect in the plots, the effect of null generating mechanism and plot choices for particular tasks.  The analysis was conducted on data that has already been collected from Amazon Turk studies conducted with lineups for studying an array of data analysis tasks.

\end {abstract}


\section{Introduction} 
Graphics are an important component of big data analysis, providing a mechanism for discovering unexpected patterns in data. Pioneering research by \citet{gelman:2004}, \citet{buja:2009} and \cite{majumder:2011} provide methods to quantify the significance of discoveries made from visualizations. 
\cite{buja:2009} introduced two protocols which bridge the gulf between traditional statistical inference and exploratory data analysis. These are the Rorschach and the lineup protocols. The Rorschach protocol helps to understand the extent of randomness. The lineup protocol places a statistical plot firmly in the hypothesis testing framework, where a plot of the data is considered to be a test statistic. Unlike the simpler numeric test statistics in classical inference, though, the plot as a test statistic is a complex entity. This plot is compared with a set of null plots, obtained from an appropriate distribution consistent with the null hypothesis. The lineup protocol places the data plot randomly among the obtained null plots, and requires a human observer to examine the plots and identify the most different plot. If this plot is that of the data, this is quantifiable evidence against the null hypothesis. 

\begin{figure}[htbp]
\centerline{\includegraphics[width=1\textwidth]{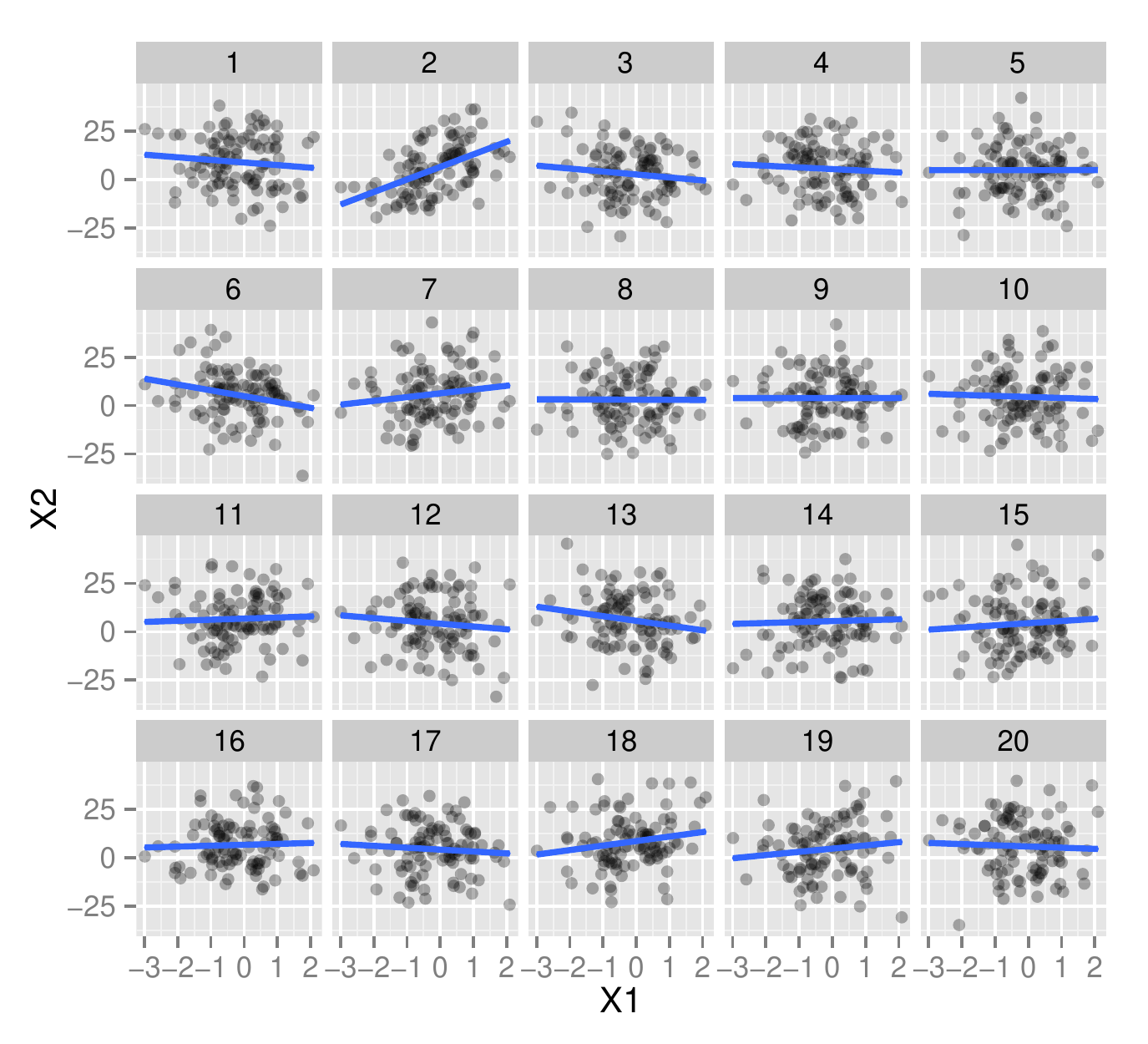}}
\caption{Lineup plot of size $m = 20$ using scatterplots with regression line overlaid. This tests $H_o: \beta_k = 0$, where covariate $X_k$ is continuous. One of the plots in the lineup is the plot of the true data. The other plots are null plots generated by simulating data from a null model that assumes that the null hypothesis is true. Can you identify the plot with the steepest slope?  }
\label{lineup-example}
\end{figure}

Figure \ref{lineup-example} is an example of a lineup. Suppose we have the following statistical model

$$Y_i = \beta_0 + \beta_1 X_{i1} + \beta_2 X_{i2} + \dots + \epsilon_i$$

\noindent and we are interested in testing the following hypothesis:

$$H_o : \beta_k = 0 \qquad \qquad \hbox{vs} \qquad \qquad H_A: \beta_k \ne 0$$

\noindent where $X_k$ is a continuous covariate. The true plot is obtained by plotting $Y$ against $X_k$ with a regression line overlaid. The null plots are obtained by simulating data from $N(X\hat{\beta}, \hat{\sigma}^2)$ and plotting using the same scatterplot method as the true data. These parameter estimates ($\hat{\beta}, \hat{\sigma}^2$) are obtained by fitting the null model to the true data. The plot of the true data is randomly placed among a set of ($m$ - 1) null plots to produce a lineup of size $m$. The human subjects are then shown this lineup and asked to identify the plot which has the steepest slope. If the human subjects can identify the plot of the true data, we reject the null hypothesis and conclude that there is a significant linear relationship between $Y$ and $X_k$. 

The lineup protocol was formally tested in a head-to-head comparison with the equivalent conventional test by \cite{majumder:2011}. The experiment utilized human subjects from Amazon's Mechanical Turk (\cite{turk}) and used simulation to control conditions. The results suggest that the visual inference is comparable to conventional tests in a controlled conventional setting. This provides support for its appropriateness for testing in real exploratory situations where no conventional test exists. Interestingly, the power of a visual test increases with the number of observers engaged to evaluate lineups, and the pattern in results suggests that the power will provide results consistent with practical significance (\cite{kirk:1996}).


%

\begin{figure}[htbp]
\centering
\mbox{\subfigure[Classical Inference]{\includegraphics[width=3in]{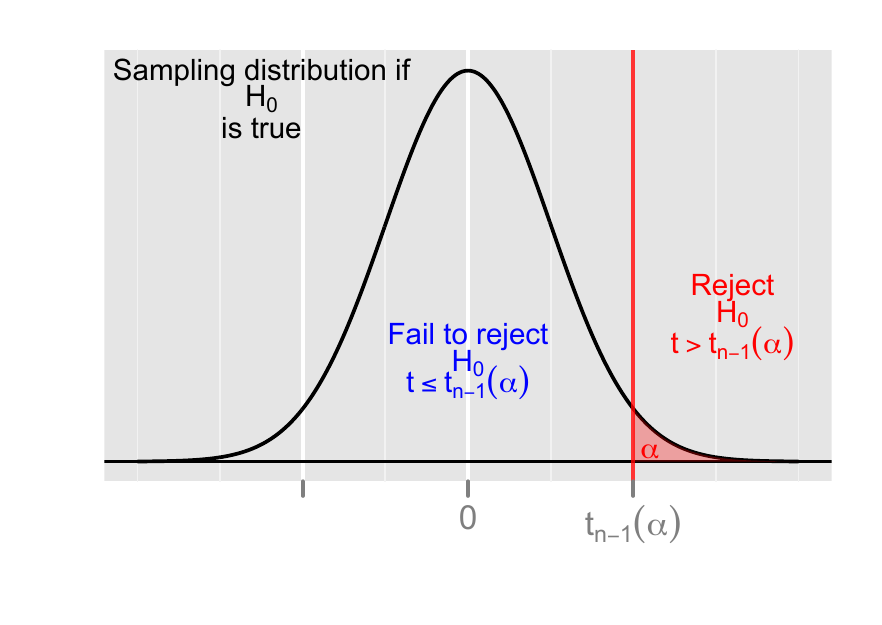}}\quad
\subfigure[Visual Inference]{\includegraphics[width=3in]{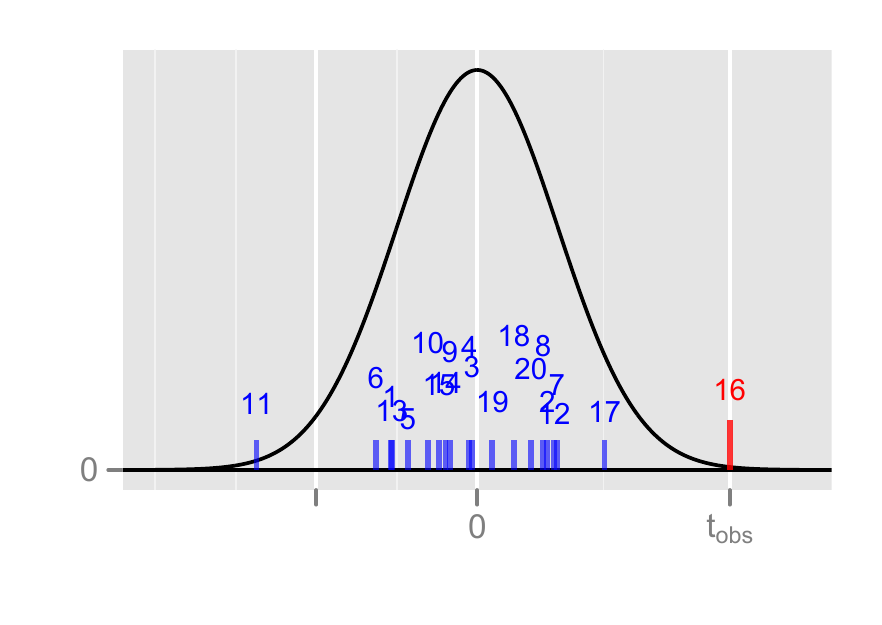} }}
\caption{If the lineup protocol was to be used instead of classical inference this is what it would look like. (a) Decision region (shaded in red) for classical inference for $H_0: \mu=\mu_0$ vs $H_a:\mu>\mu_0$  and (b) values corresponding to the true value (red) and the null plots (blue) in a single lineup of size $m=20$ that would be used to test the same null hypothesis.  The actual  data plot is extreme relative to the null plots, and observers would likely be able to pick it out, resulting in a decision to reject the null hypothesis. In practice, the lineup protocol would not be used if a classical test can be used.} 
\label{compare}
\end{figure}


In traditional hypothesis testing, the sampling distribution of a test statistic is functional and continuous. In the lineup protocol, although conceptually we may have an infinite collection of plots from the null distribution, in practice, we can only evaluate against a finite number of null plots. A human judge has a physical limit on the number of plots they can peruse. This poses one of the issues with using the lineup protocol.  Figure \ref{compare} illustrates the difference. In traditional inference, the black curve represents the sampling distribution for the $t$-distribution under the null hypothesis, and the shaded red area shows the rejection region. In visual inference, let us consider that the black curve gives the sampling distribution although the sampling distribution is essentially a distribution of null plots. Although the test statistic is not numeric, the true data plot which is the test statistic is represented using red bar and the null plots that are drawn from the null distribution are the blue bars. 
Effectively,  in visual inference the red line is compared only to these finite number of blue lines visually to make a decision, unlike classical inference where we look at the rejection region (Figure \ref{compare}) to make decisions. Even though the data plot might be extreme, it is possible by randomly selecting from the null distribution, to obtain a null plot that is more extreme, as Tukey suggested \citep{fernholz03}:

\begin{quotation}
``There [in Tukey's Data Analysis class] I discovered that [...]  a random sample is indeed a ``batch of values'' which ``fail to be utopian'' most of the time.''
\end{quotation}


This can be partially solved by having a large number of observers, who each evaluate lineups constructed with different null plots. Having some idea of the type of coverage of the sampling distribution that is provided by the lineups would be useful ahead of engaging observers and evaluating the lineups. Could we say that lineup X is expected to be ``difficult'' but lineup Y is expected to be ``easy'' then it may help in determining an appropriate number of observers? A difficult lineup is one where the data plot is similar to the null plots, and an easy lineup is where the data plot has some feature that makes it very different from the null plots. Being able to compute a plot to plot distance metric would be very helpful ahead of running a lineup protocol.

This is a two way process: As metrics are devised to measure the quality of a lineup, the lineup protocol also provides an opportunity to measure the performance of a metric. The human eye can detect patterns in a plot that just cannot be easily quantified numerically, which is why graphics provide an important tool for exploring data and finding the unexpected. Describing plots numerically, is something  of an oxymoron, it cannot be universally done. An example in past work are scagnostics \citep{tukey:1977,wilkinson2005graph} which were developed to assess the different aspects of scattered points like outliers, shape, trend, density and coherence.  If a scatterplot has just one of these structures the scagnostics are descriptive, however, they fail terribly if a plot contains more than one. The goal here is to find some distance measures that can provide some indications of the quality of a lineup, and then to use the results of observer evaluation to determine which metrics best match what people see.


Following up on choices, observers are asked to describe their reasoning. These reasons are used to obtain more information about the rejection: was it some nonlinear dependency, an outlier, clustering, that triggered the detection of the data plot? Good distance metrics may also help relate the descriptive words used with mathematically defined features. 

The article is organized as follows. Section \ref{sec:null} discusses the null generating mechanisms. Section \ref{sec:meas} defines the distance measures and discusses the choice of the measures. The distribution of the distance measures are studied in Section \ref{sec:distri}. Section \ref{sec:plot_type} describes the effect of the plot type and the question of interest on the distance measure while Section \ref{sec:eval} talks about the distance evaluations. In Section \ref{sec:nbin}, the methods to select the number of bins for the binned distance is described. Section \ref{sec:results} presents a comparison of the distance measures to the performance of human subjects in several experiments conducted by Amazon's Mechanical Turk.

\section{Experimental Data} \label{sec:expts}

There have been eleven experiments conducted using Amazon's Mechanical Turk~\citep{turk} (MTurk), with some used for evaluating the lineup protocol against classical testing, and the others using the protocol for different purposes. It is possible to try the tasks provided to Turkers\citep{majumder:2013}. For evaluating the distance metrics we used the data collected on experiments 1, 2, and 7. Table \ref{tbl:visual_stat} describes the experiments.


\begin{table*}[hbtp] 
\centering 
\caption{Overview of the different Turk experiments, from where data was taken to study distance metrics and how subjects read the plots. } 
\begin{tabular}{m{.5cm}m{2.6cm}m{2cm}m{5.5cm}} 
\hline\hline 
 Turk ID & Experiment &  Test Statistic  & Lineup question \\ [0.5ex] 
\hline 
1  & Box plot & \begin{minipage}[t]{2cm} \begin{center}	\scalebox{0.12}{\includegraphics{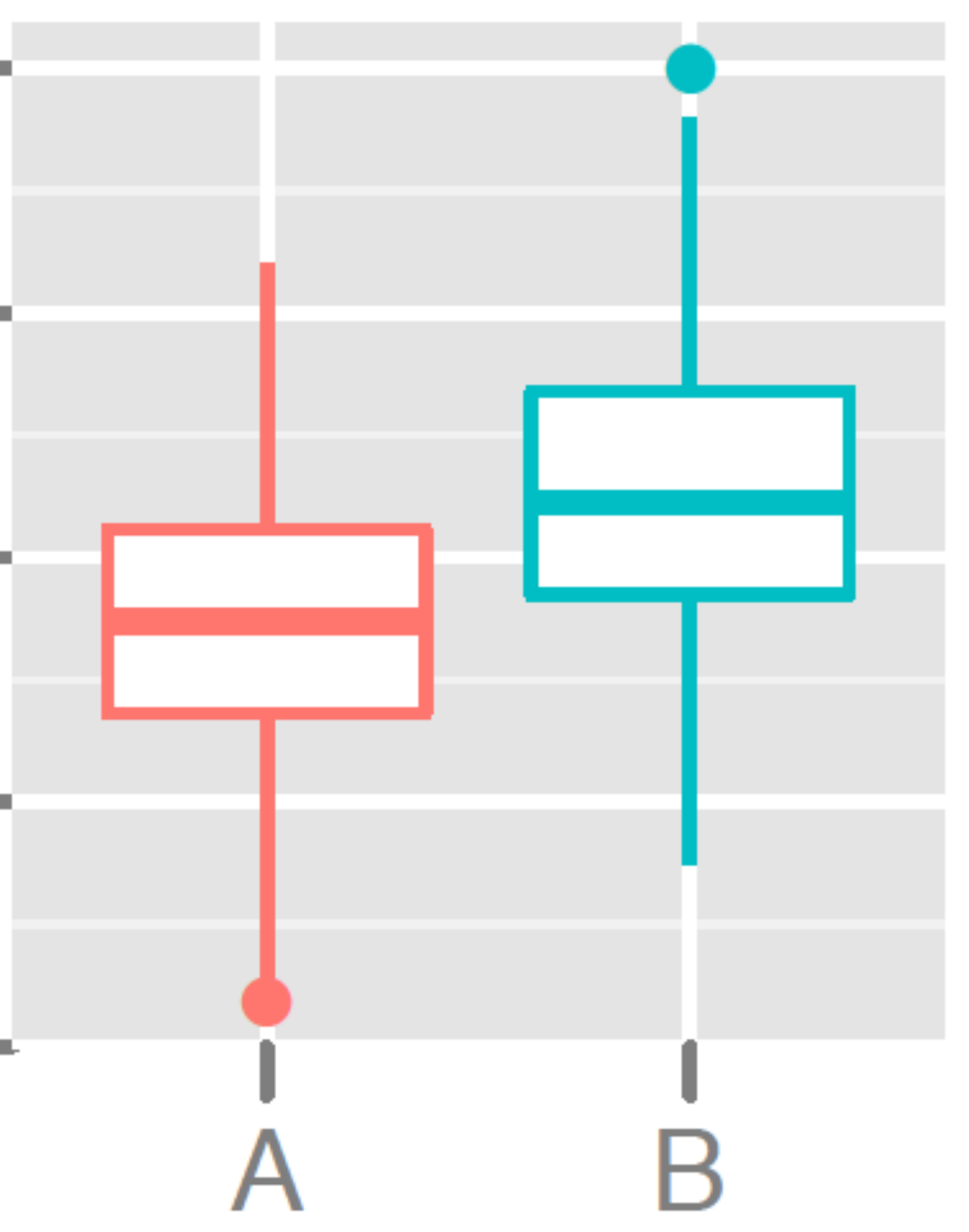}} \end{center} \end{minipage} & Which set of box plots shows biggest vertical difference 
between group A and B? \\
2 &  Scatter plot & \begin{minipage}[t]{2cm}  \begin{center} \scalebox{0.3}{\includegraphics{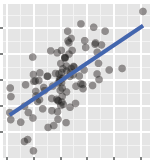}} \end{center} \end{minipage} & Of the scatter plots below which one shows data that has steepest slope? \\
  7 & Group separation & \begin{minipage}[t]{2cm} \begin{center}  \scalebox{0.4}{\includegraphics{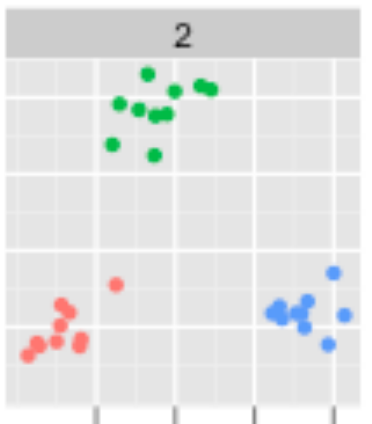}} \end{center} \end{minipage} &Which of these plots has the most separation between the coloured groups?  \\
\hline 
\end{tabular} 
\label{tbl:visual_stat} 
\end{table*}

\section{Null Generating Mechanism} \label{sec:null}

The lineup protocol embeds the true data plot among a set of null plots. The method of obtaining the data for these null plots is called the null generating mechanism. 
The null hypothesis directly affects the choice of null generating method. In the experimental data that we analyzed the null generating methods used were:
\begin{itemize}
\item Permutation: This is the most commonly used approach thus far, because it  can be used in a variety of problems. Permutation is used to break association between two or more variables, and thus is appropriate when the null hypothesis is that there is no association. Consider two variables $X_1$ and $X_2$. Either $X_1$ or $X_2$ is permuted keeping the other variable fixed. Any association between $X_1$ and $X_2$ is broken in the process. The marginal distribution of $X_1$ and $X_2$ remains the same while the joint distribution is altered. The method works in situations where one or both the variables are continuous or categorical. Let us consider a case where we have one categorical variable, say, Group and a continuous variable. Let us assume that the variable Group has two levels (say, A and B) and we want to test whether there is any significant difference between the two groups, i.e. $H_o: \mu_A = \mu_B$. To generate the null data, the values of the variable Group are permuted keeping the continuous variable fixed. If there is a difference between the two groups, this difference is broken by the permutation, and any difference observed in the permuted data is consistent with random variation.
\item Simulation under a null model: Sometimes there is a model underlying the problem being studied. In this situation simulating from the model will be the null generating mechanism. Assuming that the null hypothesis is true, the model is fitted to the true data. The parameter estimates are obtained from the fitted model and then the data is generated using the parameter estimates. Let us consider that we are interested in testing whether there is any significant linear relationship between two continuous variables $X_1$ and $X_2$. Hence we test for $H_o : \beta_1 = 0$ versus $H_a: \beta_1 \ne 0$. Under the null hypothesis, we fit the following model to the data:
$$Y = \beta_0 + \varepsilon$$
where $\varepsilon \sim \hbox{Normal}(0, \sigma^2)$. The parameter estimates of $\beta_0$ and $\sigma^2$ are obtained and the null data is generated from $\hbox{Normal}(\widehat{\beta_0}, \widehat{\sigma}^2)$. 
\end{itemize} 


\section{Distance Measures} \label{sec:meas}

%
%

%

By calculating the ``distance'' between plots we may be able to determine if a lineup should be easy -- the the actual data plot is detectably different from the null plots -- and also to better understand what aspect of the plot people use to make their choice. It is not an easy task to measure the difference between plots. Here we examine several possibilities. 

The problem could be tackled by considering the data as a reference distribution, and compare all of the null sets with this reference. Comparing data with a reference probability distribution or comparing two datasets are common statistical tasks. For example, the Kolmogorov-Smirnov test \citep{stephens:1974} sorts values in two samples, computes the empirical distribution function of each and compares these two, to determine if the two samples are likely to have come from the same distribution.  The Anderson-Darling \citep{stephens:1974} and Shapiro-Wilk \citep{shapiro:1965} tests compare datasets with normal probability distributions. These measure differences between univariate distributions which limits their applicability to distances between plots, generally. 

Hausdorff distance \citep{huttenlocher:1993} has been successfully used for comparing images. It effectively matches points between sets and computes the distances between the matched points. When permutation is the null generating mechanism, Hamming distance \citep{hamming:1950} can be used to calculate how different the permutations are, by measuring the minimum number of substitutions it takes to get from one permutation to another. 

Alternatively, interpoint distance metrics might be adapted to measure distances between plots. For example, when the purpose is differences between groups in a single plot, like side-by-side boxplots, a distance metric that focuses on group separation calculated on each data set might be useful. Bhattacharyya distance \citep{bhattacharyya:1946} is widely used in image processing, for feature extraction. The R package {\tt fpc} \cite{hennig:2010} contains many different ways to calculate distances between groups. 

Ultimately, a very simple binned distance was used in the analyses of the MTurk data, which worked fairly well in most circumstances. However, it was clear immediately that plot design, and the question asked, has a large impact on how a plot is read, and specific distance metrics designed for specific plot types is needed. Below is a summary of distance metrics used. For all of the distance measures below, let $X$ denote the true dataset with one or two variables. Let $Y$ denote the null dataset obtained from $X$ using an appropriate null generating mechanism.



\begin{itemize}

\item Binned Distance (BN):
Let $X_1$ and $X_2$ be two continuous variables. Let $X_1$ be divided into $p$ bins and $X_2$ divided into $q$ bins. 
 Let $C(X_1,X_2)$ be defined as a $p \times q$ matrix. Each $(i,j)$-th element of the matrix represents the number of points falling in the $(i,j)$-th cell, where $i = 1, \dots, p$ , $j = 1, \dots, q$.
The distance is then defined as
\begin{eqnarray*}
d_{\hbox{BN}}^2 (X, Y) &:=& ||C_X(X_1, X_2) - C_Y(X_1,X_2)||^2 \\ &=& \sum_{i=1}^p \sum_{j=1}^q (C_X(X_{1i},X_{2j}) - C_Y(X_{1i},X_{2j}))^2.
\end{eqnarray*}


\begin{figure*}[hbtp]
\centering
\subfigure[ Dataset $X$ with two variables $X_1$ and $X_2$]{
\includegraphics[scale=0.55]{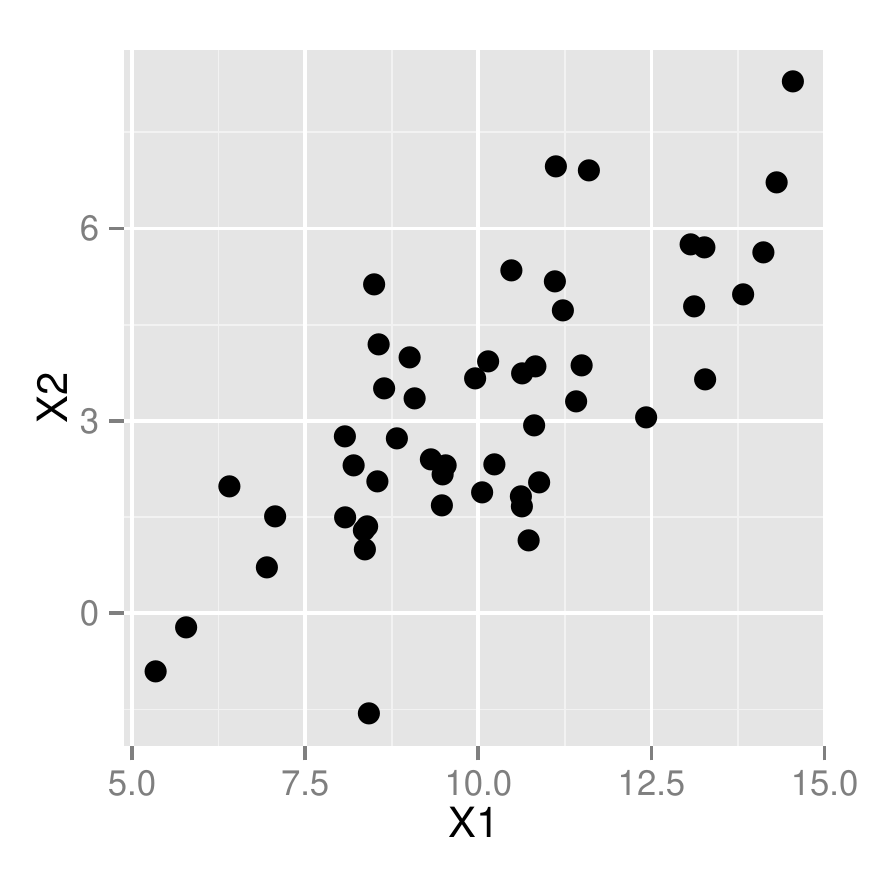}
\includegraphics[scale=0.55]{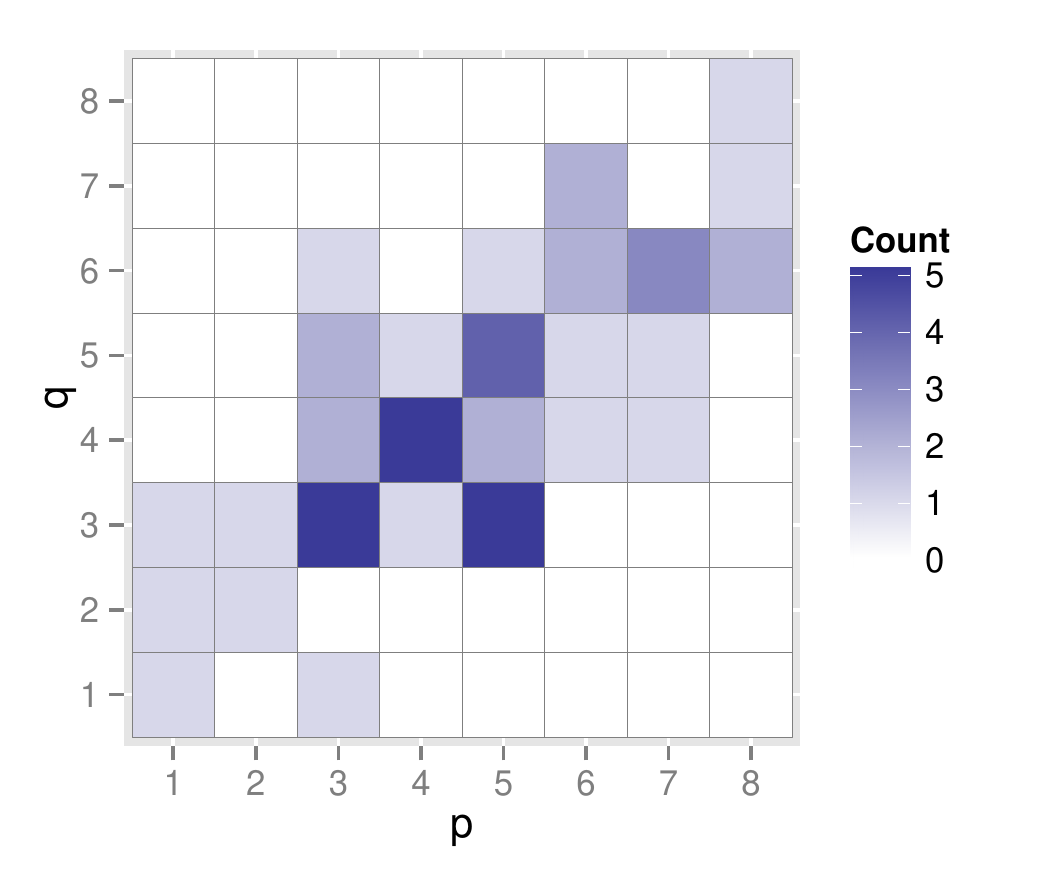}
\includegraphics[scale=0.55]{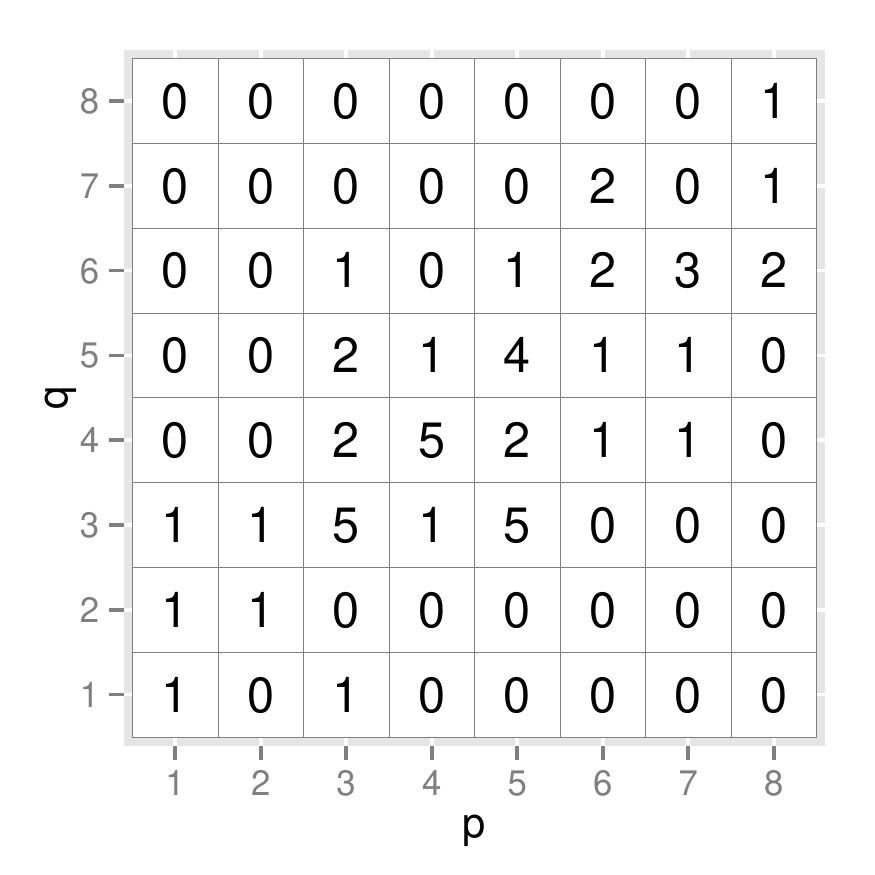}

\label{type_1}
}
\subfigure[ Dataset $Y$ with permuted $X_1$ and original $X_2$]{
\includegraphics[scale=0.55]{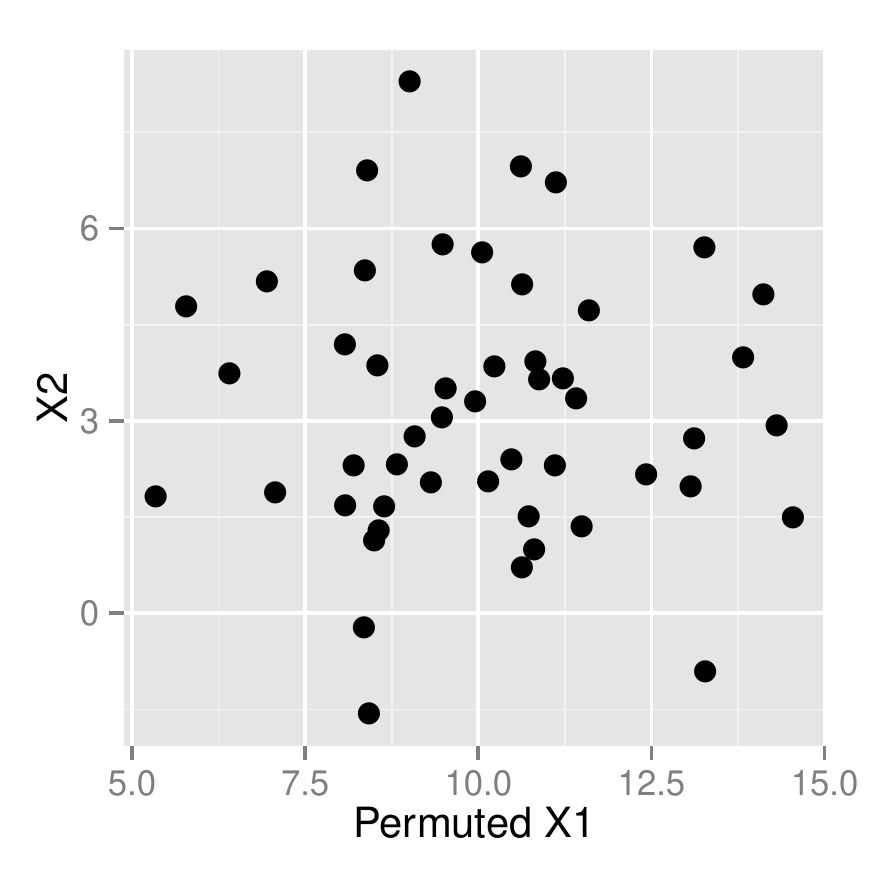}
\includegraphics[scale=0.55]{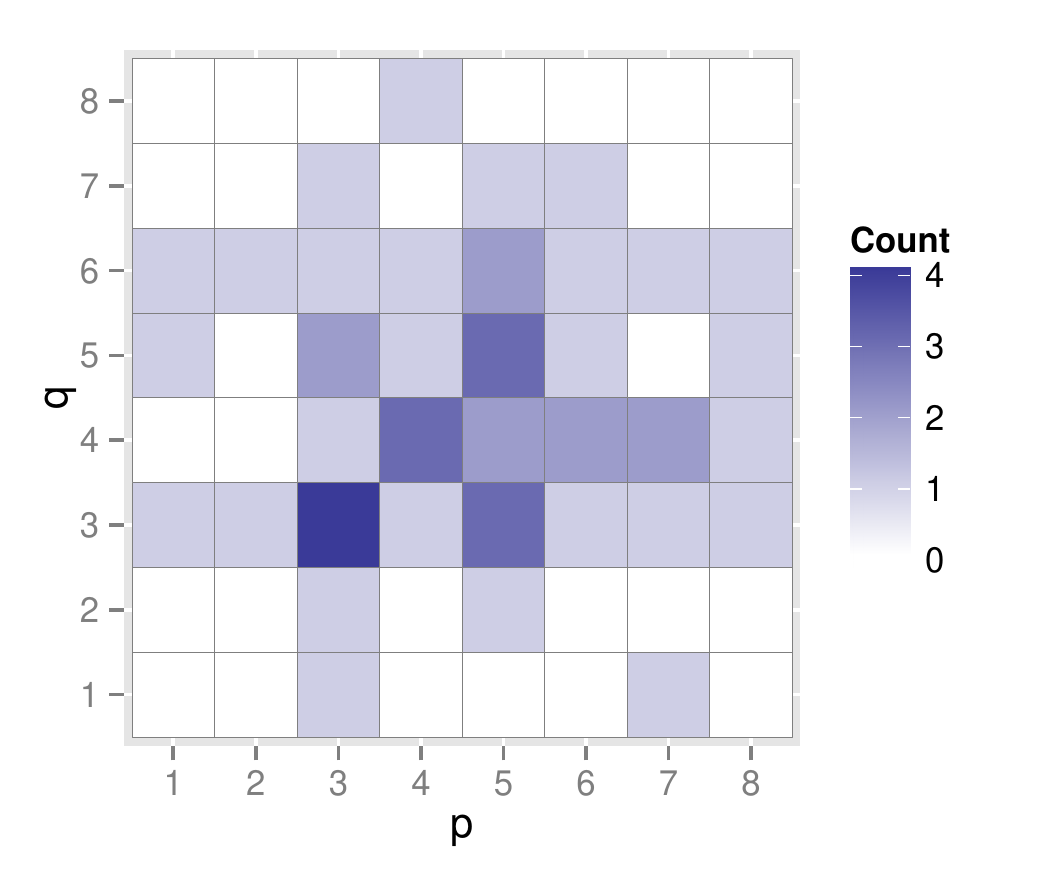}
\includegraphics[scale=0.55]{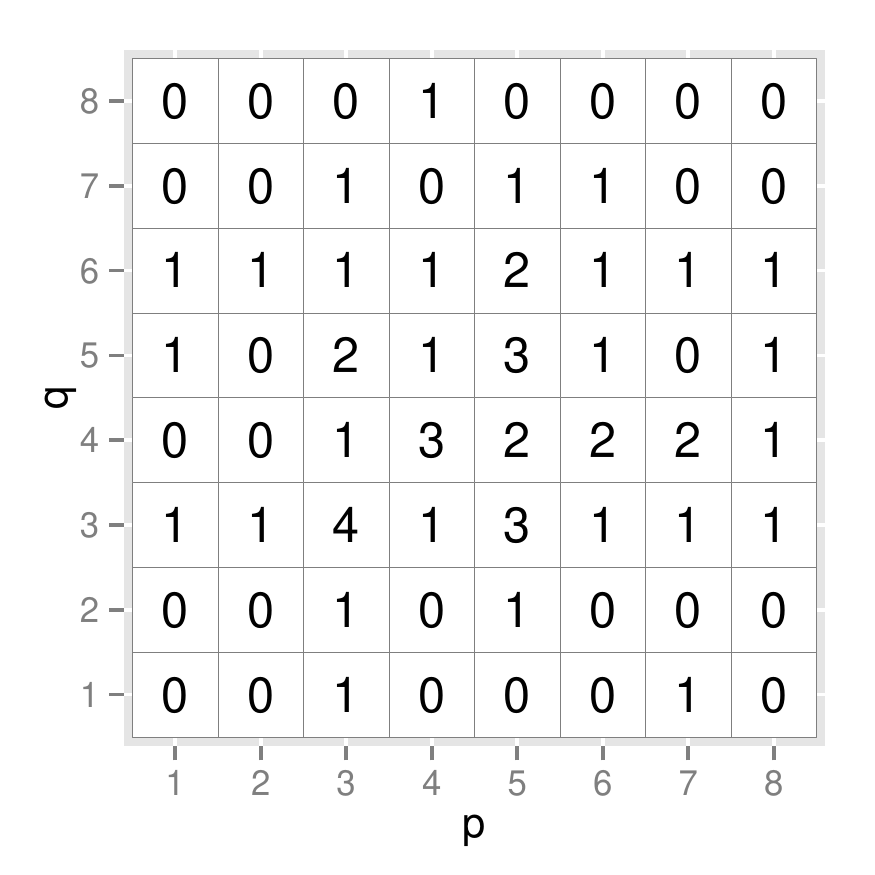}

\label{type_2}
}
\label{plottype}
	\vspace{-.1in}
       \caption{Illustration of binned distance, for data with strong association (a), and the same data where one variable has been permuted (b). The scatterplot of the data is shown (left) along with the binned view of the data (center) and the number of points in each cell (right). Binned distance is the euclidean distance of these counts. The binned distance between these plots is 6.4807. }
\end{figure*}

This distance can be calculated for univariate continuous data, bivariate data with two categorical variables, or data with one continuous and one categorical variable. For the categorical variable, the number of bins would equal to the number of categories.

Binned distance is highly susceptible to small differences in values and depends on the number of bins as well the anchor positions. It is necessary to find the optimal number of bins in each direction. For our purposes optimal was defined number of bins that produced the most detectable observed data plot. The distance was measured between the observed data plot, and the closest null, and compared with the biggest distance between any pair of null plots. Details of these choices on various different data sets is in the Appendix.

Variations on this distance are possible, using kernel density estimates, or using a different power than the square, or using transformations on the counts. Hausdorff distance \citep{huttenlocher:1993} was examined as a generic distance metric as well, but the binned distance was faster to calculate and performed as well as the Hausdorff as a rough, generic measure of similarity of plots.




\item Distance based on boxplots (BX): Let $X_1$ be a categorical variable representing the groups in the data and $X_2$ be a continuous variable. Then the distance metric is given by
 \[
d^2_{\hbox{BX}}(X, Y) := || d_q(X) - d_q(Y)||^2 = \sum_{i=1}^3 ((d_q(X))_i - (d_q(Y))_i)^2
\]

where $d_q(.)$ is a vector giving the absolute difference of the first quartile, median and the third quartile of $X_2$ between the two groups in $X_1$. This distance measure works specifically for the boxplots using only the graphical elements. This is based on the assumption that after the boxplots have already been constructed, the subjects only look at the difference in the boxes to make the distinction. Variations on this might include adding whiskers ends, outliers, or even removing the absolute value. 

\item Distance based on the regression line (RG): Let $X_1$ and $X_2$ be two continuous variables. $X_1$ and $X_2$ are plotted in a scatterplot and assume that the scatterplot is binned vertically into $b$ bins. In each vertical bin, a linear regression model is fitted and the regression coefficients i.e. the estimated intercept and the estimated slope are noted. The distance metric based on the regression coefficients is given by
 \[
d^2_{\hbox{RG}}(X, Y) := \hbox{tr} (B(X) - B(Y))' (B(X) - B(Y)) = \sum_{i=1}^b ((b_0(X))_i - (b_0(Y))_i)^2 + \sum_{i=1}^b ((b_1(X))_i - (b_1(Y))_i)^2
\]

where $b_0$ and $b_1$ denote the vector of the intercept and slope respectively while $b$ is the number of bins. $B(.)$ is a $b \times 2$ matrix of the regression coefficients where each row represent the  intercept and the slope obtained from each bin. The number of bins have a significant effect on the distance measure. It can be seen that it works best for smaller number of bins like 1 or 2. With larger number of bins (i.e. smaller bin sizes), the regression coefficients are affected by the skewness of the data. Variations might include using slope alone, or absolute value of slope.

\item Distance based on separation between multiple groups (MS, AS, WL): Let $X_1$ and $X_2$ be two continuous variable. Let $X_3$ be a categorical variable providing the groups associated with each variable. $X_1$ and $X_2$ are plotted in a scatterplot colored by the group variable $X_3$. The separation can be described in a number of different ways (\cite{hennig:2010}). Two versions are used in this paper. Let us define, 

(i) $s_{m}(.)$ be a vector of cluster wise minimum distance between a point in the cluster to the points in other clusters for $g$ clusters. The distance metric based on separation is defined as
\[
d^2_{\hbox{MS}}(X, Y):= ||s_m(X) - s_m(Y)||^2 = \sum_{i = 1}^g ((s_m(X))_i - (s_m(Y))_i)^2
\]

(ii)  $s_{a}(.)$ be a vector of cluster wise average distances of all the points in the cluster to all point of other clusters for $g$ clusters. The distance metric based on separation is defined as
\[
d^2_{\hbox{AS}}(X, Y):= ||s_a(X) - s_g(Y)||^2 = \sum_{i = 1}^g ((s_a(X))_i - (s_a(Y))_i)^2
\]


Figure \ref{sep-dist} illustrates the difference between the different distance metrics separation. In practice, many possible metrics could be used to measure the separation, such as those readily available in the {\tt fpc} package.


\end{itemize}

\begin{figure}[hbtp]
\centering
\subfigure[Minimum Separation]{
\includegraphics[scale=0.45]{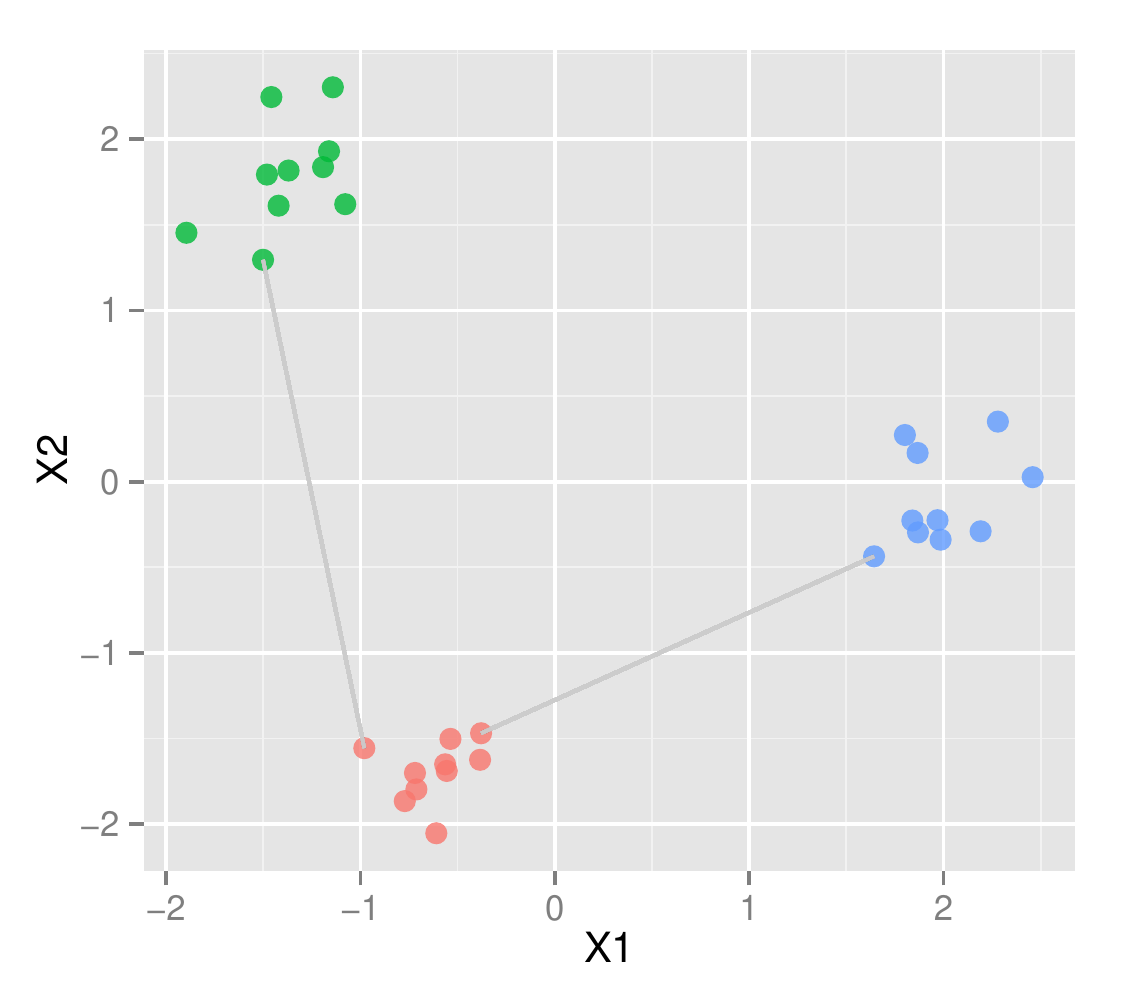}
\label{type_1}
}
\subfigure[Average Separation]{
\includegraphics[scale=0.45]{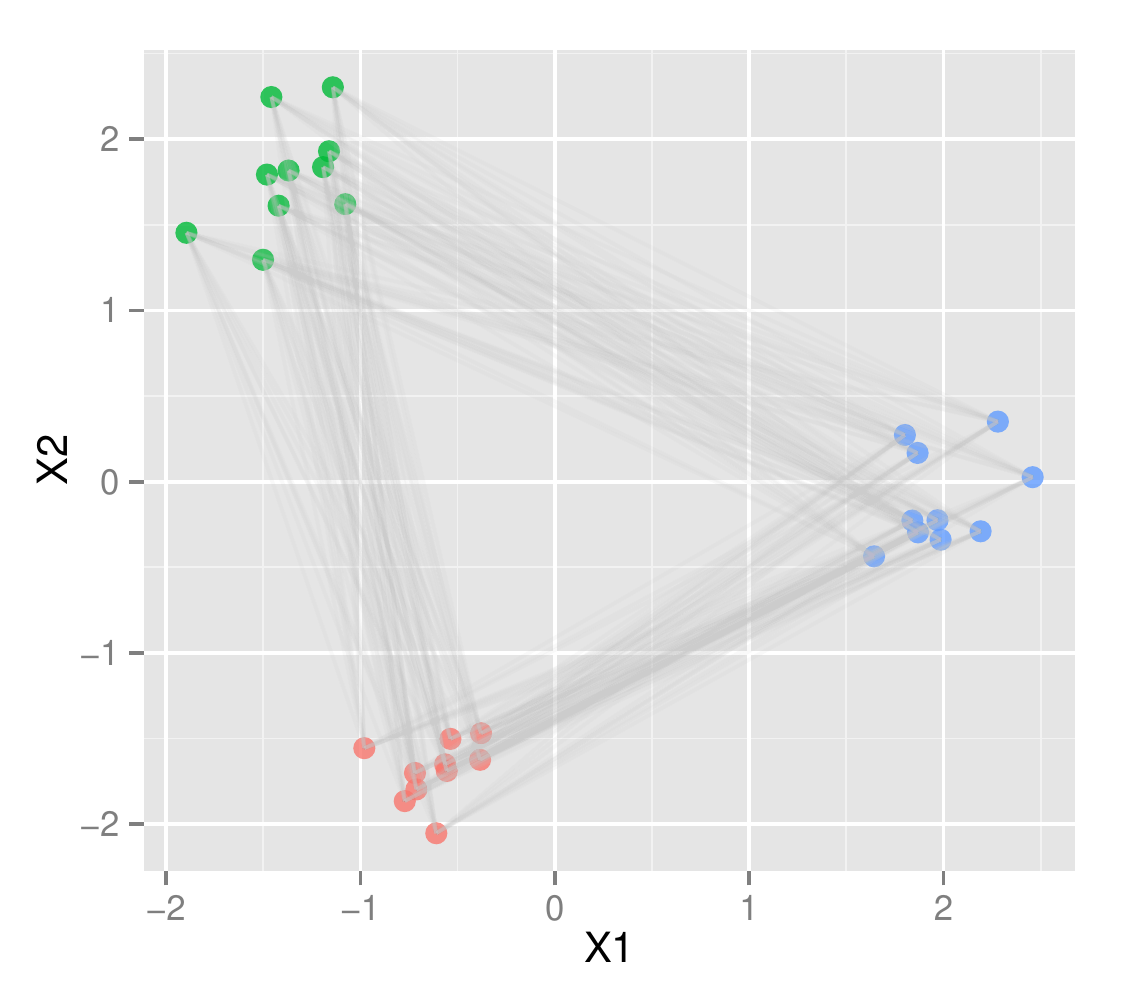}
\label{type_2}
}
\subfigure[Cluster Mean Separation]{
\includegraphics[scale=0.45]{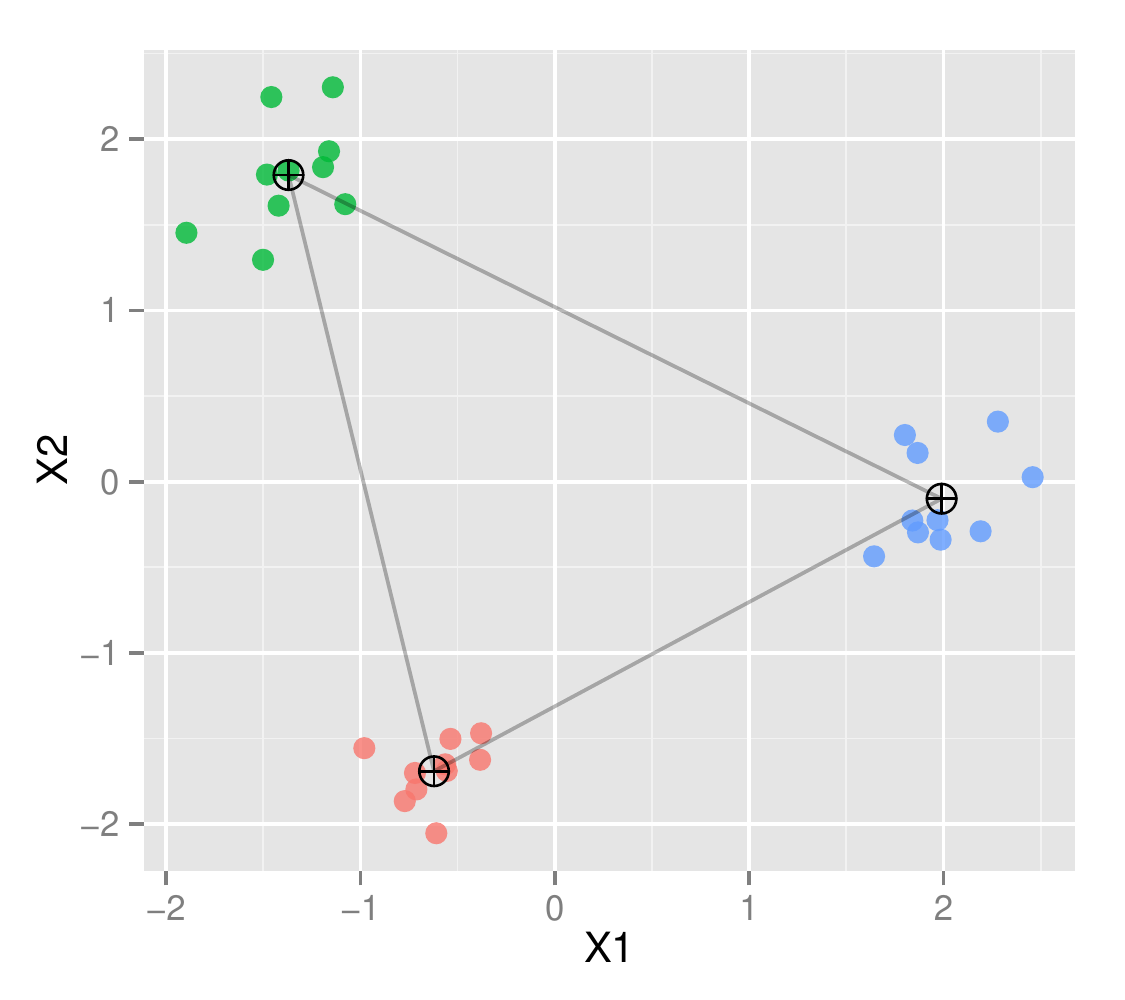}
\label{type_2}
}
	\vspace{-.1in}
\caption{Illustration of three different distance metrics based on separation. Two dimensional projections are plotted with 3 groups. Minimum Separation (in (a)) calculates the minimum distance between points of each cluster from the other clusters. Average separation (in (b)) calculates the average distance of each point  in a cluster to the other clusters. In (c), the cluster mean distance calculates the distance between the means of each cluster. }
\label{sep-dist}
\end{figure}

How well each distance measure matches the observers' responses depends some on the question of interest. But, in general, this should not be a problem because the question which is typically asked is ``Which plot among these is different ?''. In the MTurk experiments, more focused questions were asked because the purpose was very specific, to compare visual inference with classical inference.

\section{Distance Metric Distribution} \label{sec:distri}

For a given lineup of size $m$, the empirical distribution of distance metrics is obtained using the following algorithm:

\begin{enumerate} \itemsep 0in
\item Calculate the distance between the true data and all the null datasets and take the average of these distances. 

\item For each of the ($m$ - 1) null datasets, calculate the distance between the null data and all the other ($m$ - 2) null datasets and obtain the average distance. Hence obtain ($m$ - 1) distances corresponding to each null plot.

\item Generate a null dataset from the true dataset by the null generating mechanism. Consider this null dataset as ``true'' dataset. From this ``true'' dataset, generate ($m$ - 2) null datasets and calculate the average distance between the ``true'' data and the null datasets. Hence obtain a distance corresponding to this ``true'' dataset. Repeat this procedure a large number of times ( order of 1000 or 10000).

\item Plot these 1000 distances to obtain the empirical distribution of the distance metric. Plot the distances for the true plot and the null plots to show the distances for the lineups. 
\end{enumerate}

The empirical distribution of the distance measures is obtained by calculating the distances between the null plots among themselves. One null data is generated from the true data set using the null generating mechanism. Assuming this null data to be the ``true'' data set, a number of null data sets are obtained from this null data and the distances between these datasets are calculated. One single distance value is obtained by averaging all these distances. This process is repeated a large number of times, say, $N$ where $N$ is a large number of the order $10^3$ or $10^4$. Finally $N$ mean distances or average distances are obtained which gives the empirical distribution of the distance. 

The empirical distribution of the distance works as the $t$-distribution in the classical setting. In the classical setting, the test statistics follows a $t$-distribution under the null hypothesis. The observed test statistic is then compared to this distribution, as shown in Figure \ref{compare}. In visual inference, the mean distances of the null plots gives the empirical distribution. The mean distance of the true plot from the null plots in the lineup acts as the observed test statistic. Unlike the $t$-distribution, the empirical distribution is generally skewed.

The mean distance between the true plot and the null plots in a lineup of size $m = 20$ is calculated by averaging over the distances between the true plot and each of the  $(m - 1)$ null plots. The mean distances for the $(m - 1)$ null plots in the lineup are calculated by taking the mean of the distances of the particular null plot and the other $(m - 2)$ null plots. The mean distances for the true dataset and the null datasets are plotted on the empirical distribution. If the mean distance of the true plot is larger than any of the null plots, the lineup would be regarded as``easy''. Otherwise, it is a ``difficult'' lineup. 

\begin{figure}[hbtp]
\centering
\subfigure[]{
\includegraphics[scale=0.55]{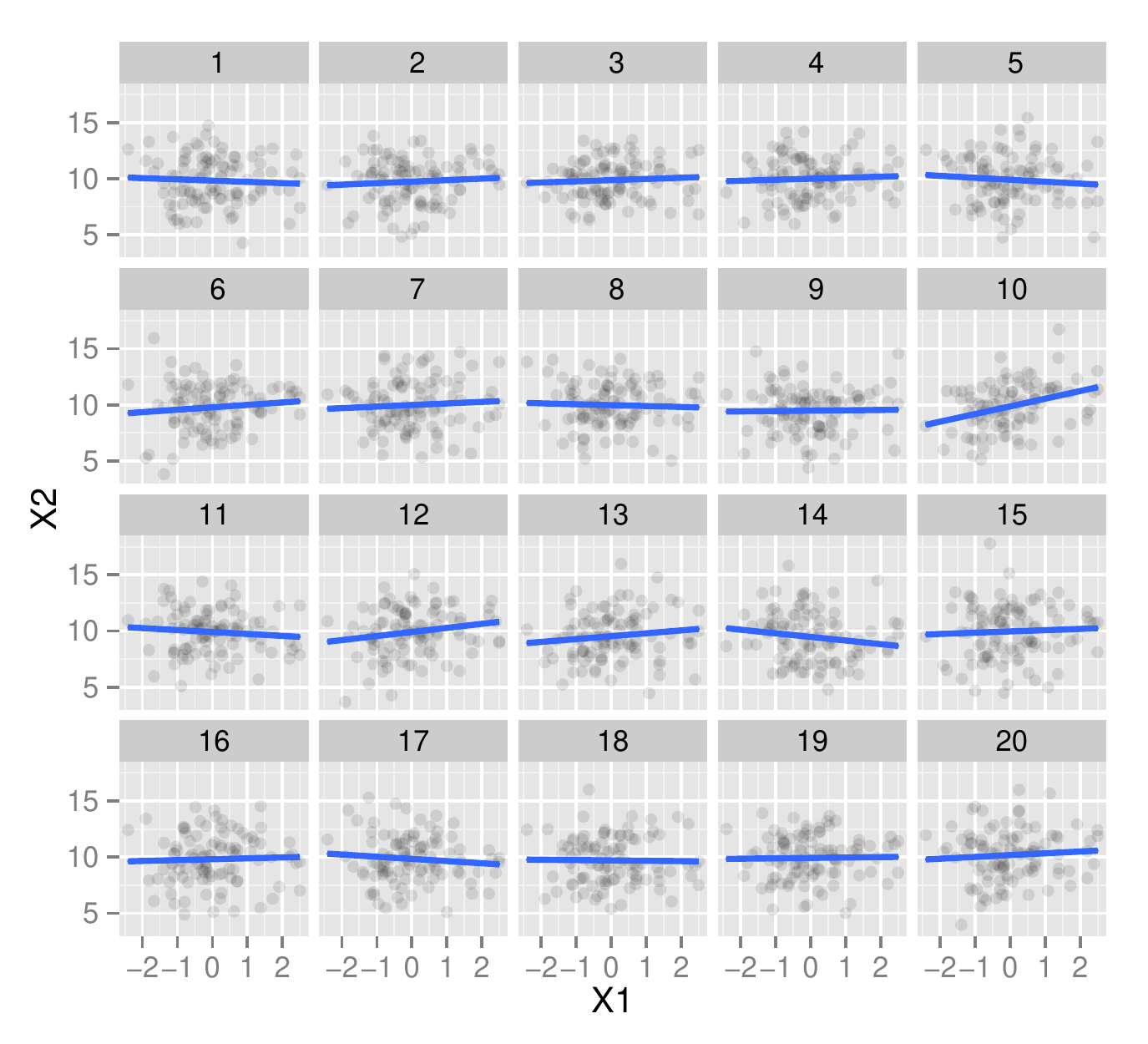}
\label{dist_1}
}
\subfigure[]{
\includegraphics[scale=0.7]{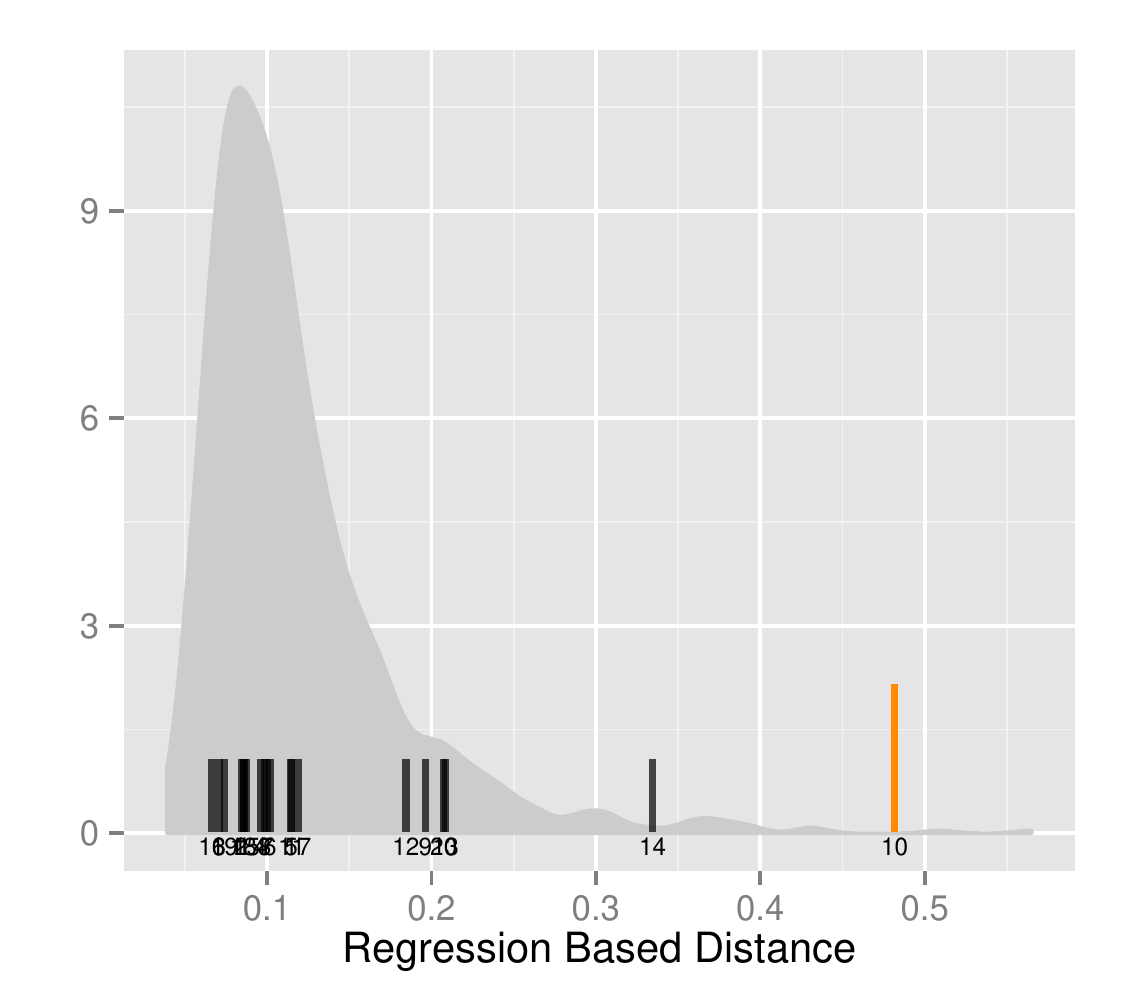}
\label{dist_2}
}
\label{dist}
	\vspace{-.1in}
\caption[Optional caption for list of figures]{Illustration of the behavior of a distance metric with a lineup plot in (a) and the distribution of regression based distance metric in (b). A lineup of size $m$ = 20 is shown (left) for testing whether there exists a significant linear relationship between $X_1$ and $X_2$. The 19 null plots are obtained by simulating from the null model.  The empirical distribution of the distance metric is shown (right). The distances for the true plot and the null plots are shown in orange and black respectively.  }
\end{figure}

The empirical distribution of the distance based on regression is shown in Figure \ref{dist}. To generate this distribution, $N = 1000$ and $m = 20$ was used. Figure \ref{dist_1} shows the lineup plot for $m = 20$ for testing whether there exists a significant linear relationship between $X_1$ and $X_2$. The 19 null plots are generated by fitting the null model and generating from the null model. Figure \ref{dist_2} shows the empirical distribution of the distance with the mean distances for the true plot (in orange) and the null plots (in black) for the particular. The true plot is easy to be identified in the lineup (Figure \ref{dist_1}). It can also be seen in Figure \ref{dist_2} as the orange line is extreme compared to the black lines. 

\begin{figure}[hbtp]
\centering
\subfigure[]{
\includegraphics[scale=0.55]{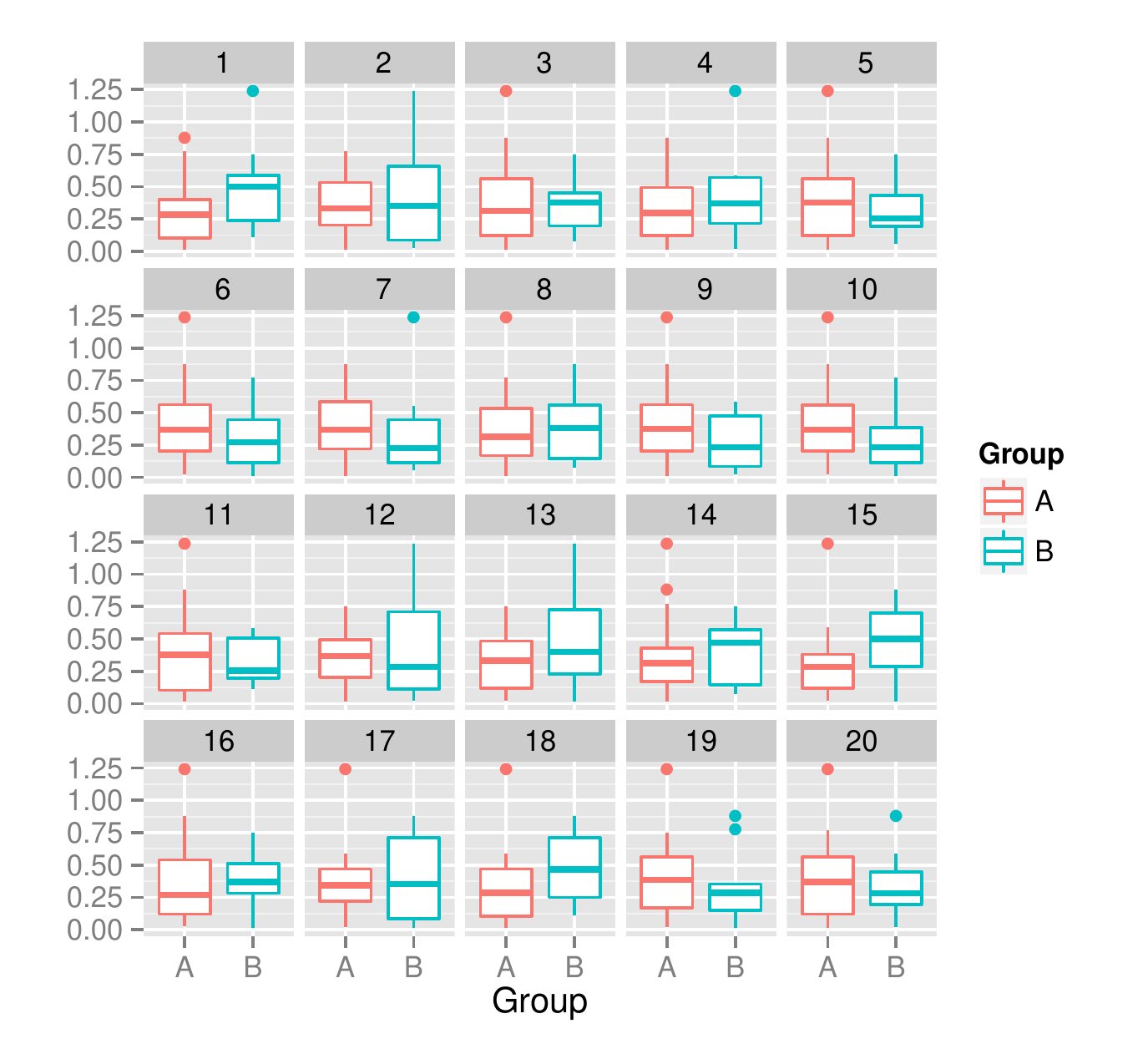}
\label{dist2_1}
}
\subfigure[]{
\includegraphics[scale=0.7]{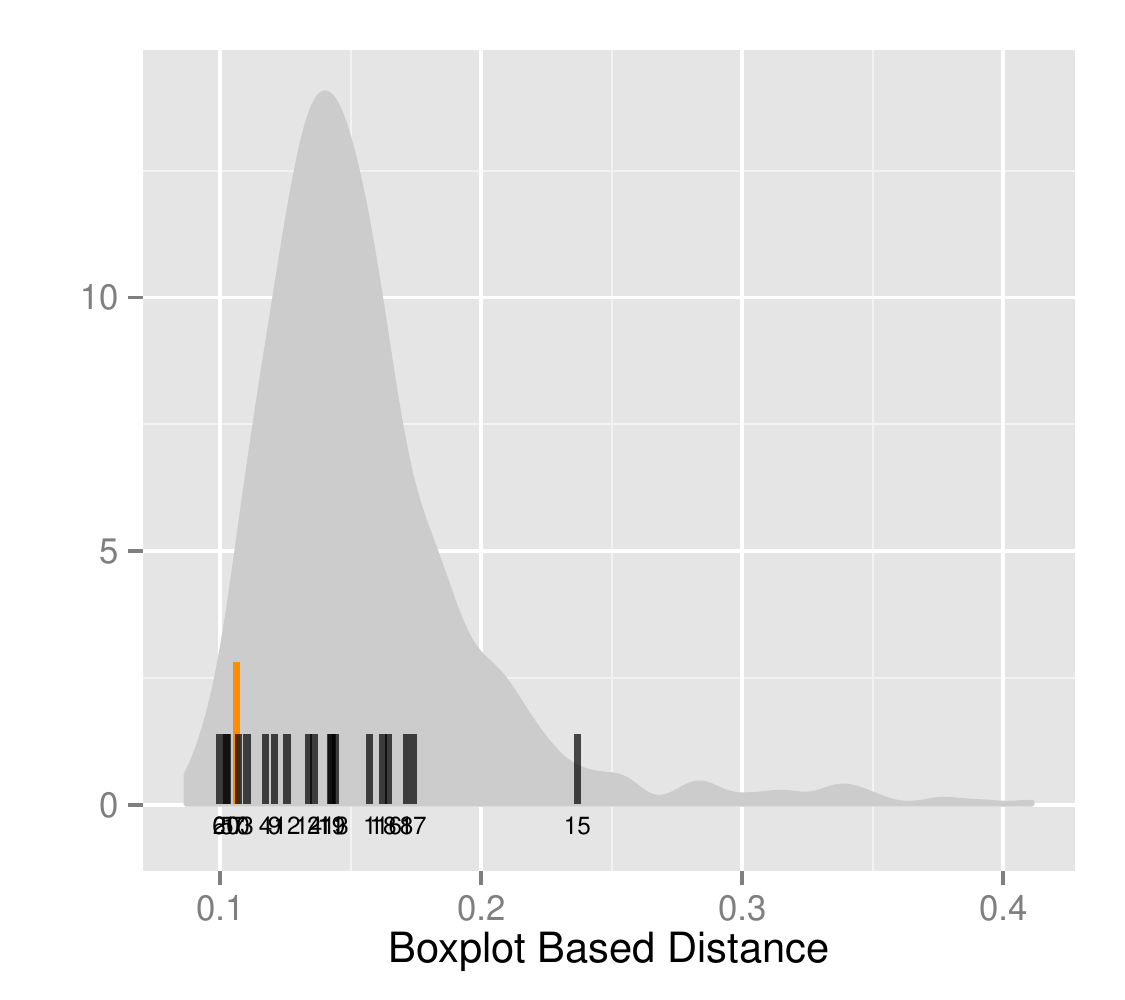}
\label{dist2_2}
}
\label{dist2}
	\vspace{-.1in}
\caption[Optional caption for list of figures]{Illustration of the behavior of a distance metric with a lineup plot in (a) and the distribution of boxplot based distance metric in (b). A lineup of size $m$ = 20 is shown (left) for testing whether there exists a significant linear relationship between $X_1$ and $X_2$. The 19 null plots are obtained by simulating from the null model.  The empirical distribution of the distance metric is shown (right). The distances for the true plot and the null plots are shown in orange and black respectively.}
\end{figure}

Figure \ref{dist2_1} shows the lineup plot for $m = 20$ for testing whether there exists a significant difference between the two groups A and B. The 19 null plots are generated by permuting the group variable keeping the other variable fixed. Figure \ref{dist2_2} shows the empirical distribution of the distance based on the boxplots with the mean distance for the true plot (in orange) and the null plots (in black). The true plot is hard to be identified from the lineup which is also evident in the distribution since many black lines are to the right of the orange line.

\section{Effect of Plot Type and Question of Interest } \label{sec:plot_type}

Previous studies have suggested that the type of plot used in the lineup have an effect on the response of the subjects \citep{zhao:2012}. For example the subjects find it easier to identify the true plot for a large sample data when a box plot is used in the lineup instead of a dot plot. Similarly the distance metric should also be altered according to the plot type. The distance metric should account for the additional information provided by the graphical elements in the lineup. The graphical elements, like the presence of a box or a regression line overlaid on a scatterplot may influence the response of the subject. Figure \ref{plottype} illustrates this idea. 

\begin{figure}[hbtp]
\centering
\subfigure[]{
\includegraphics[scale=0.55]{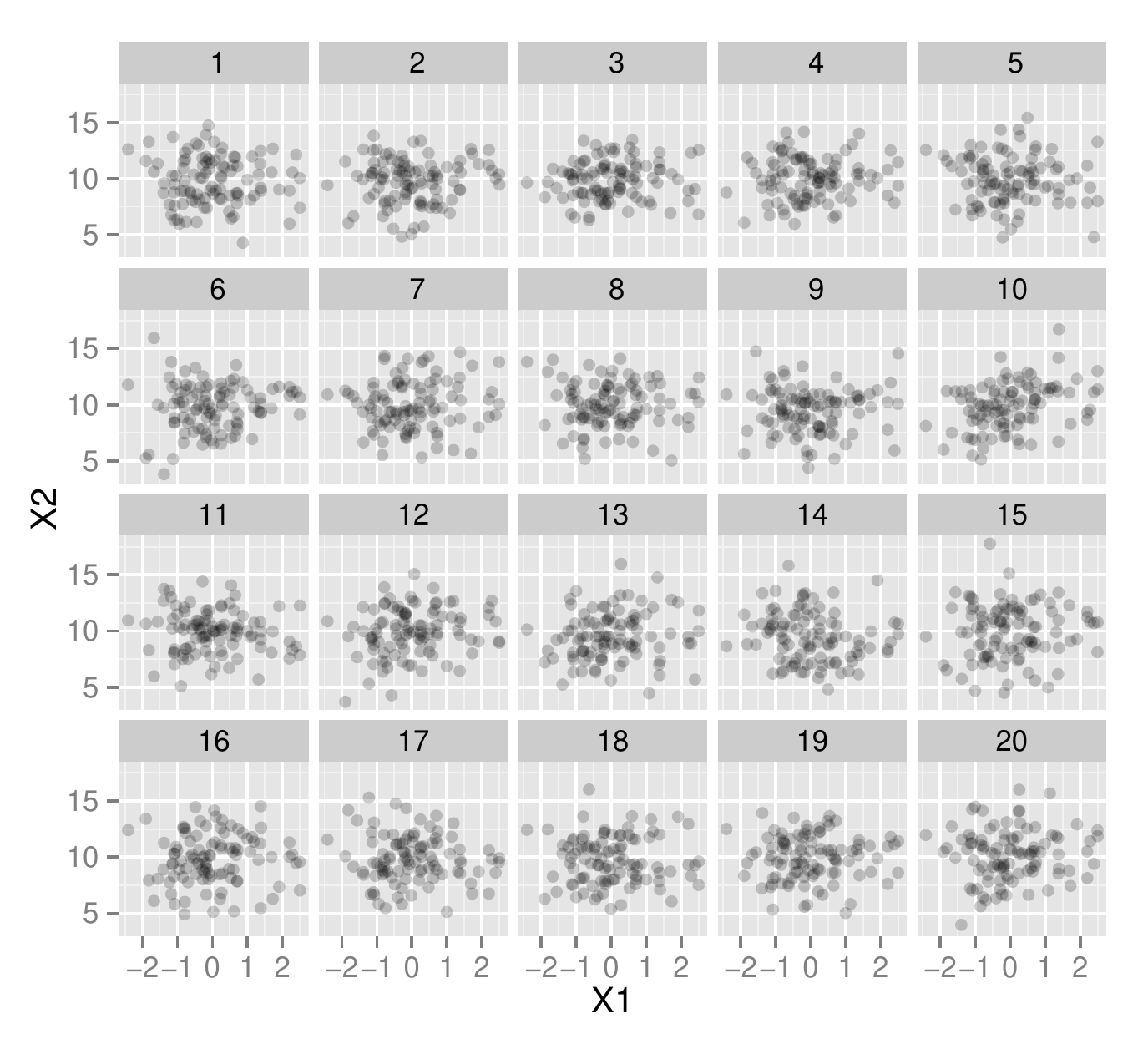}
\label{type_1}
}
\subfigure[]{
\includegraphics[scale=0.55]{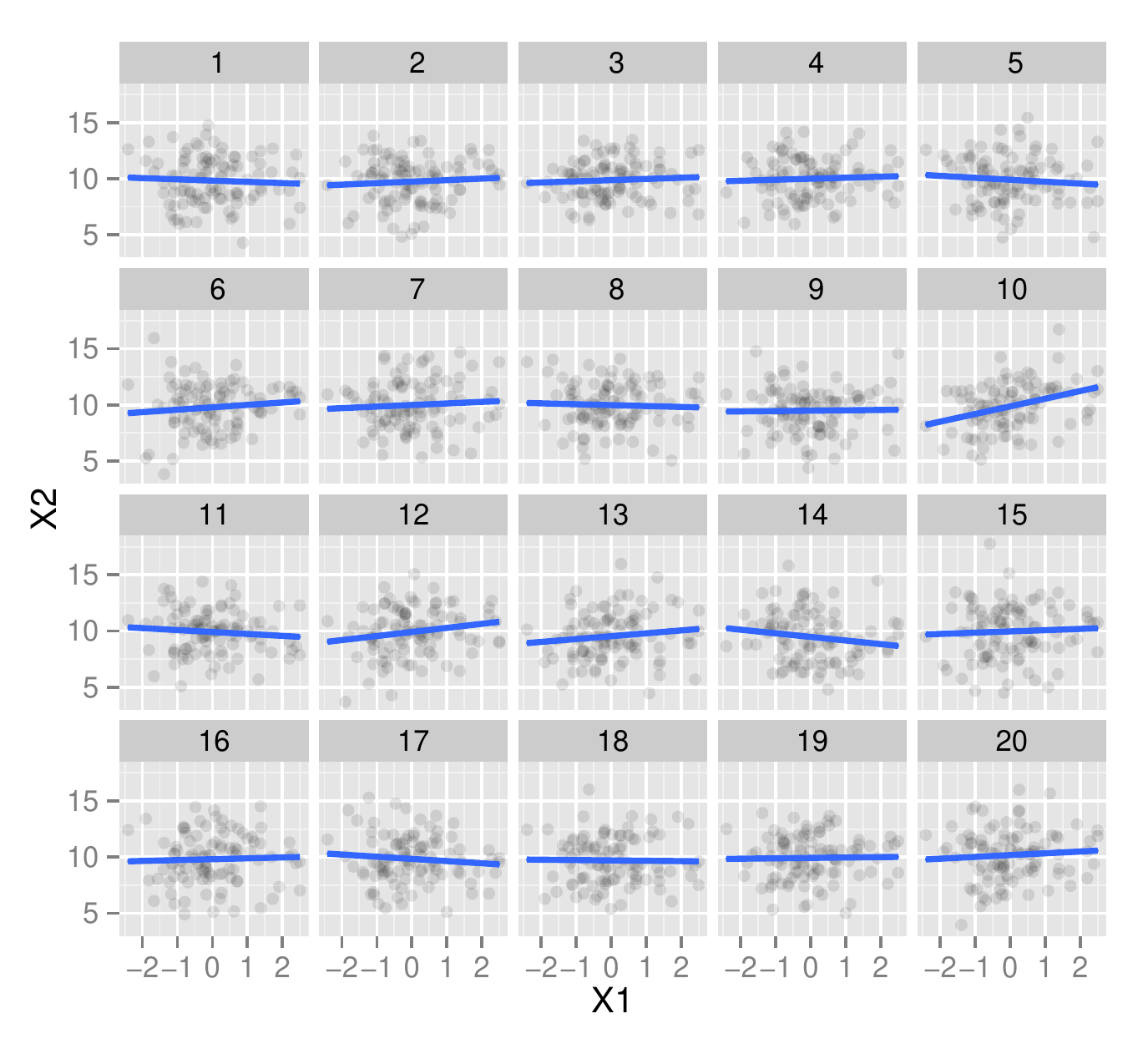}
\label{type_2}
}
	\vspace{-.1in}
\caption[Optional caption for list of figures]{Comparison of two lineups: scatterplots in (a) and scatterplots with a regression line overlaid in (b). The raw data is same for the lineups. The human subjects are shown the lineups and asked to identify the plot with the steepest slope. The presence of regression line may affect the decision of human subjects.  }
\label{plottype}
\end{figure}

Figure \ref{type_1} shows a lineup of scatterplots with 100 points between two variables $X_1$ and $X_2$. Figure \ref{type_2}, on the other hand, gives a lineup of the same scatterplots with the regression line overlaid. Showing Figure \ref{type_1}, if the subjects are asked to identify the plot which has the steepest slope, then the subjects probably will face some difficulty in identifying the true plot. But in Figure \ref{type_2}, the regression line overlaid makes it easier for the subjects to identify the true plot. A different distance metric should be used in each case to correctly measure the quality of the lineup.

The question asked to the subjects plays an important role to identify the true plot in the lineup. A minor change in the question can change the response of the subject. In Figure \ref{dist2_1}, if the subjects are asked to identify the plot in which the green group has a larger vertical difference than the red group, the subjects should pick Plot 6. If the subjects are asked which plot has the largest vertical difference between the two groups, the subjects should pick Plot 15. A distance metric should also take into account the question of interest. But, in general, the question of interest is which plot among these is different.

\section{Metric Evaluation} \label{sec:eval}

For a lineup of size $m = 20$, the distance for the true plot is compared to the 19 null plots. This comparison can sometimes complicate things. A logical solution can be to look at one statistic for one lineup. Such a statistic can be defined as the difference between the mean distance of the true plot and maximum of the mean distances for the null plots. Hence we define, 
\begin{enumerate}
\item Difference: the difference between the mean distance for the true plot and the maximum of the mean distances for the null plots. Mathematically,
$$\delta_{\hbox{lineup}} = \bar{d}_{\hbox{true}} - \max_j \bar{d}_{\hbox{null}_j}$$
for $j = 1, \dots, (m  - 1).$
 A positive difference would indicate that the mean distance of the true plot is greater than the maximum of the mean distances of the null plots. Hence the true plot is extreme compared to all the null plots. Similarly a negative difference indicates that there is at least one null plot which is extreme compared to the true plot based on the distance.
 
The issue with this statistic is that $\delta_{\hbox{lineup}}$ indicates an ``easy" or ``difficult" lineup only on the basis of whether it is positive or negative, although it may be really close to 0. The statistic does not imply how many null plots are more extreme than the true plot. So we define,
\item Larger than the true plot: the number of null plots which have larger mean distances than the mean distance of the true plot is noted. Mathematically,
 $$\gamma_{\hbox{lineup}} = \sum_{j = 1}^{m - 1} a_j$$ where 
\begin{equation}
a_j =
\begin{cases}
1 & \text{if } \bar{d}_{\hbox{null}_j} > \bar{d}_{\hbox{true}} ,
\\
0 & \text{otherwise}
\end{cases}
\end{equation}
$\gamma_{\hbox{lineup}}$ takes all values between 0 and $(m - 1)$. A large value of this measure would indicate that there are a number of null plots more extreme than the true plot and hence it is hard to identify the true plot in the lineup.
\end{enumerate}


\section{Results} \label{sec:results}


The performance of the distance metrics was evaluated with comparing the distances with the response of the subjects. A number of experiments were done in Amazon Mechanical Turk \citep{turk}. Subjects were recruited through Amazon Mechanical Turk \citep{turk} and were shown a sequence of lineups. In each experiment, they were asked specific questions. Their responses were recorded along with other demographic informations. The details about the design of experiments can be found in \cite{majumder:2011} and \cite{roychowdhury:2013}. 

\subsection{Turk Experiment 1 -- Side by Side Boxplots}

\begin{figure}[htbp]
\centering
\includegraphics[width=.95\textwidth]{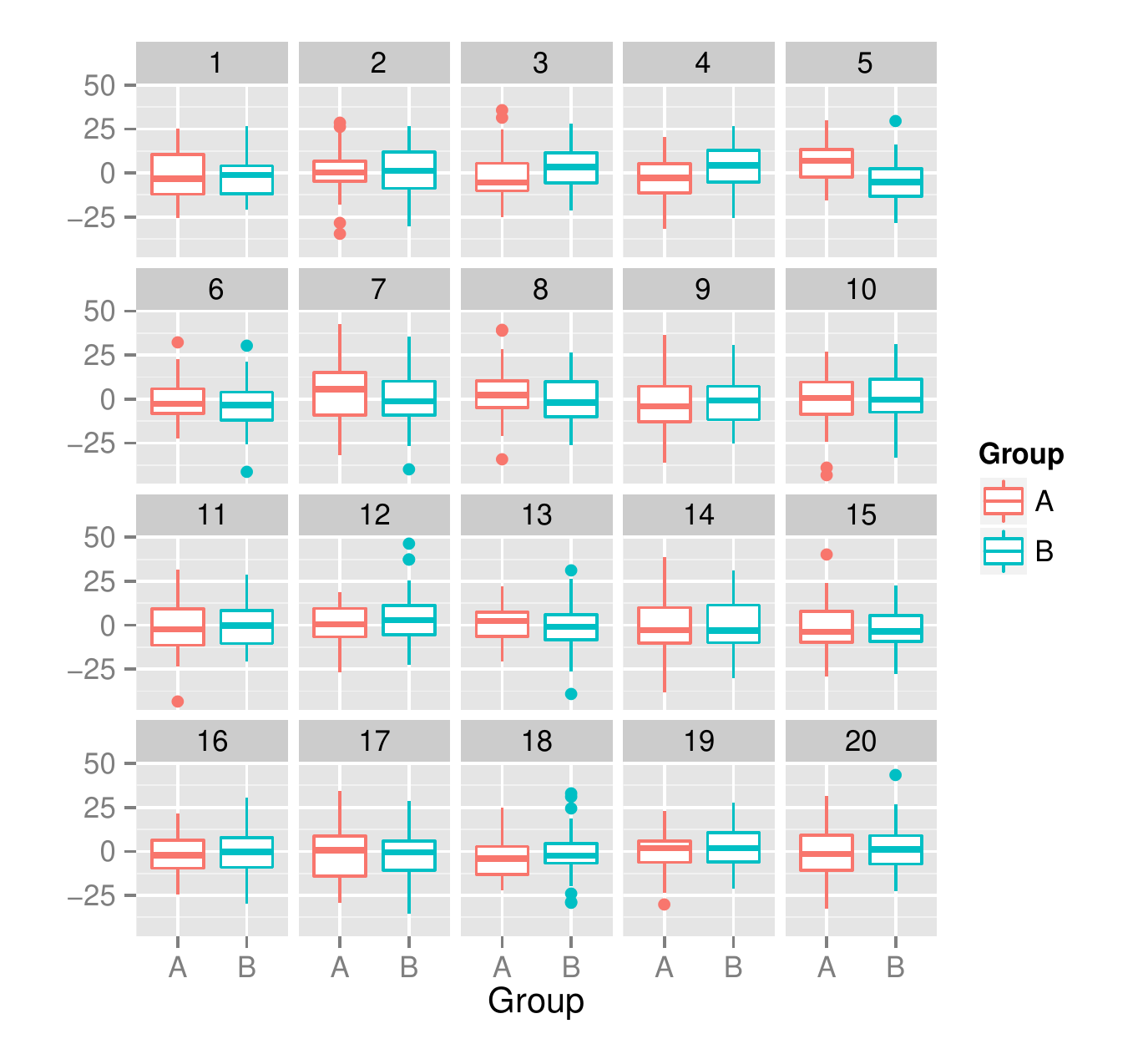}
\caption{An example lineup from Turk Experiment 1. The lineup has $m = 20$ plots of which one is the observed data plot and the remaining $m - 1$ are the null plots generated assuming that the null hypothesis is true. Subjects were asked to identify the plot which has the largest vertical difference between the two groups. Can you identify the observed data plot?}
\label{turk1}
\end{figure}

In this experiment, all the lineups generated had a side by side boxplot as the test statistic. Assuming that the null hypothesis is true, the null plots were generated by assuming that there is no difference between the two distributions. The subjects were shown a few lineups and were asked to identify the plot which has the largest vertical difference between group 1 and group 2. Figure \ref{turk1} gives such a lineup. \\

\begin{figure}[hbtp]
\centering
\subfigure[]{
\includegraphics[scale=0.75]{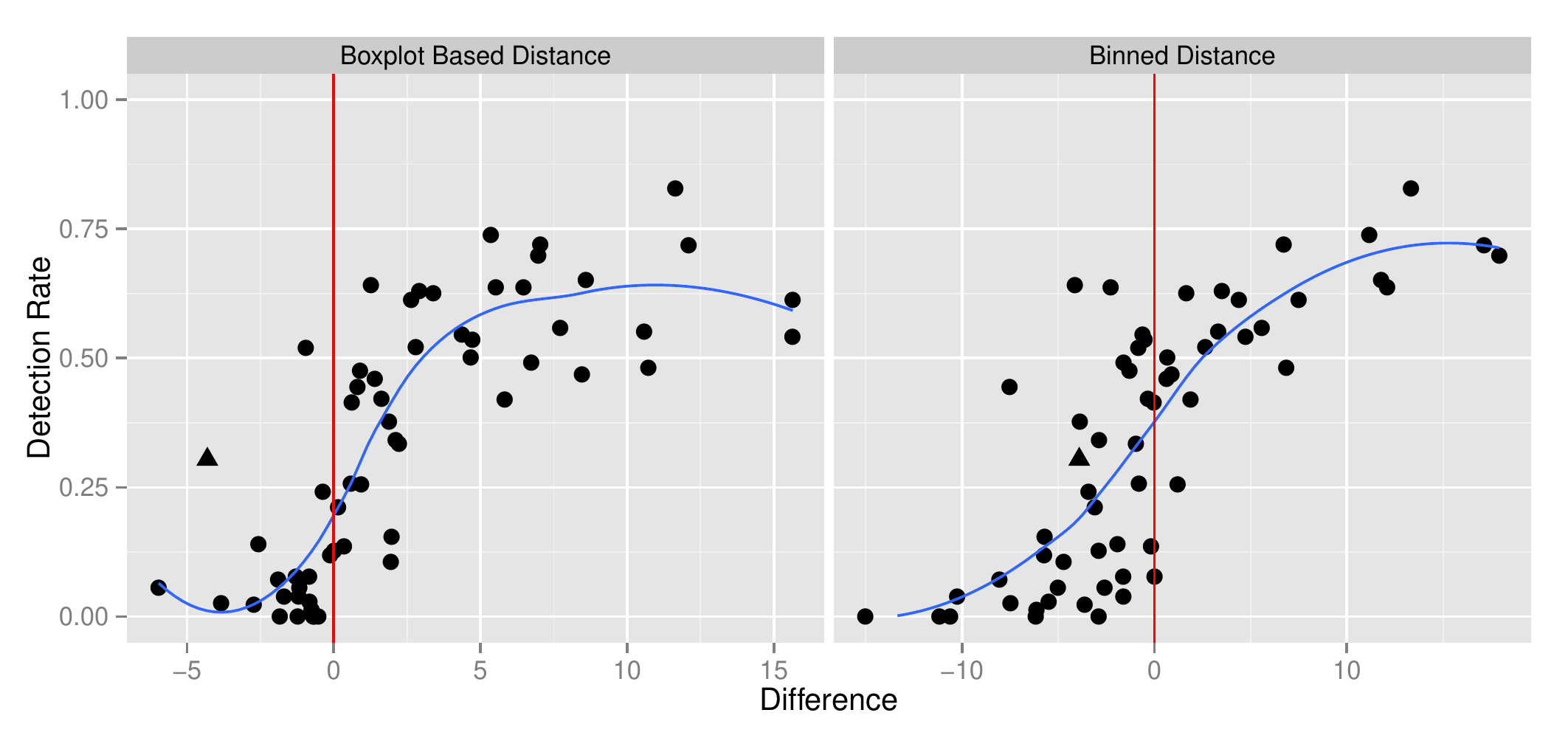}
\label{t1comp_1}
}
\subfigure[]{
\includegraphics[scale=0.75]{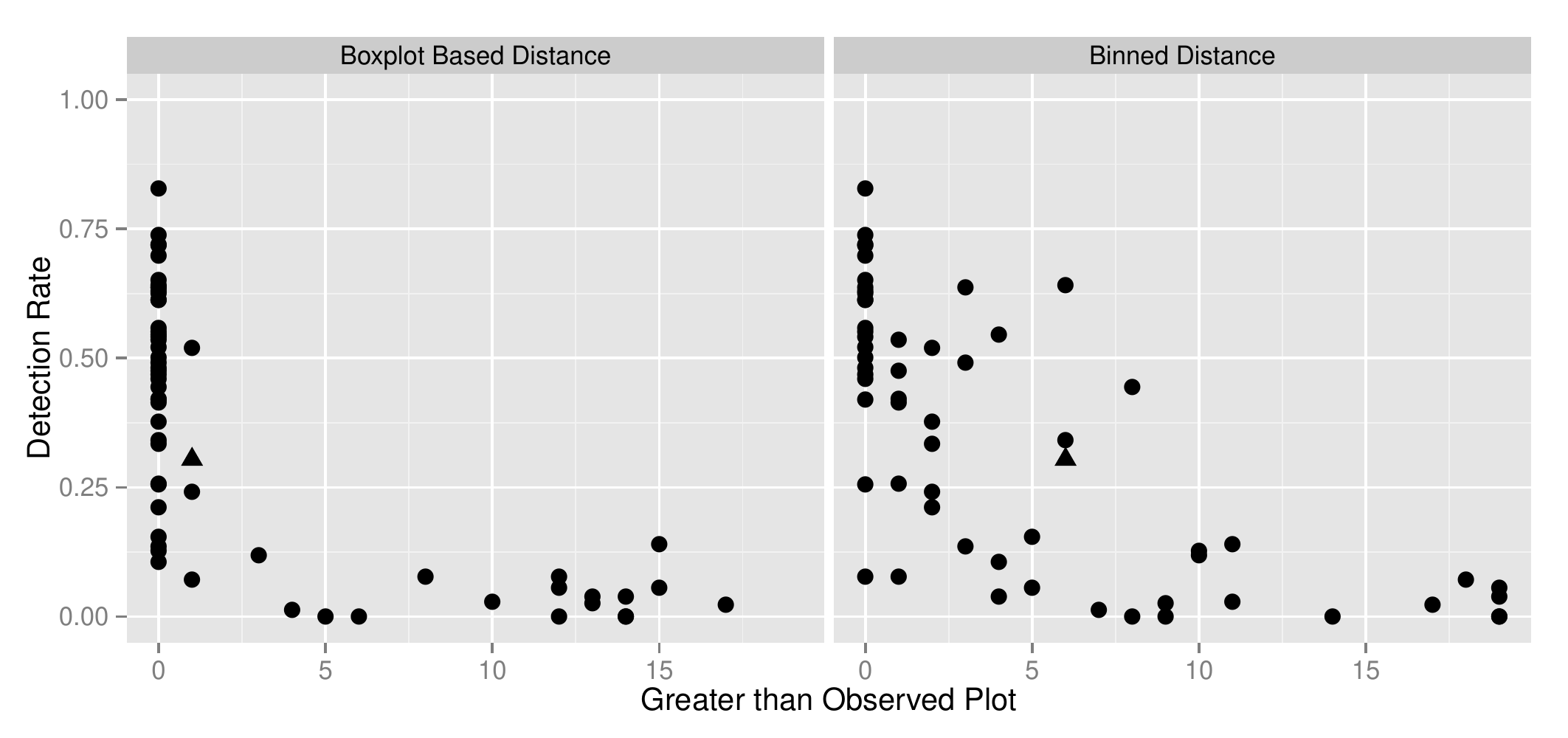}
\label{t1bin_1}
}
	\vspace{-.1in}
\caption[Optional caption for list of figures]{Comparison of distance metrics for side-by-side boxplots. Detection Rate (a) and the number of plots greater than the observed (b) are plotted against the difference based on the boxplot and binned distance. The vertical line represents the difference equal to 0 when there is at least one null plot similar to the observed plot. The detection rate increases with the difference. As the number of plots with distance greater than the observed increases, the detection rate decreases.  The triangle represents a lineup which has high detection rate but negative difference. This particular lineup is examined in Figure \ref{turk1-exp}. }
\label{turk1comp}
\end{figure}

The response of the subjects were noted and the proportion of correct response was calculated for each lineup. The distances between the plots in each lineup were computed using both the distance based on boxplots ($d_{\hbox{box}}$) and the binned distance ($d_{\hbox{bin}}$). The mean distance for the true plot and the null plots were calculated and $\delta_{\hbox{lineup}}$ and $\gamma_{\hbox{lineup}}$ are obtained. The proportion of correct response was plotted against each of the two statistics. Figure \ref{turk1comp} shows the detection rate against the difference for $d_{\hbox{box}}$ and $d_{\hbox{bin}}$ and the number of null plots greater than the observed plot for the two distance measures.

In Figure \ref{turk1comp}, the detection rate is plotted against the difference. The red vertical line represents difference equal to 0 indicating that the mean distance of the true plot is equal to the maximum of the mean distance of the null plots i.e. the mean distance of the true plot is equal to at least one of the mean distance of the null plots. It can be seen that as the difference increases, the detection rate increases. So the subjects do better in the easier lineups than the hard ones. The binned distance was calculated using 8 bins on both the axes. Figure \ref{turk1comp} also shows the relation between detection rate and the number of null plots larger than the true plot. It can be seen that as there are more extreme null plots compared to the observed plot, the subjects find it difficult to pick the observed plot. It is interesting to see that the subjects can pick the observed plot with one or two extreme null plots. 

Though the distance based on the boxplots works better, the binned distance does a decent job in this case. According to the binned distance, there are a few lineups which has a negative difference but the proportion correct is above 60\%, which can be also be seen in Figure \ref{turk1comp}. It should be noted that the binned distance does not take into account the graphical elements of the plot (e.g. boxplot) and calculates the distance solely based on the data. So an outlier may have a huge effect on the binned distance but does not effect the distance based on the boxplots. Hence it is advisable to use a distance based on the graphical elements since that is exactly what the subjects look at in the lineup.

\begin{figure}[hbtp]
\centering
\includegraphics[scale=0.75]{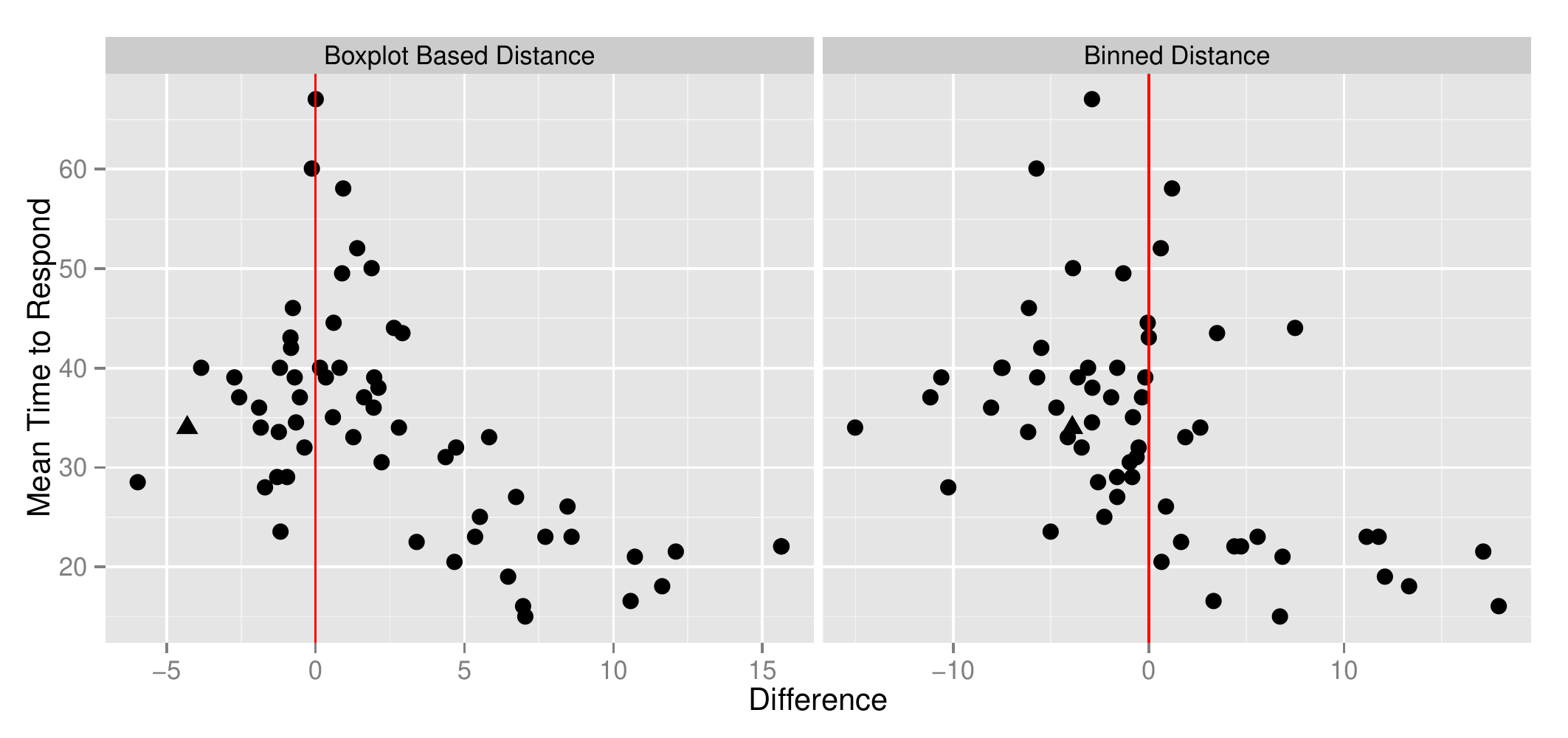}
	\vspace{-.1in}
\caption[Optional caption for list of figures]{Comparison of distance metrics for side-by-side boxplots. Mean time to respond is plotted against the difference based on the boxplot and binned distance. The vertical line represents the difference equal to 0 when there is at least one null plot similar to the observed plot. The mean time taken decreases with the difference. The triangle represents a lineup which is examined in Figure \ref{turk1-exp}. }
\label{turk1-mtime}
\end{figure}

The time taken to respond by the subjects is another measure of difficulty of the lineups. Due the presence of some huge outliers, the mean time taken by the subjects for each lineup is looked at and plotted against the difference for both the distance measures. Figure \ref{turk1-mtime} shows the plots. It can be clearly seen that when the difference is below 0, there is no real trend in the mean time and there is a huge variability, indicating that the time taken depends on the subjects. But when the difference is above 0, the mean time decreases rapidly as the difference increases. Hence the subjects can pick the true plot quickly if the true plot is extreme compared to the null plots.
 

\begin{figure}[hbtp]
\centering
\subfigure[]{
\includegraphics[scale=0.7]{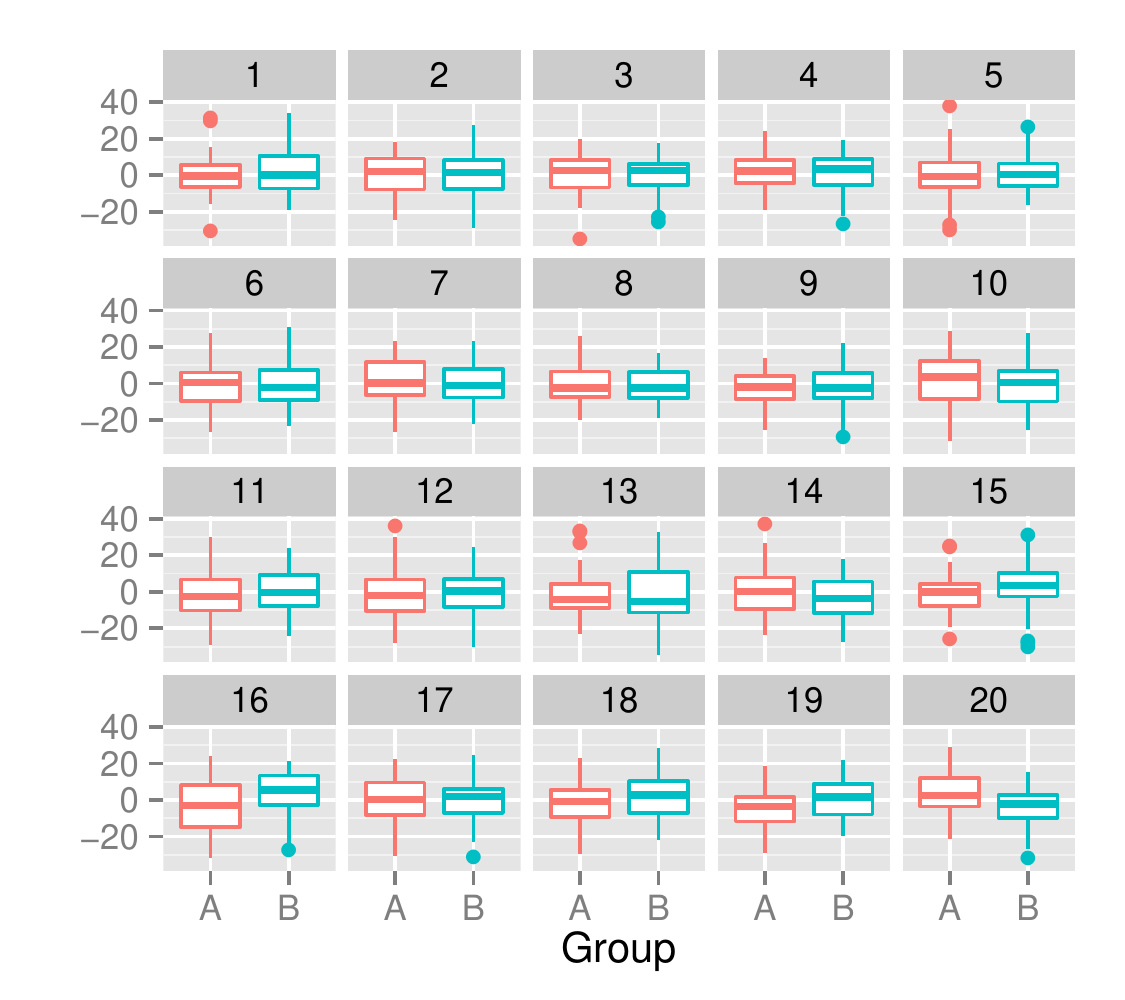}
\label{t1comp_1}
}
\subfigure[]{
\includegraphics[scale=0.7]{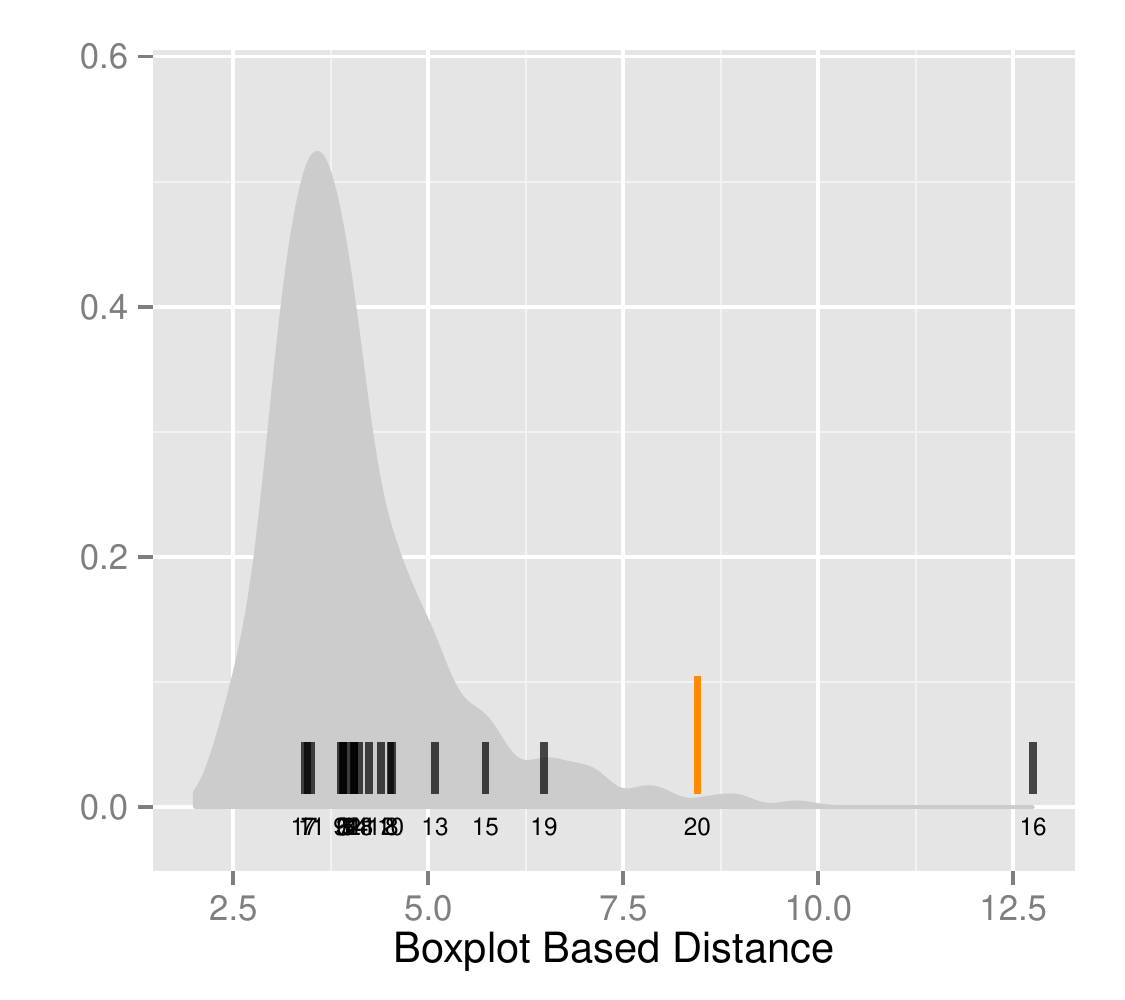}
\label{t1bin_1}
}
\subfigure[]{
\includegraphics[scale=0.7]{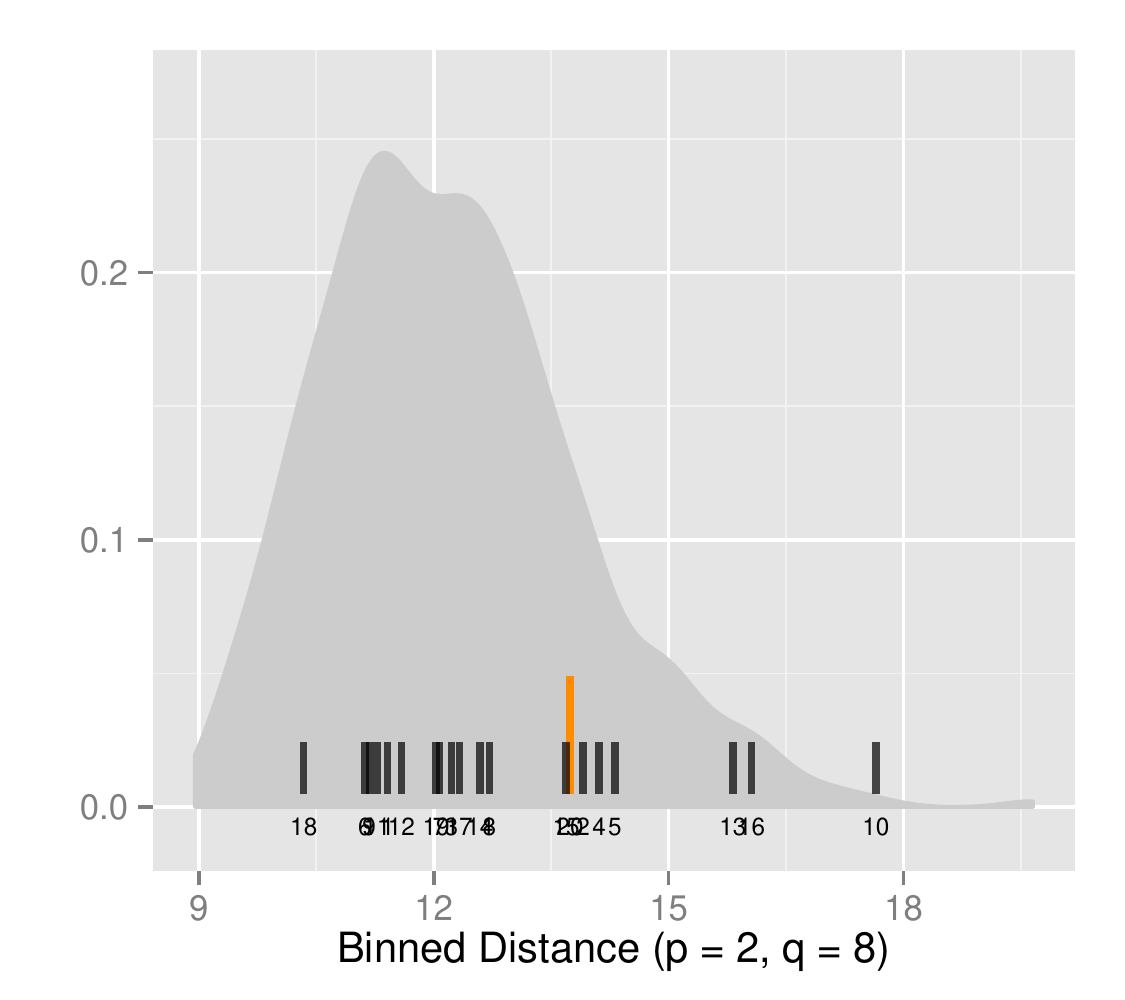}
\label{t1bin_1}
}
\subfigure[]{
\includegraphics[scale=0.7]{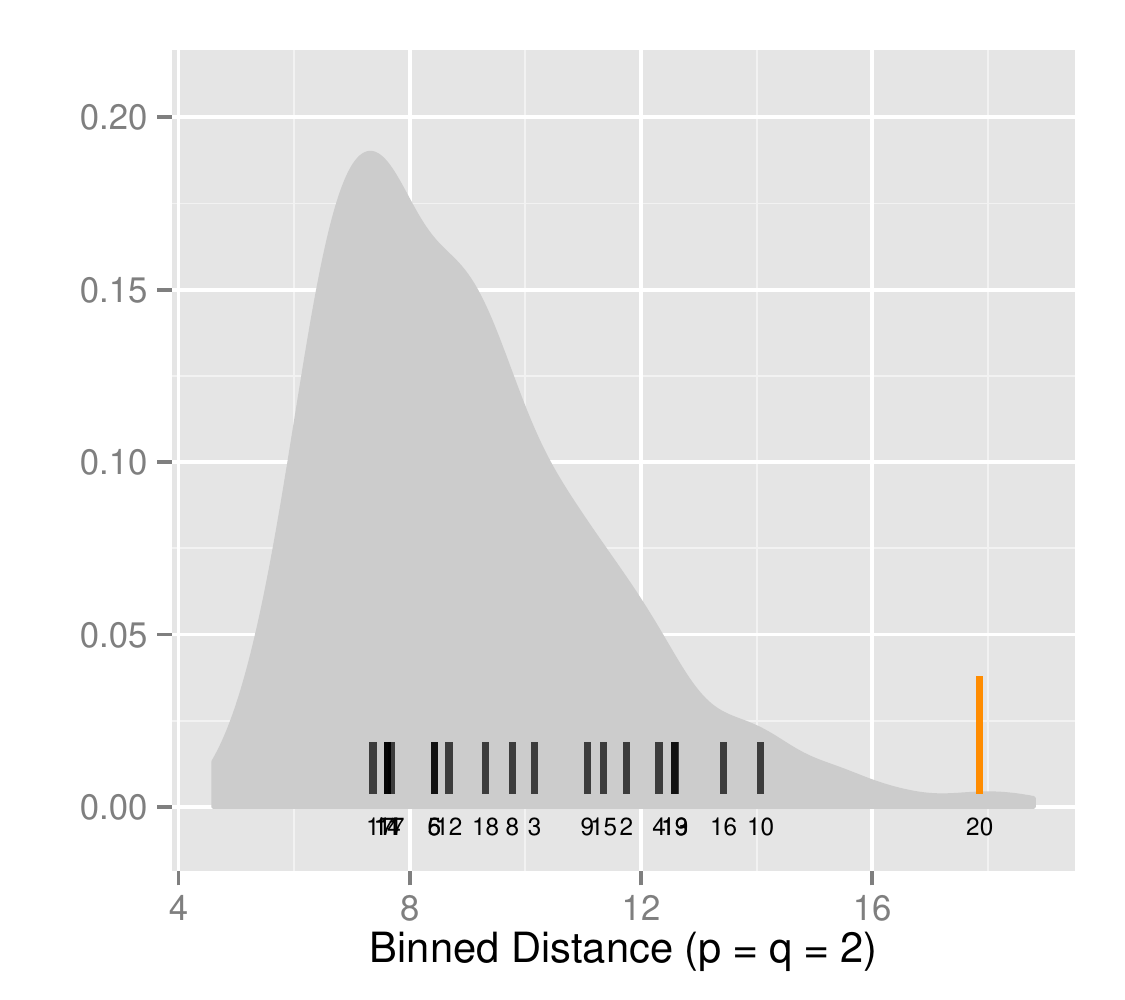}
\label{t1bin_1}
}
	\vspace{-.1in}
\caption[Optional caption for list of figures]{Illustration of the behavior of the different distance metrics. The lineup is shown in (a) and the distributions of different distance metrics based on this lineup is shown in the other plots: boxplot based distance in (b), binned distance with 2 and 8 bins on x and y axis in (c) and binned distance with 2 bins in both axes in (d). The lineup corresponds to the point marked with a triangle in difference vs. detection rate plot in Figure \ref{turk1comp}. }
\label{turk1-exp}
\end{figure}

It can be noticed in Figure \ref{turk1comp} that for some of the lineups, the detection rate is high but the difference using distance metric is negative suggesting that the lineup is difficult. One such lineup is marked using a triangle in Figure \ref{turk1comp}. It would be interesting to look into the lineup closely to identify what made the people pick the actual plot as different. Figure \ref{turk1-exp} shows the lineup and the distribution of the distance metrics. 

The lineup in Figure \ref{turk1-exp} is a lineup of side-by-side boxplots. The observed data plot is Plot 20 but there are other candidates who can be picked easily. Plot 19 and Plot 16 seems to have large differences between the quartiles. Specifically in Plot 16, the difference between the  
first quartiles for the two groups is very large but the differences between the medians and the third quartiles are small. The huge difference of the first quartiles may have affected the huge mean distance of Plot 16 from all the other plots.   

\subsection{Turk Experiment 2 -- Scatterplots with an Overlaid Regression Line}

\begin{figure}[htbp]
\centering
\includegraphics[width=.95\textwidth]{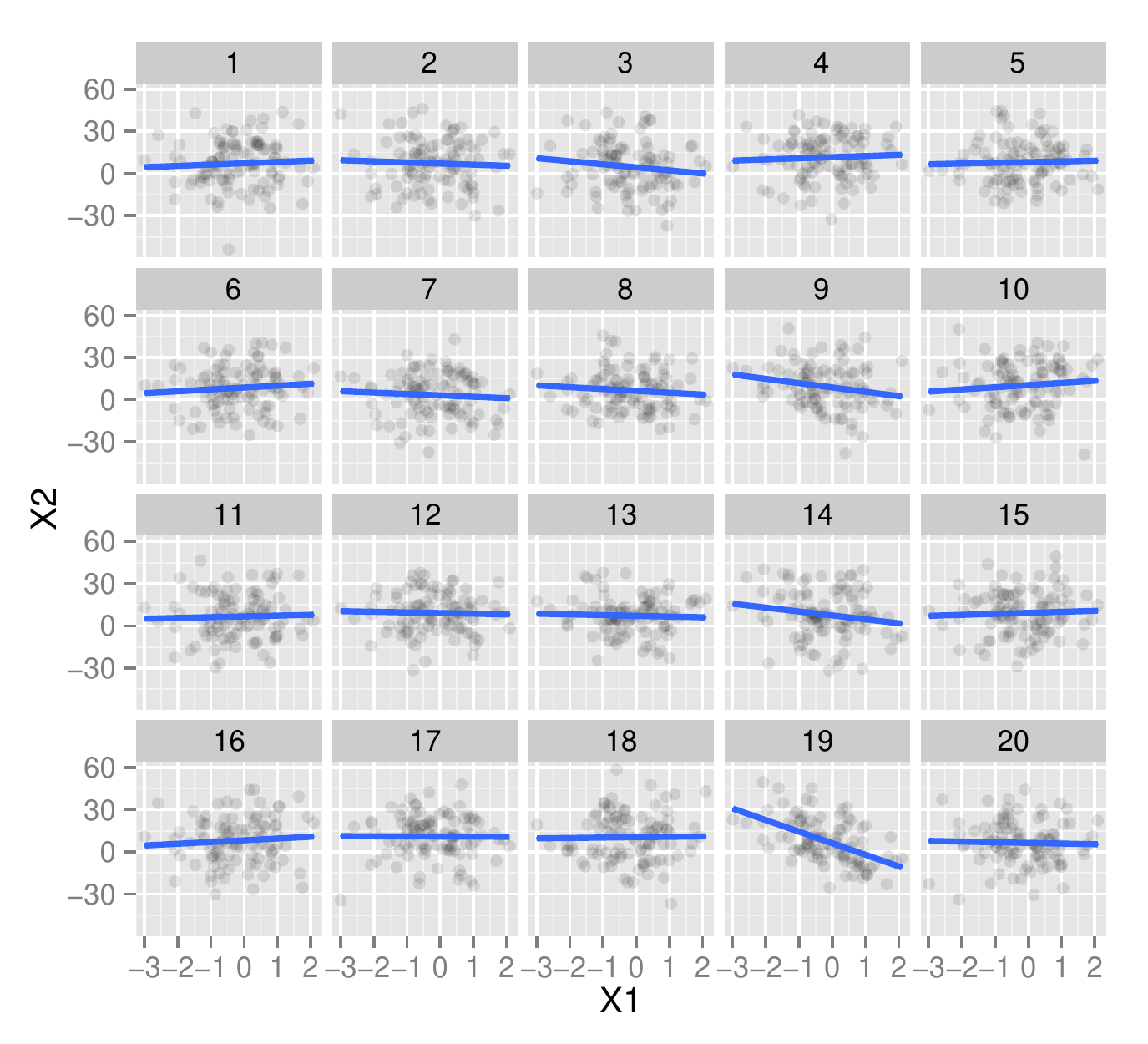}
\caption{An example lineup from Turk Experiment 2. In this lineup, one of the plots is the observed plot and the other 19 plots are the null plots generated assuming that the null hypothesis $H_o : \beta = 0$ is true. Subjects were asked to identify the plot with the steepest slope. Can you identify the observed plot ?}
\label{turk2}
\end{figure}

In this experiment, the test statistic is a scatterplot with the regression line overlaid. Assuming that the null hypothesis is true, the null plots are generated by assuming that there is no significant linear relationship between the two variables. The subjects were shown a few lineups and were asked to identify the plot which has the steepest slope. Figure \ref{lineup-example,dist,turk2} show example lineups. 

\begin{figure}[hbtp]
\centering
\subfigure[]{
\includegraphics[scale=0.75]{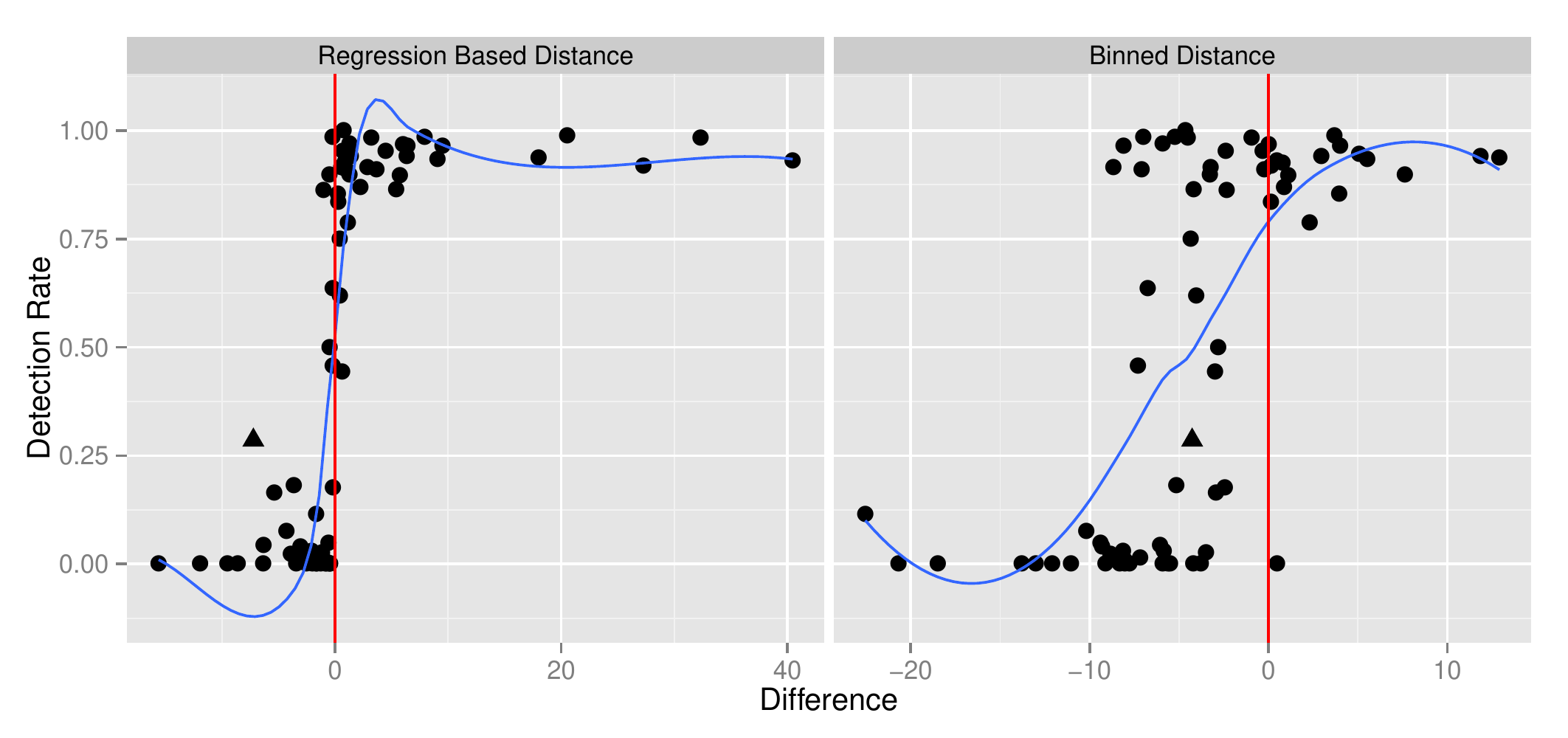}
\label{t2comp_1}
}
\subfigure[]{
\includegraphics[scale=0.75]{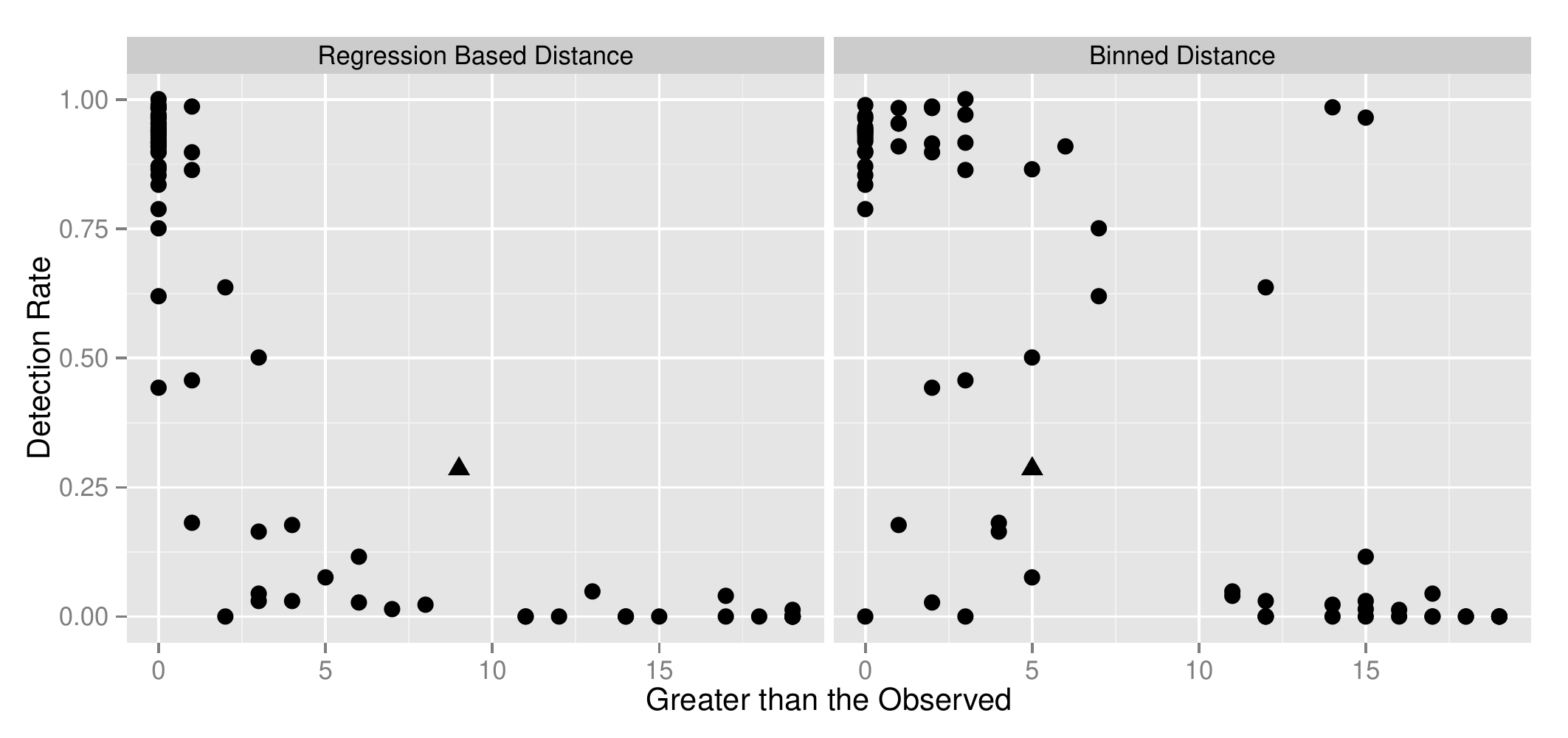}
\label{t2comp_2}
}
	\vspace{-.1in}
\caption[Optional caption for list of figures]{Comparison of distance metrics for scatteplots with a regression line overlaid. Detection Rate (a) and the number of plots greater than the observed (b) are plotted against the difference based on the regression based and binned distance. The vertical line represents the difference equal to 0 when there is at least one null plot similar to the observed plot. The detection rate increases with the difference. As the number of plots with distance greater than the observed increases, the detection rate decreases.  The triangle represents a lineup which has high detection rate but negative difference. This particular lineup is examined in Figure \ref{turk2-exp}.}
\label{turk2comp}
\end{figure}

The distances between the plots in this experiment were computed using both the distance based on regression line ($d_{\hbox{reg}}$) and the binned distance ($d_{\hbox{bin}}$) with a small number of bins. The proportion of correct response for each lineup was calculated from the response of the subjects and plotted against $\delta_{\hbox{lineup}}$ and $\gamma_{\hbox{lineup}}$.  Figure \ref{turk2comp} shows the results for the distance based on the regression line and the binned distance against $\delta_{\hbox{lineup}}$.

Figure \ref{turk2comp} shows the detection rate against the difference. The vertical line represents difference equal to 0. It can be seen that as the difference increases, the detection rate increases. So the subjects do better in the easier lineups than the hard ones. The distance based on regression works well in capturing the complexity of the lineups. The difference is positive for lineups with large differences and negative for lineups with small difference. A few lineups have difference close to zero for which the detection rate is close to 50\%.

But the binned distance fails terribly. Although the detection rate increases with difference, the detection rate is high for values with negative difference. This is a classic scenario where a graphical element affects the response. The presence of the overlaid regression line on almost transparent points of the scatterplot affected the response of the subjects. One other reason may be the use of the same number of bins (2 $\times$ 2) in this case for all the lineups. The relationship may improve if different number of bins can be used for different lineups.

Figure \ref{turk2comp} also shows the detection rate against the number of null plots greater than the observed plot. As the number of plots greater than the observed plot increases, the detection rate decreases.  Hence as there are more extreme null plots compared to the observed plot, the subjects find it difficult to pick the observed plot. For a few lineups, almost all the subjects identify the observed plot although there is one more extreme null plot. Though from Figure \ref{turk2comp}, it can be seen that the extremeness is marginal in most cases. 

\begin{figure}[hbtp]
\centering
\includegraphics[scale=0.75]{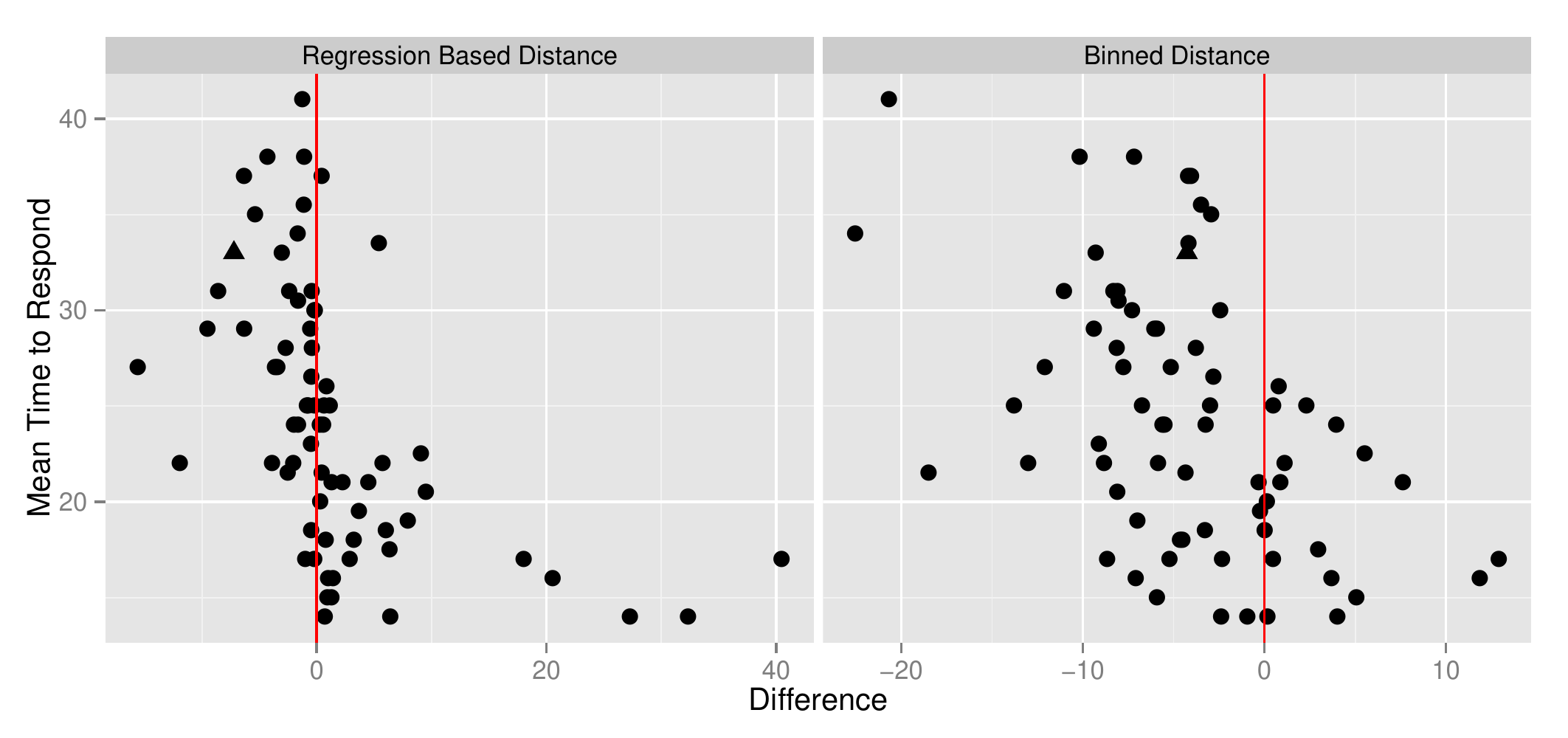}
	\vspace{-.1in}
\caption[Optional caption for list of figures]{Comparison of distance metrics for scatterplots with a regression line over laid. Mean time to respond is plotted against the difference based on the regression based and binned distance. The vertical line represents the difference equal to 0 when there is at least one null plot similar to the observed plot. The mean time taken decreases with the difference. The triangle represents a lineup which is examined in Figure \ref{turk2-exp}.   }
\label{turk2-mtime}
\end{figure}

Figure \ref{turk2-mtime} shows the relationship between the mean time taken to respond and the difference for both the distances. It can be clearly seen that there is a strong negative association showing that as the difference increases, the subjects take less time to respond. Also the variability of the mean time is higher for smaller difference. In case of binned distance, the relationship is negative though the variability is higher for the above mentioned reasons.

\begin{figure}[hbtp]
\centering
\subfigure[]{
\includegraphics[scale=0.55]{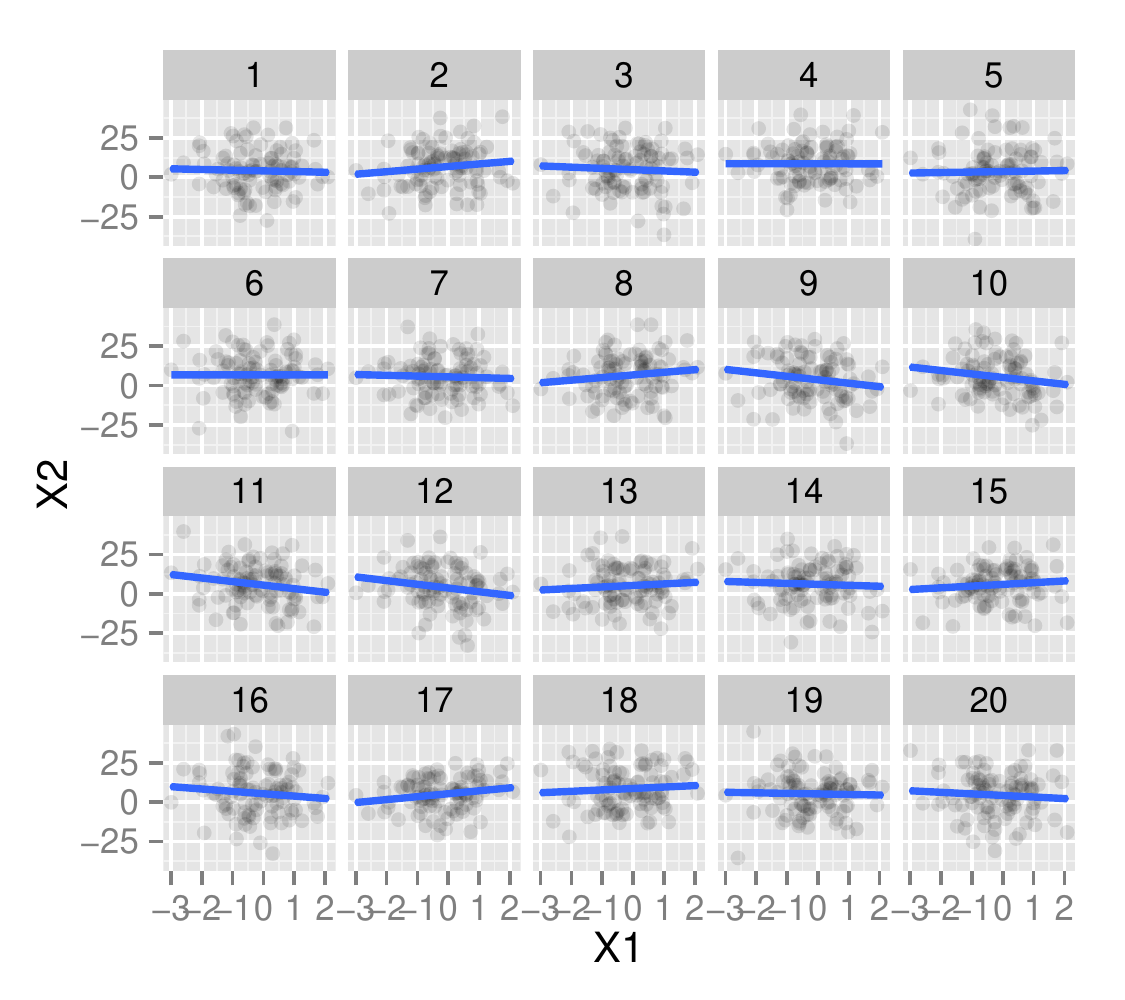}
\label{t2comp_1}
}
\subfigure[]{
\includegraphics[scale=0.55]{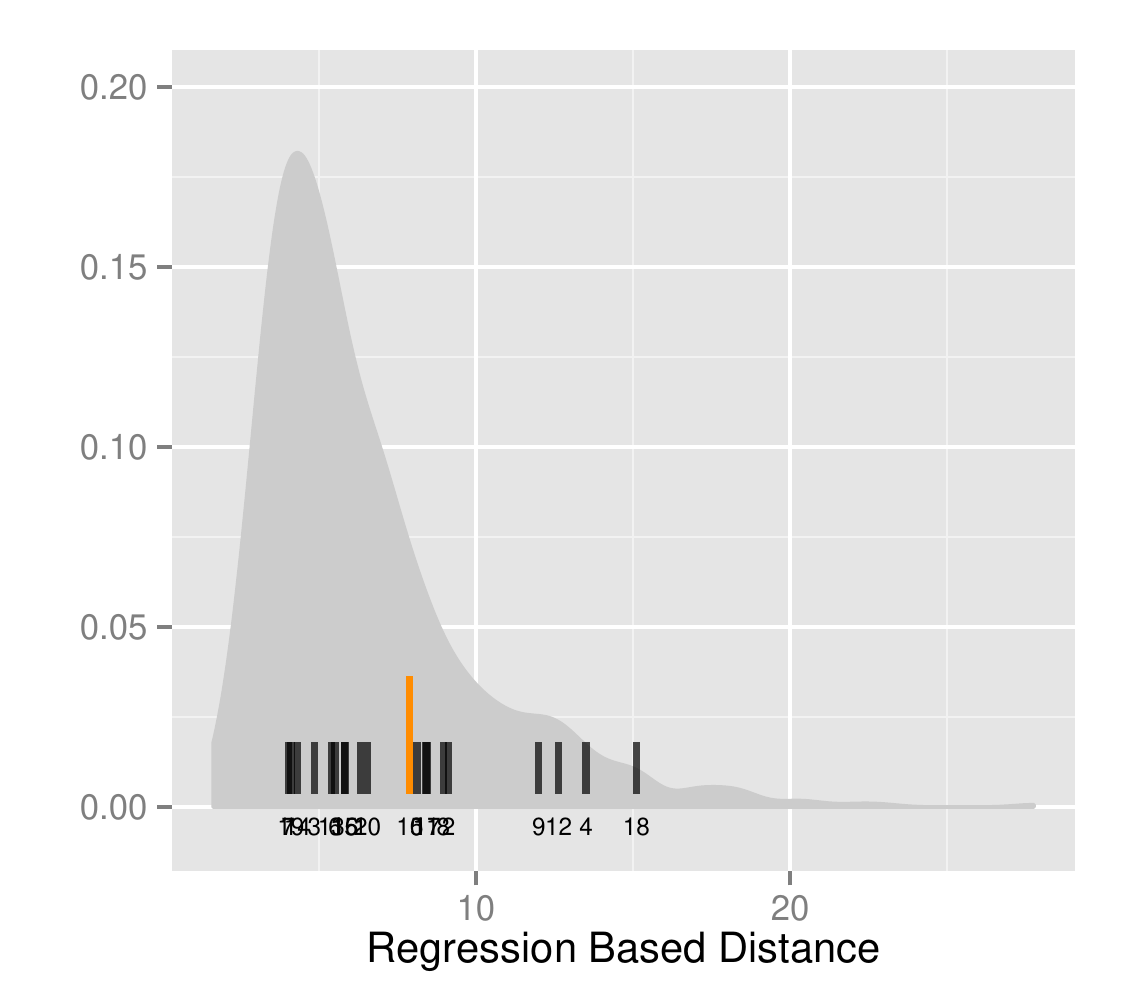}
\label{t2comp_1}
}
\subfigure[]{
\includegraphics[scale=0.55]{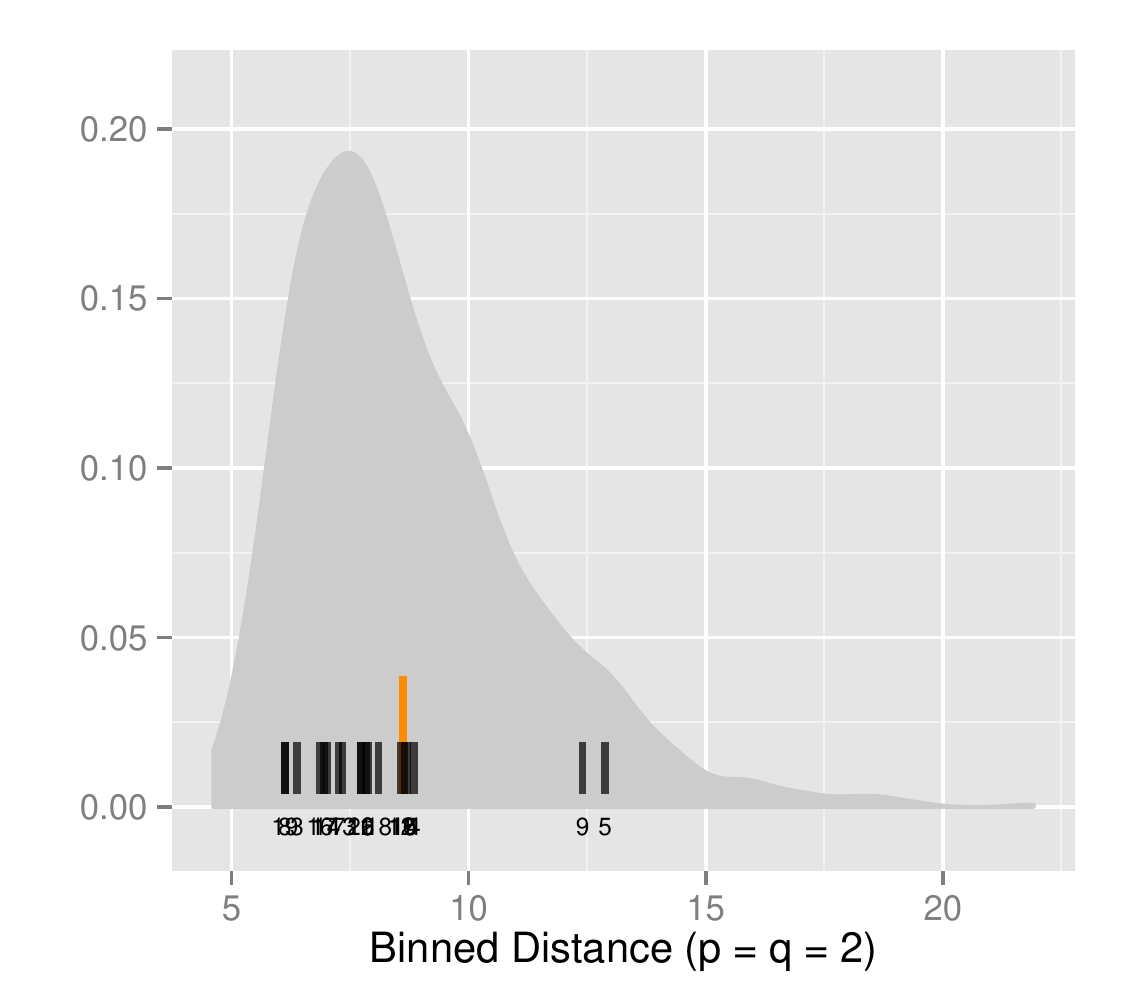}
\label{t2comp_2}
}
\subfigure[]{
\includegraphics[scale=0.55]{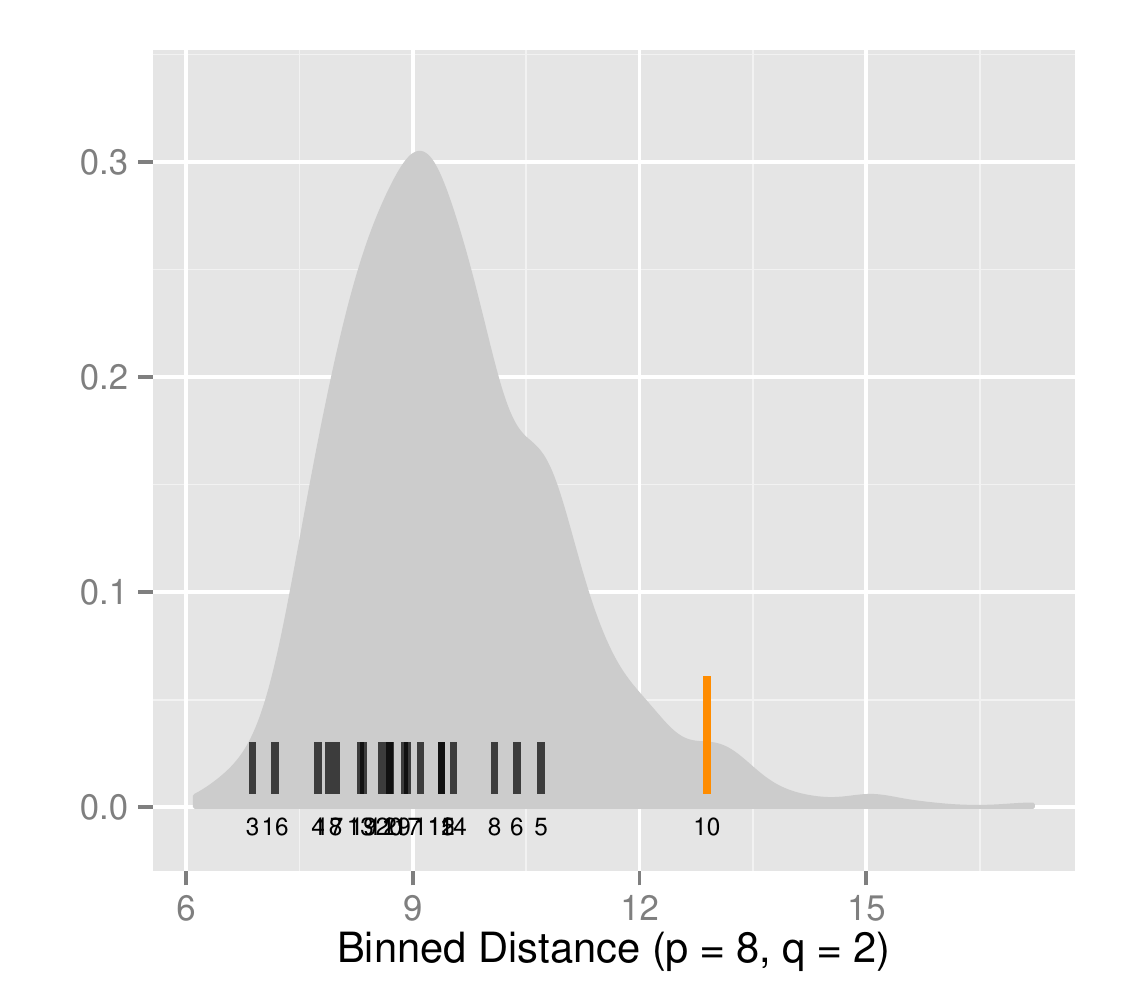}
\label{t2comp_1}
}
\subfigure[]{
\includegraphics[scale=0.55]{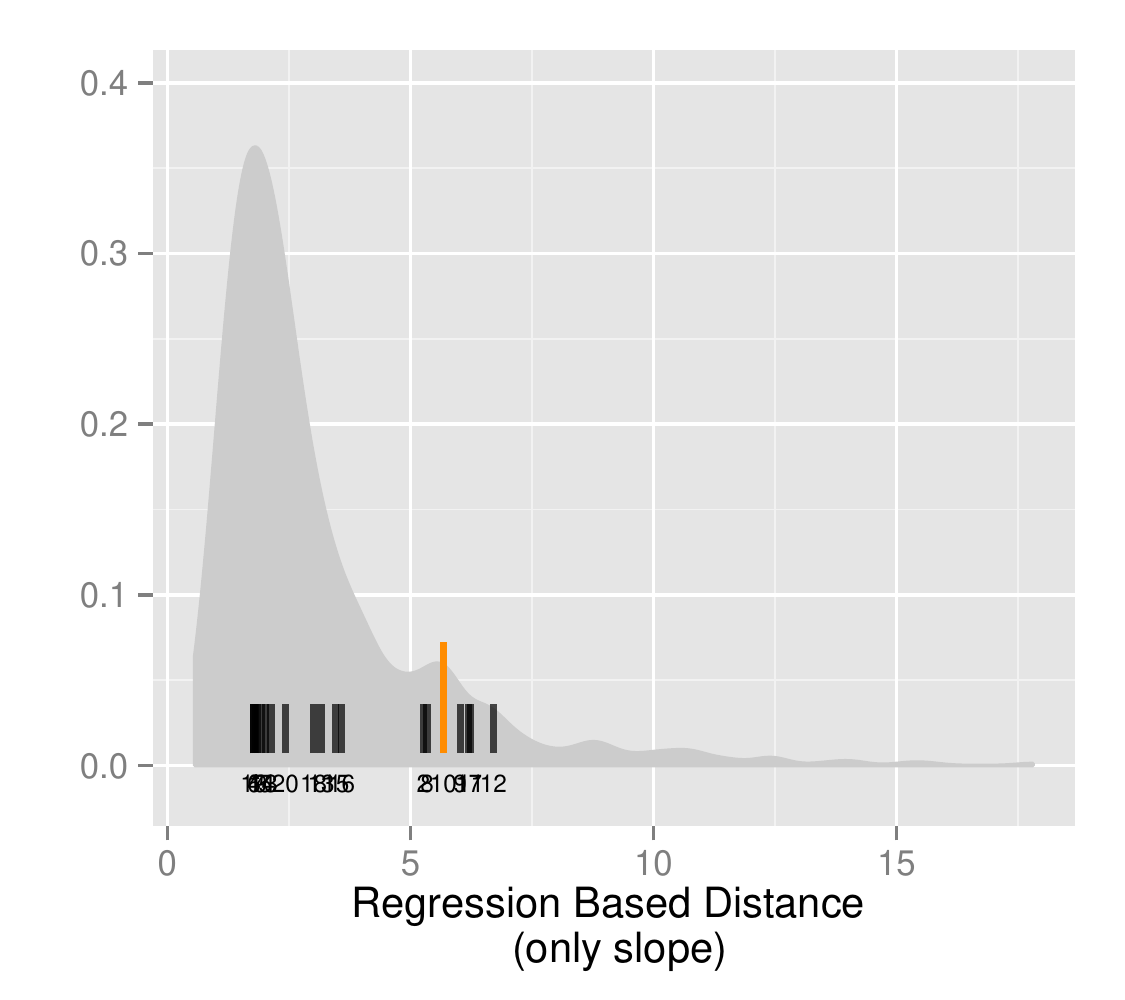}
\label{t2comp_1}
}
\subfigure[]{
\includegraphics[scale=0.55]{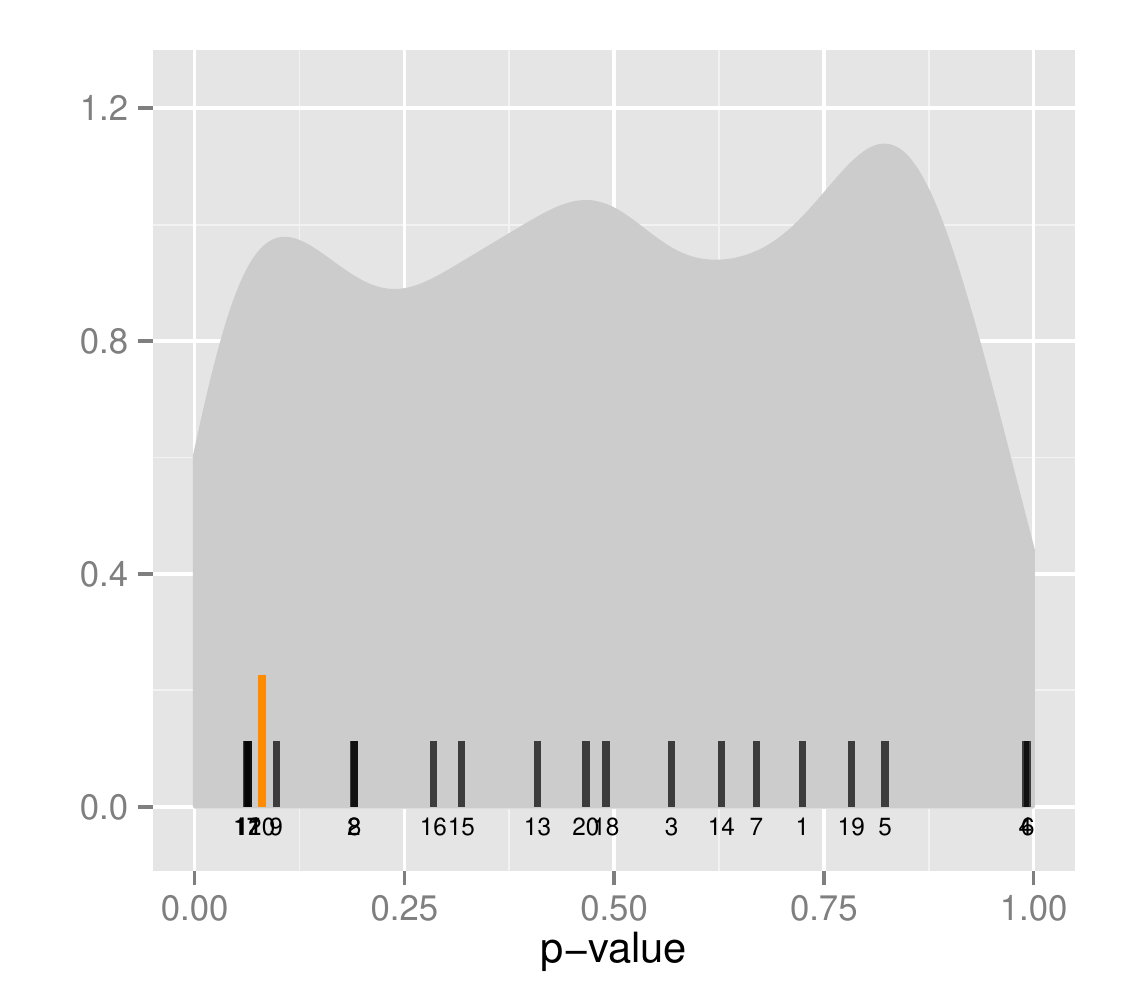}
\label{t2comp_1}
}
	\vspace{-.1in}
\caption[Optional caption for list of figures]{Illustration of the behavior of different distance metrics. The lineup is shown in (a) and the distributions of different distance metrics using this lineup is shown in the other plots ((b) - (e)): regression based distance in (b), binned distance with 2 bins on each axes in (c), binned distance with 8 and 2 bins in x and y axis respectively in (d) and regression distance based only on slope in (e). In (f), the distribution of the conventional $p$-values are plotted with $p$-values for the lineups marked on the distribution. The lineup corresponds to the point marked with a triangle in difference vs. detection rate plot in Figure \ref{turk2comp}.}
\label{turk2-exp}
\end{figure}

Although the regression based distance seems to efficiently identify the quality of the lineup, there is one lineup (marked by a solid triangle in Figure \ref{turk2comp}) which had a negative difference although people identified the true plot with reasonable success.  Figure \ref{turk2-exp} shows the lineup and the distributions of different distance metrics. 

The lineup in Figure \ref{turk2-exp} is a difficult one as suggested by the distribution of the distance metrics based on regression. Although around 28\% of the people identified the true plot correctly, the conventional $p$-value for testing the slope equal to 0 is 0.085, which shows that the relationship is not significant. The binned distance with 2 bins on each axes also shows the same. However the binned distance using the optimal number of bins (8 on the x-axis and 2 on the y-axis) by the optimal number of bins selection method identifies the true plot as different from the others. 

%

\subsection{Turk Experiment 7 -- Large $p$, Small $n$ Data}

\begin{figure}[htbp]
\centering
\includegraphics[width=.95\textwidth]{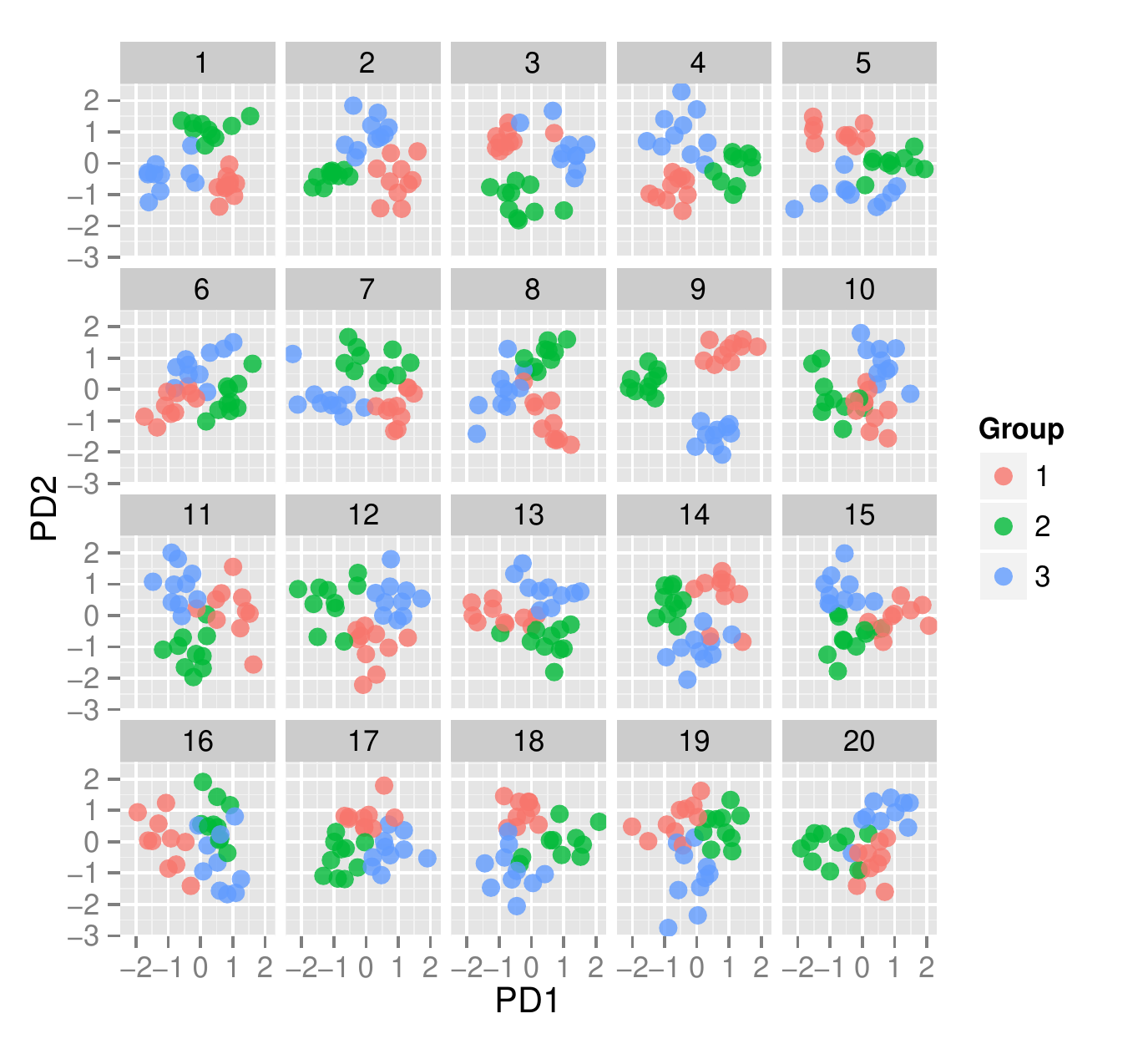}
\caption{An example lineup from Turk Experiment 7. Here two dimensional projections of the PDA index are plotted for a data with separation having $p = 20$ dimensions and $n = 30$ observations. The subjects were asked to identify the plot with the most separated colors. Can you identify the observed data plot?}
\label{largep}
\end{figure}

The motivation behind this experiment is to study the effect of large dimensions in a data with complete noise and some real separation. Data was simulated with different dimensions and fixed sample size. Data was divided into two or three groups. A projection pursuit with Penalized Discriminant Analysis Index was used and the one and two dimensional projections were obtained. The one or two dimensional projections were then plotted which resulted in the observed data plot. To generate the null data, the group variable in the data was permuted and the projection pursuit was applied. The subjects were shown these lineups and were asked to identify the plot with the most separated colored groups. Figure \ref{largep} gives an example of such a lineup with two dimensional projections with 3 colored groups. 

\begin{figure}[hbtp]
\centering
\subfigure[]{
\includegraphics[scale=0.75]{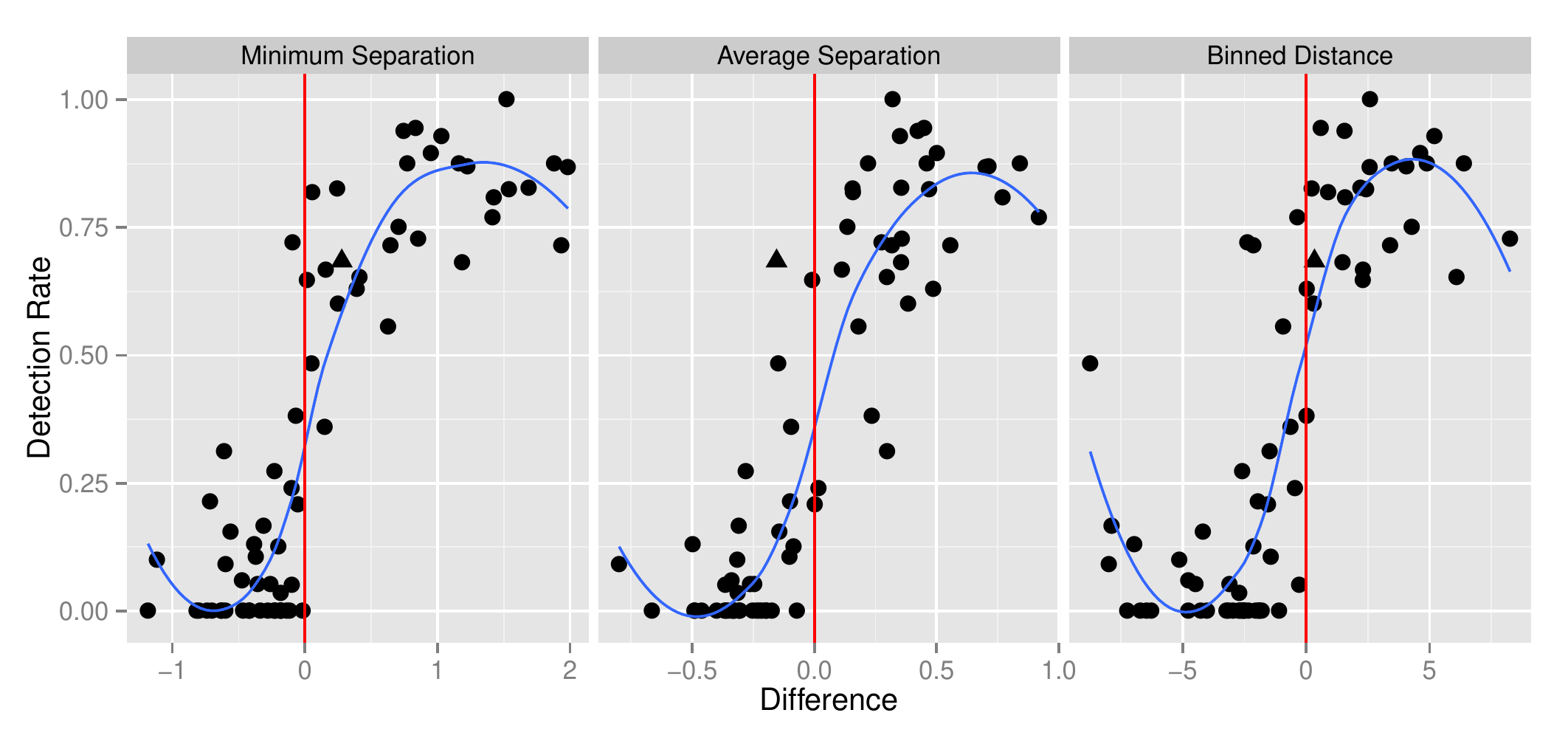}
\label{lpcomp_1}
}
\subfigure[]{
\includegraphics[scale=0.75]{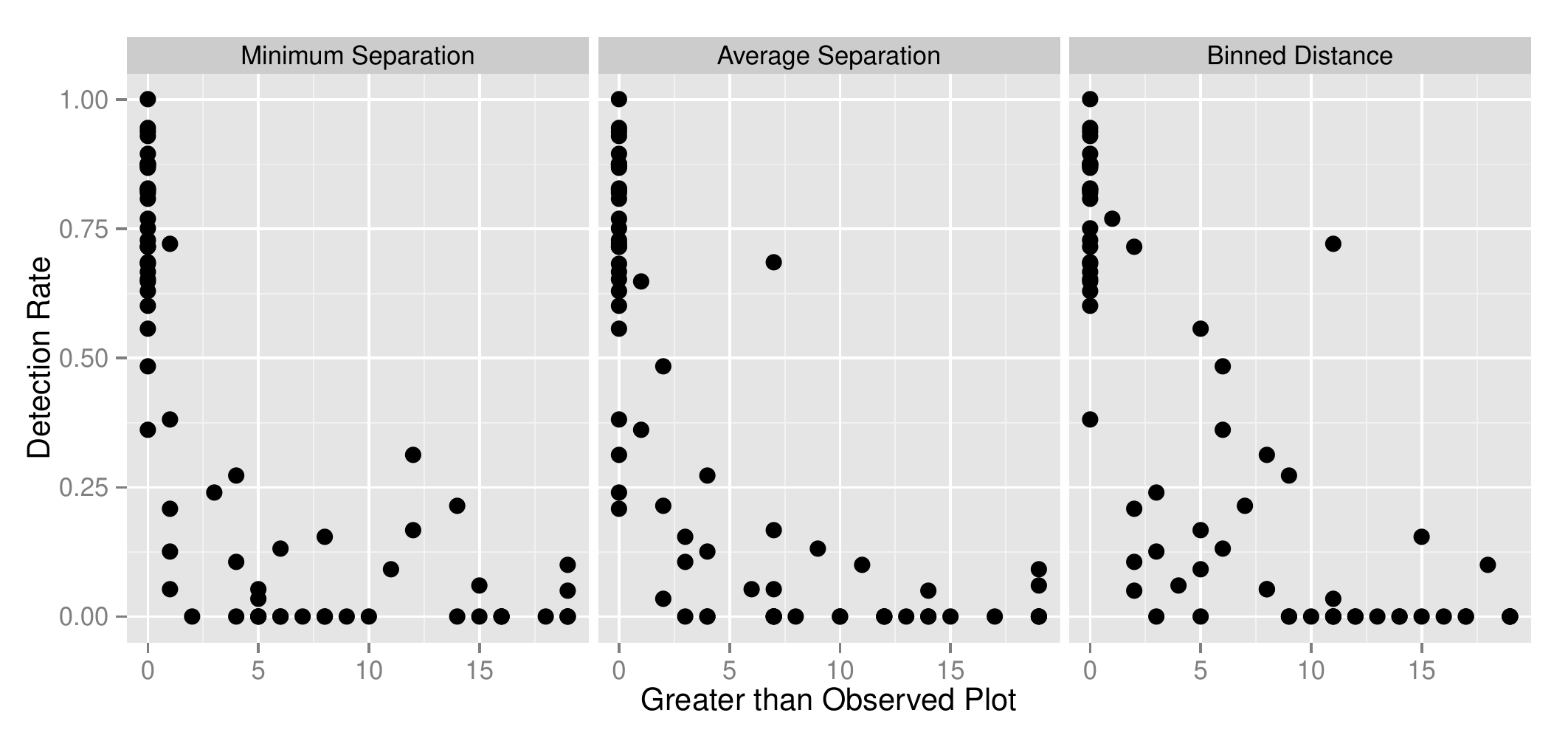}
\label{lpcomp_2}
}
	\vspace{-.1in}
\caption[Optional caption for list of figures]{Comparison of distance metrics for the scatterplot with clusters. Detection rate (a) and the number of plots greater than the observed (b) are plotted against the difference based on the minimum separation, average separation and binned distance. The vertical line represents the difference equal to 0 when there is at least one null plot similar to the observed plot. The detection rate increases with the difference. As the number of plots with distance greater than the observed increases, the detection rate decreases. The triangle represents a lineup with high detection rate and negative difference based on the average separation distance. This is examined in Figure \ref{lp-exp}. }
\label{lp-comp}
\end{figure}

The distances between the plots in this experiment were computed using the distance based on minimum separation and average separation of the clusters and also the binned distance. The number of bins used for the lineups with one dimensional projections is larger (10 in this case) but for the lineups with two dimensional projections, the number of bins used is 5. The proportion of correct response is plotted against $\delta_{\hbox{lineup}}$ and $\gamma_{\hbox{lineup}}$ for both the distances. Figure \ref{lp-comp} shows the results.

In Figure \ref{lp-comp}, the detection rate is plotted against the difference for distance based on minimum separation, average separation and the binned distance. The red vertical line shows difference equal to 0.  It can be seen that as the difference increases, the detection rate increases and both the distances do a good job in capturing the response of the subjects.  In (b) it can be seen that as there are more extreme null plots compared to the observed plot, the subjects find it difficult to pick the observed plot. For a few lineups, a large number of the subjects identify the observed plot although there is more extreme null plots. 

\begin{figure}[hbtp]
\centering
\includegraphics[scale=0.75]{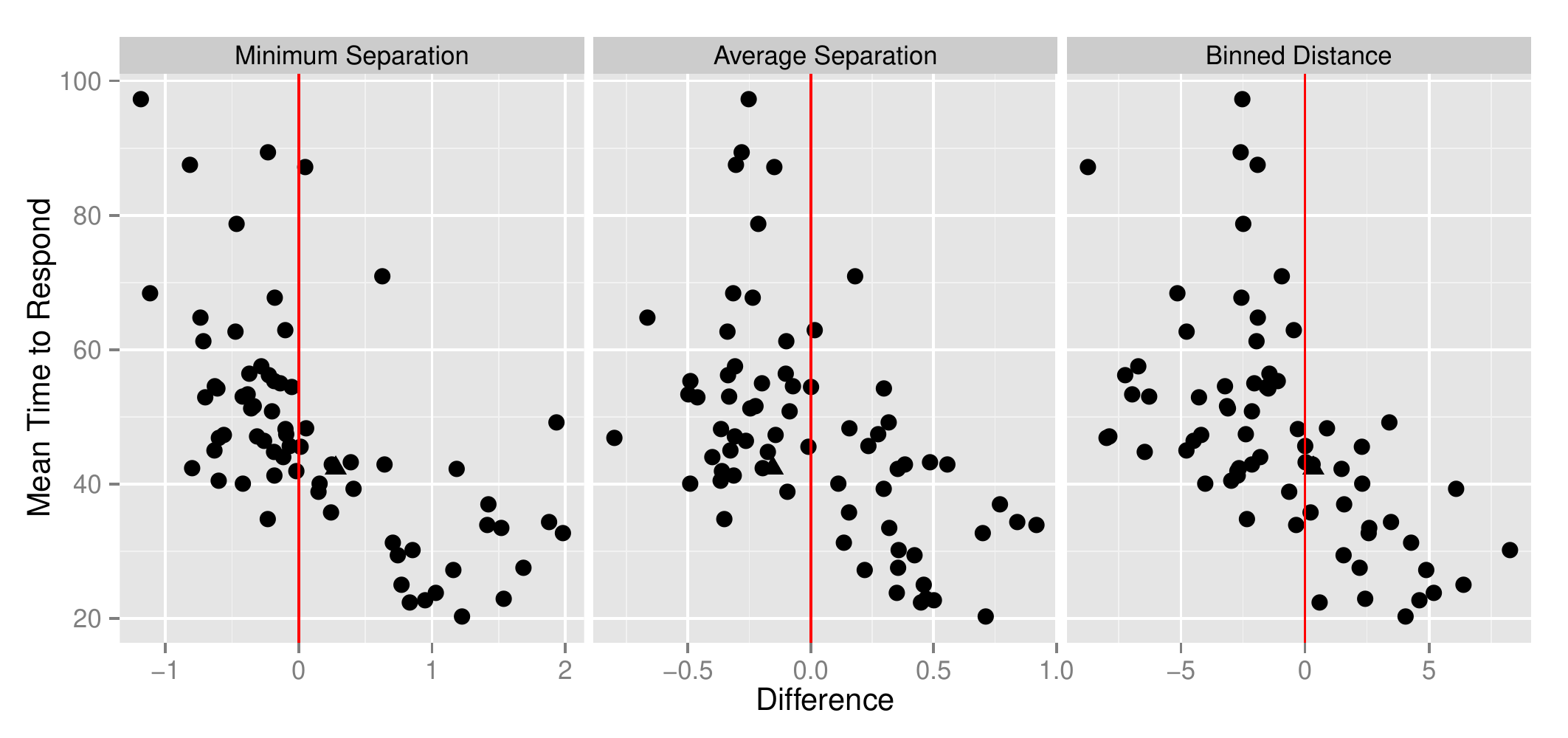}
	\vspace{-.1in}
\caption[Optional caption for list of figures]{Plot showing the mean time to respond by the subjects against the difference based on the minimum separation distance, average separation and binned distance. The vertical line represents the difference equal to 0 when there is at least one null plot similar to the observed plot. The mean time decreases as the difference increases.  }
\label{lp-mtime}
\end{figure}

Figure \ref{lp-mtime} shows the relationship between the mean time taken to respond and the difference for the three different distances. It can be clearly seen that there is a strong negative association showing that as the difference increases, the subjects take lesser time to respond. Also the variability of the mean time is higher for smaller difference. In case of binned distance, the relationship is negative though the variability is higher for all differences.

\begin{figure}[hbtp]
\centering
\subfigure[]{
\includegraphics[scale=0.7]{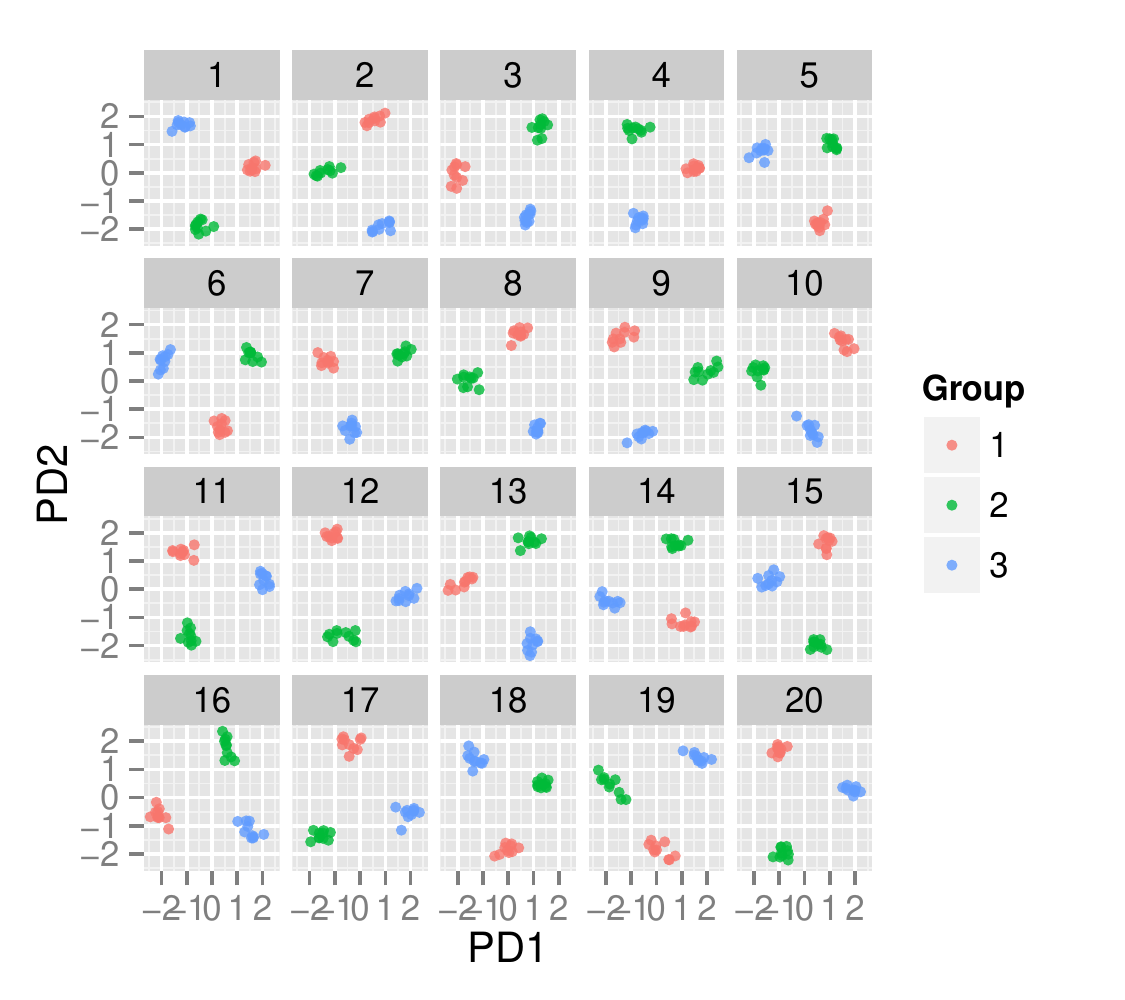}
\label{t2comp_1}
}
\subfigure[]{
\includegraphics[scale=0.7]{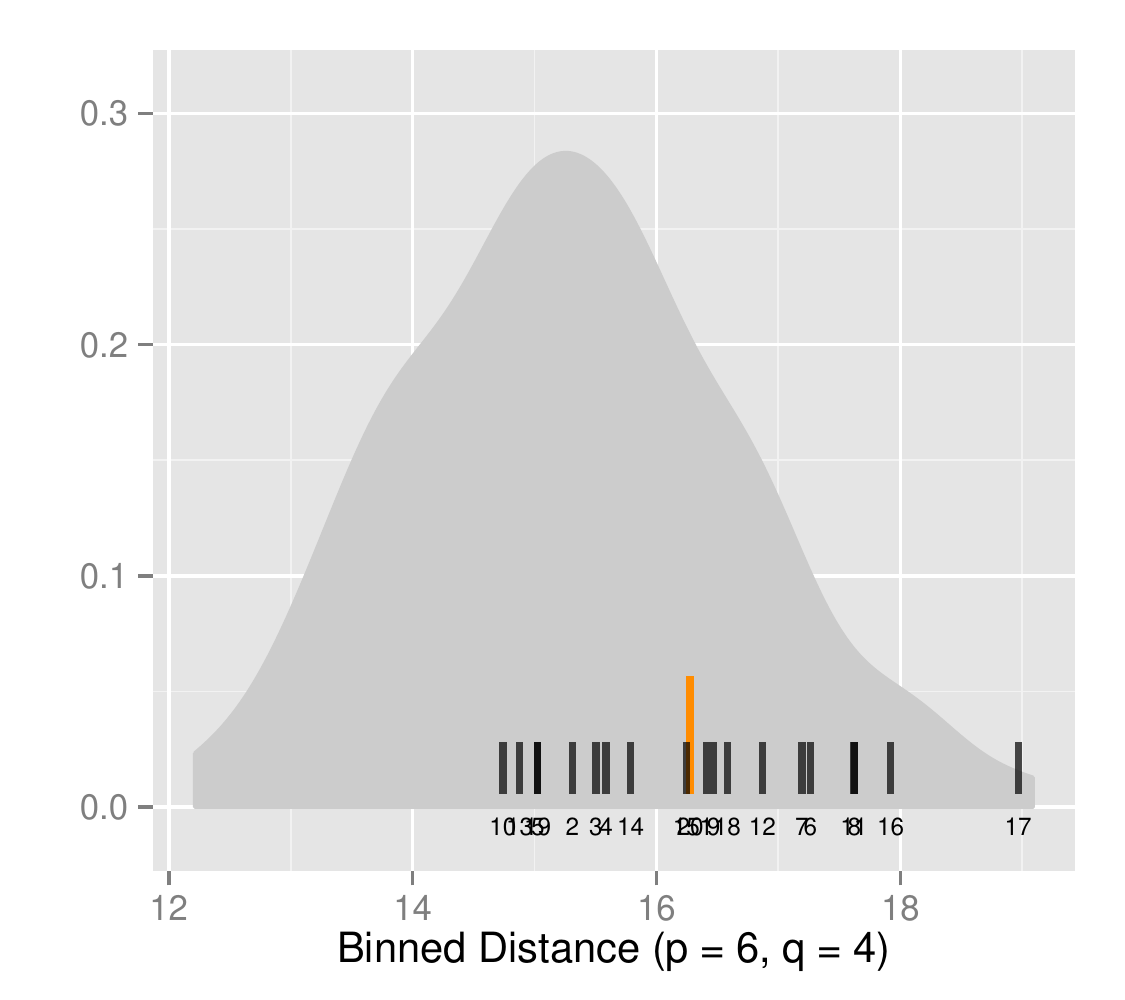}
\label{t2comp_1}
}
\subfigure[]{
\includegraphics[scale=0.7]{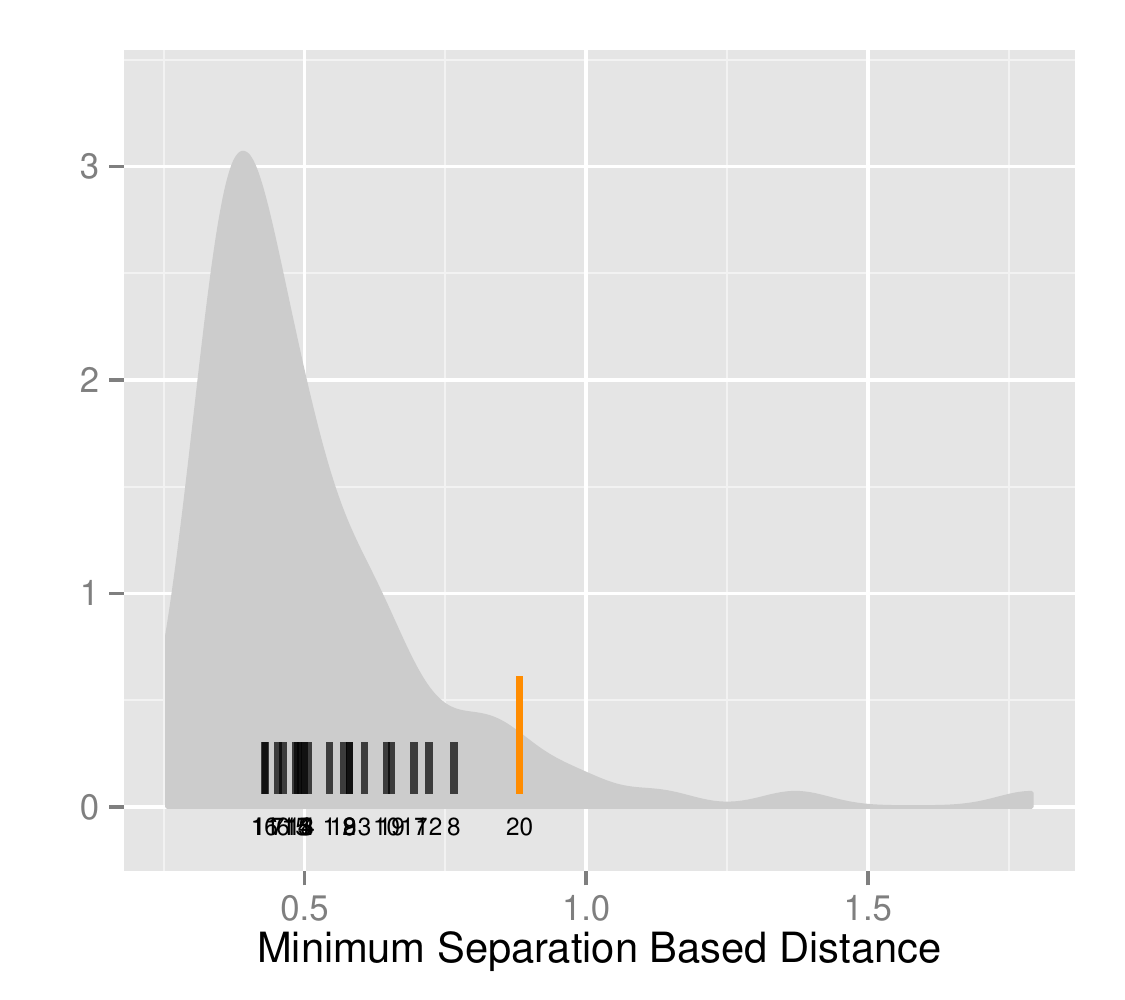}
\label{t2comp_2}
}
\subfigure[]{
\includegraphics[scale=0.7]{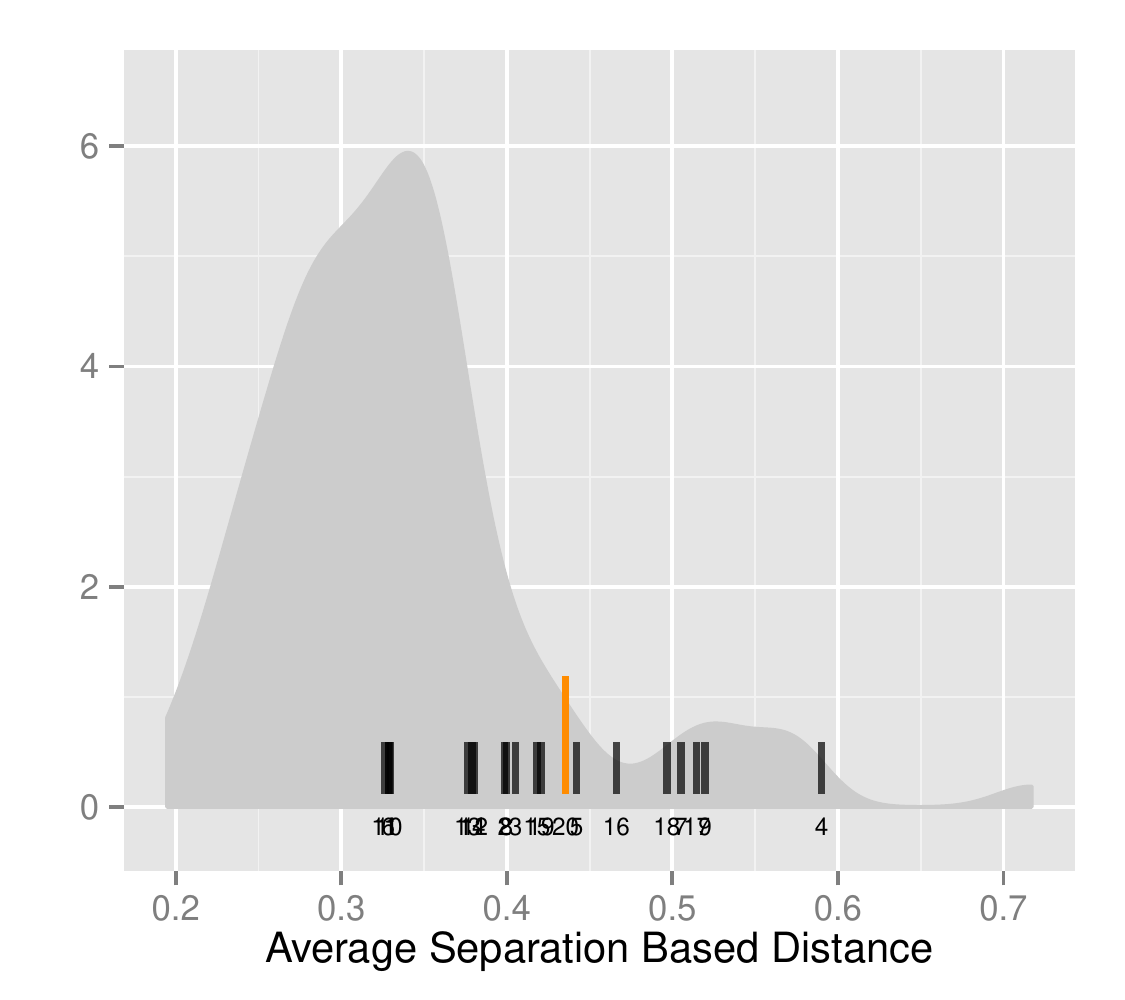}
\label{t2comp_1}
}

	\vspace{-.1in}
\caption[Optional caption for list of figures]{Illustration of the behavior of different distance metrics. The lineup is shown in (a) and the distributions of different distance metrics are shown in the other plots:  binned distance with 6 and 4 bins in x and y axis respectively in (b), distance based on average separation in (c) and distance based on minimum separation in (d). }
\label{lp-exp}
\end{figure}

Figure \ref{lp-exp} shows the lineup in a high dimension, low sample size setting. The number of dimensions used is 100 and two of the dimensions have some separation. Plot 20 shows the two-dimensional projections of the original data. The null plots are obtained by permuting the group variable and plotting the two dimensional projections obtained from a projection pursuit with PDA index (\cite{lee:2009}). Since the true plot has real separation, it is expected that the subjects would be able to identify the plot. The distance based on average separation yields a negative difference showing that the lineup is difficult, while the distance based on minimum separation yields a positive difference. The distance metrics identifies different characteristics in a plot. The average separation looks at the average of the distances of the points in a cluster to the points in other clusters. The presence of an outlier point in the opposite side of the other clusters affects this distance considerably. On the other hand, the minimum separation looks at the minimum of the distances. Hence it is not affected by the outlier point.

\section{Conclusion}

Distance metrics are compared to the response of human subjects on lineups. They are comparable to a certain extent except in certain situations where they disagree. There seems to be various reasons behind the disagreement. When people look at a lineup, they may identify a plot as different from the others due to various reasons. But the distance metrics are constructed such that it takes into specific properties of the plot. 

Distance metrics can be used to measure the quality of a lineup before showing the lineups to human subjects. Hence the distance metrics allows us to provide a range of lineups to the human subjects to evaluate.

In classical inference, the test statistic under null hypothesis follows a certain distribution. Similarly the null plots in visual inference can also be assumed to be random samples from a sampling distribution. Though theoretically this is true, practically it is impossible to investigate such a distribution. The distribution of the distance metrics approximates such a sampling distribution for a given distance metric. The value of the distance metric for the actual plot can be compared to all the other plots using such a distribution.

The reason of choice can provide a way of evaluating the performance of a distance metric. For example, for a lineup of scatterplots with regression line overlaid, if the choice of reason for majority is steepest slope, the regression based distance may work better than the binned distance. Similarly if the reason of choice is presence of outliers, the binned distance with large number of bins on both axes may be the best distance metric. This can be a probable future work.

\paragraph{Acknowledgement:}

This work was funded by National Science Foundation grant DMS 1007697. All plots are done with the {\tt ggplot2} \citep{hadley:2009} package in R.

\newpage

\bibliographystyle{plainnat}
\bibliography{references}

\section*{Appendix}

\subsection*{Selection of the Number of Bins} \label{sec:nbin}

Binned distance works for any type of data and for any null generating mechanism. It does not take into account the graphical elements in the plot, and the raw data is used. Binned distance can be used in situations where no distance measure is known for the particular plot type and hence it can be regarded as universal. But the choice of number of bins or the bin size highly affects the distance. A wrong choice may produce erroneous or conflicting results. Hence the choice of the number of bins is important.

The choice of number of bins or bin sizes is investigated with different types of data. Different null generating mechanisms are also used for the same data type. Null datasets are obtained for a true data using a null generating mechanism and hence a lineup is constructed. Mean binned distance is calculated between the true data and the null datasets and also among the null datasets. The number of bins for the binned distance are varied from 2 to 10 on both $x$ and $y$ direction and $\delta_{\hbox{lineup}}$ is calculated for each combination. Table \ref{tbl:bin1} and Table \ref{tbl:bin2} shows the type of data, the observed plot, the null generating mechanism, a typical null plot, the difference $\delta_{\hbox{lineup}}$ and also the maximum value of $\delta_{\hbox{lineup}}$, the $x$-bin and $y$-bin for which the maximum was obtained. The minimum $\delta_{\hbox{lineup}}$ is also reported to get an idea of the range of values.

%
%

\begin{table*}[hbtp]
\caption{Preferable number of bins for different types of observed data to calculate the binned distance.}
\centering 
\begin{tabular}{p{1.5cm}l c  p{2cm} c cc l c p{4cm}} 
\hline
 Type of Data & Observed Plot && Null Generating Mechanism & A typical null plot && & Difference && (x-bin, y-bin, Max; Min) \\ 
 \hline
 Linear association & \begin{minipage}[h]{1.5cm} \begin{center} \scalebox{0.25}{\includegraphics{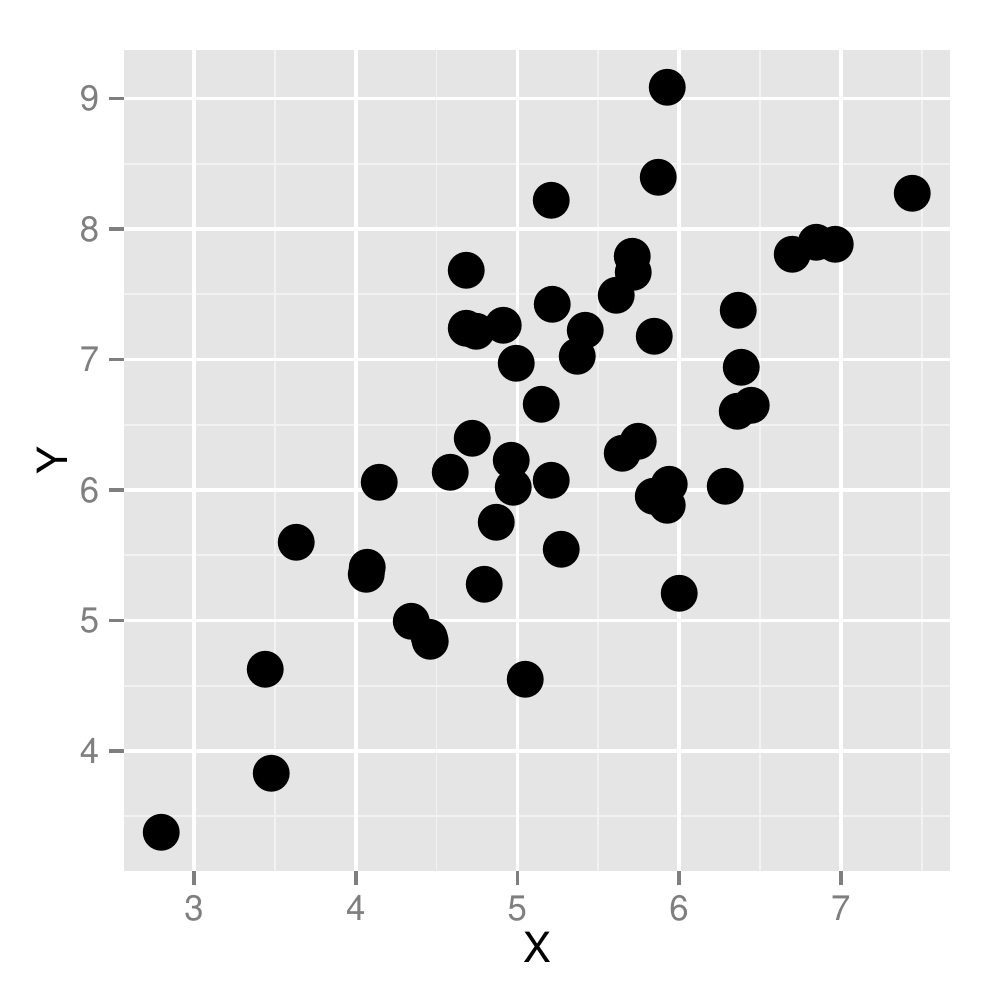}} \end{center} \end{minipage} && Permutation &  \begin{minipage}[h]{1.5cm} \begin{center} \scalebox{0.25}{\includegraphics{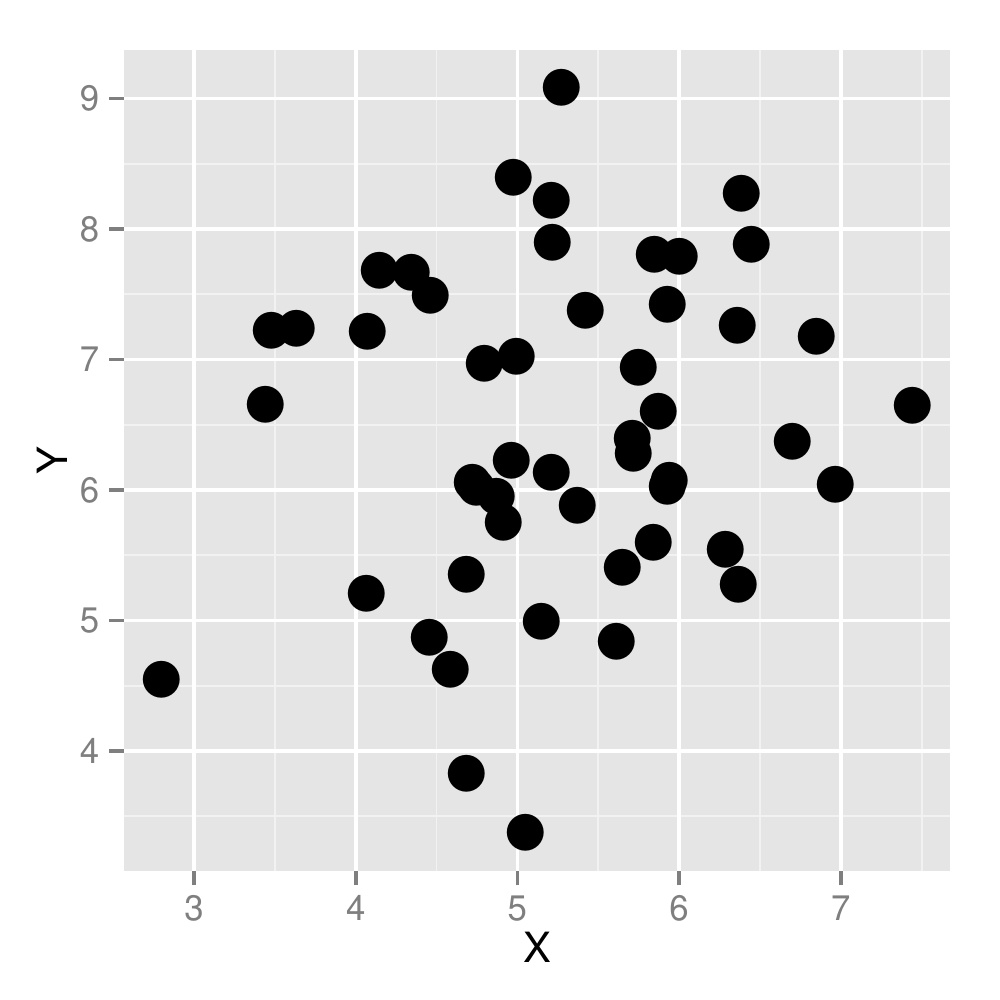}} \end{center} \end{minipage} &&&  \begin{minipage}[h]{1.5cm} \begin{center} \scalebox{0.25}{\includegraphics{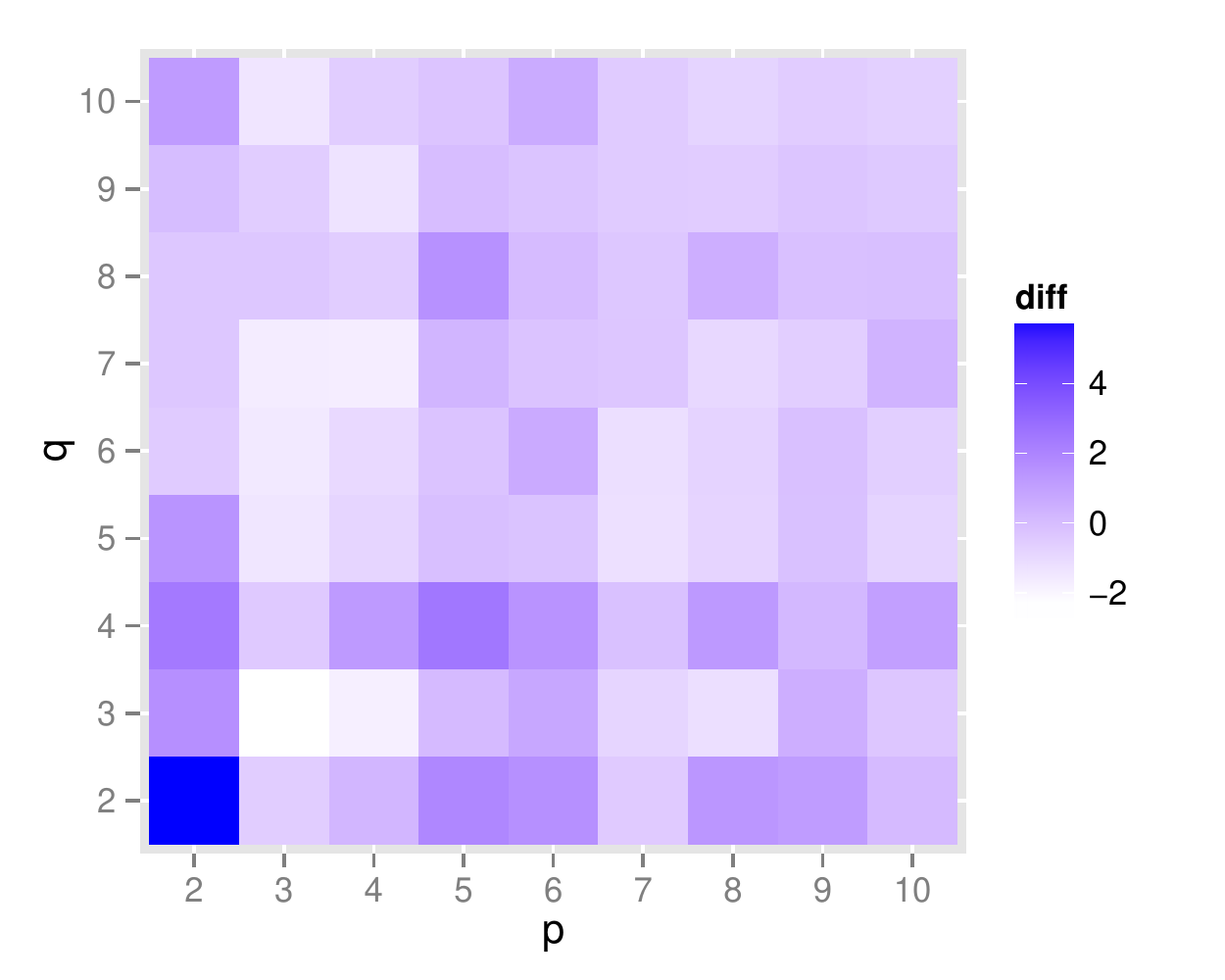}} \end{center} \end{minipage} &&           \hspace{0.8cm} (2, 2, 5.7 ; - 2.5)\\
 \hline
Nonlinear relationship & \begin{minipage}[h]{1.5cm} \begin{center} \scalebox{0.25}{\includegraphics{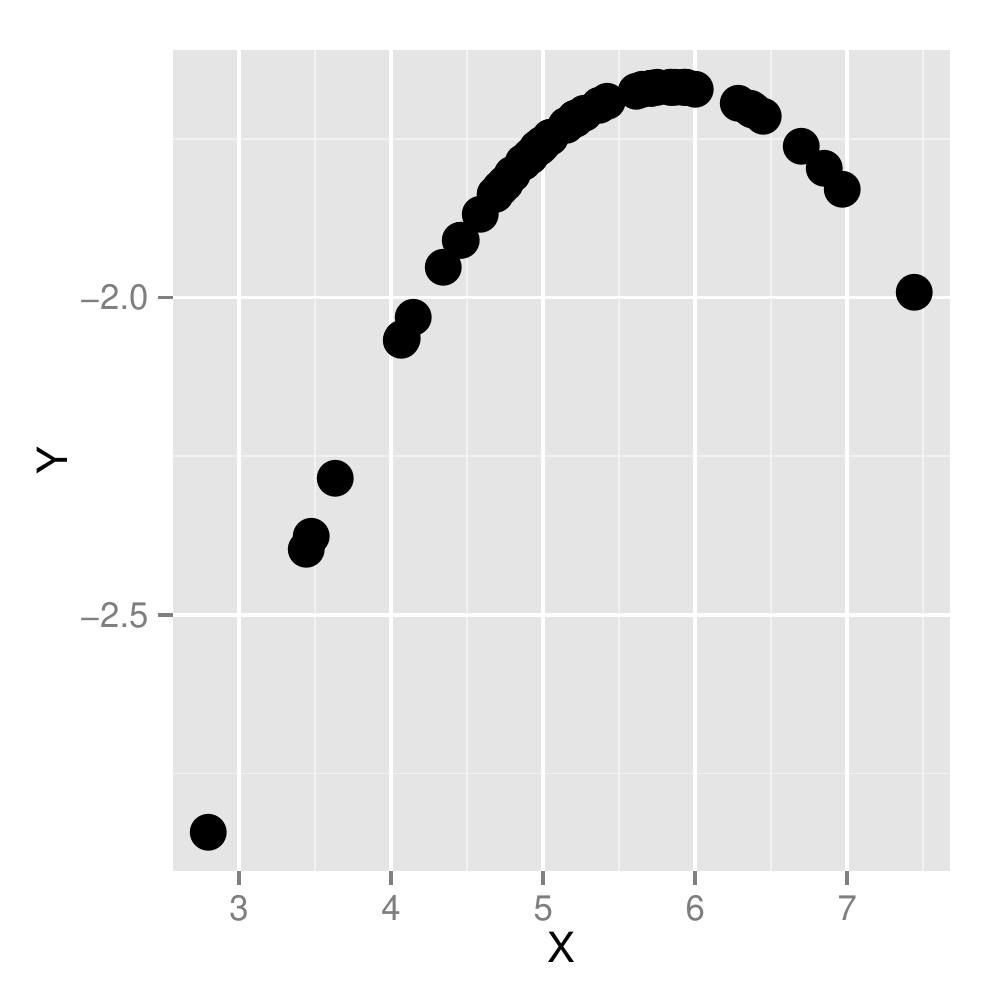}} \end{center} \end{minipage} && Permutation &  \begin{minipage}[h]{1.5cm} \begin{center} \scalebox{0.25}{\includegraphics{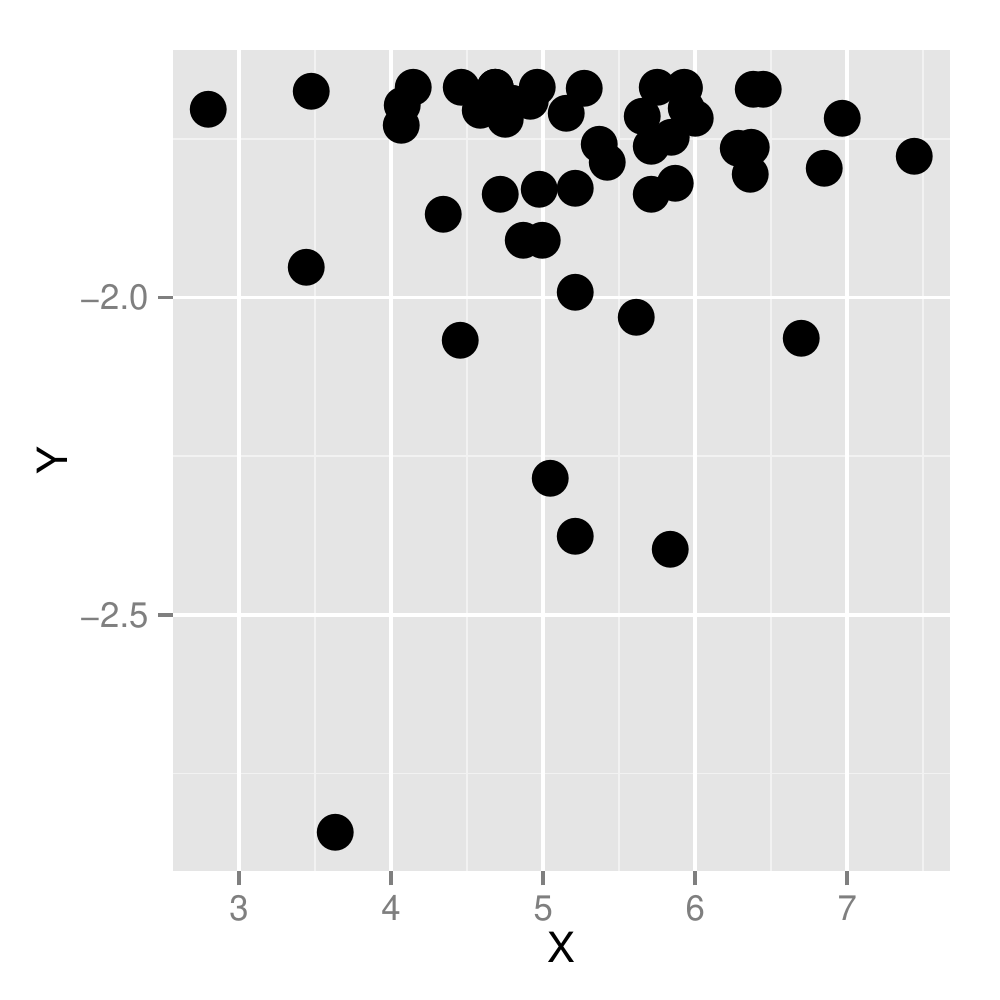}} \end{center} \end{minipage} &&&  \begin{minipage}[h]{1.5cm} \begin{center} \scalebox{0.25}{\includegraphics{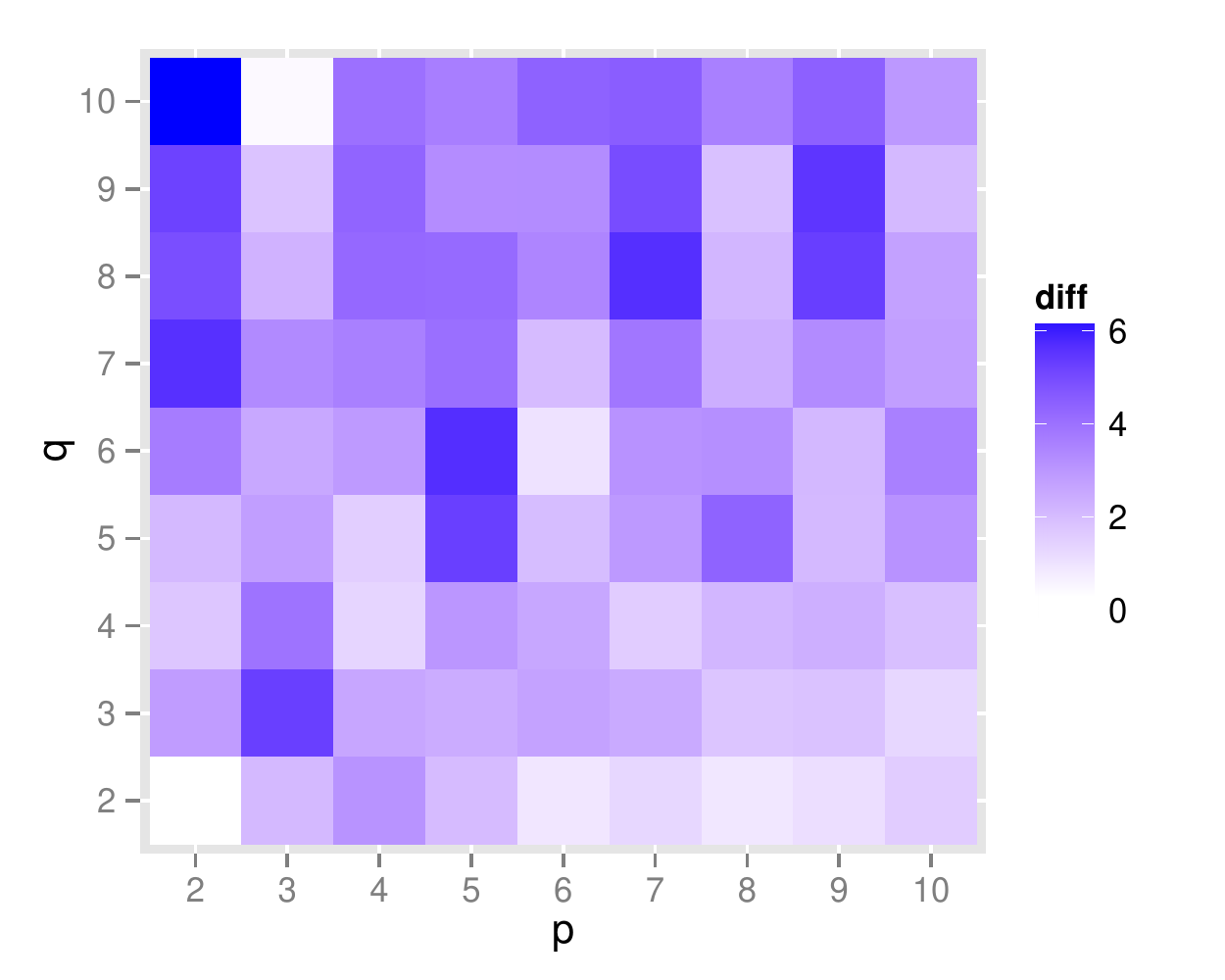}} \end{center} \end{minipage} &&           \hspace{0.8cm} (2, 10, 6.2 ; - 0.0)\\
 \hline
 Linear relation with outliers & \begin{minipage}[h]{1.5cm} \begin{center} \scalebox{0.25}{\includegraphics{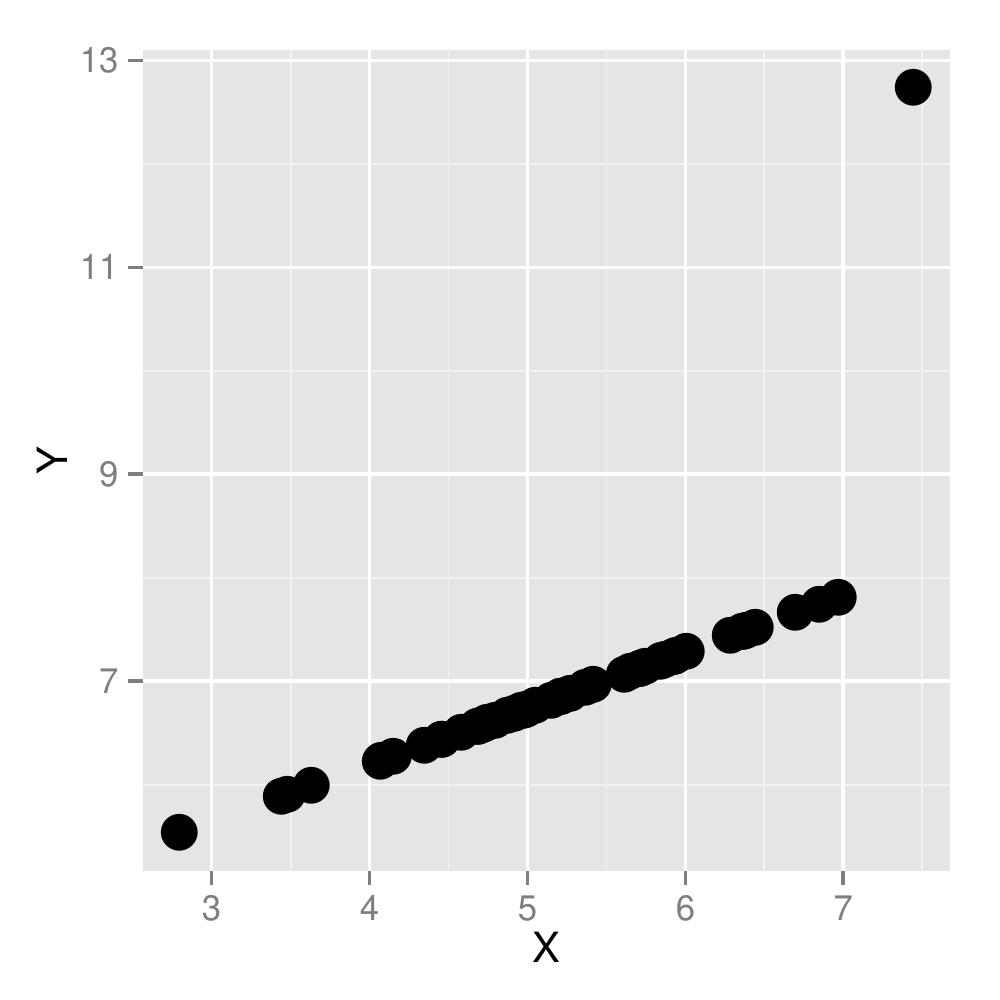}} \end{center} \end{minipage} && Permutation &  \begin{minipage}[h]{1.5cm} \begin{center} \scalebox{0.25}{\includegraphics{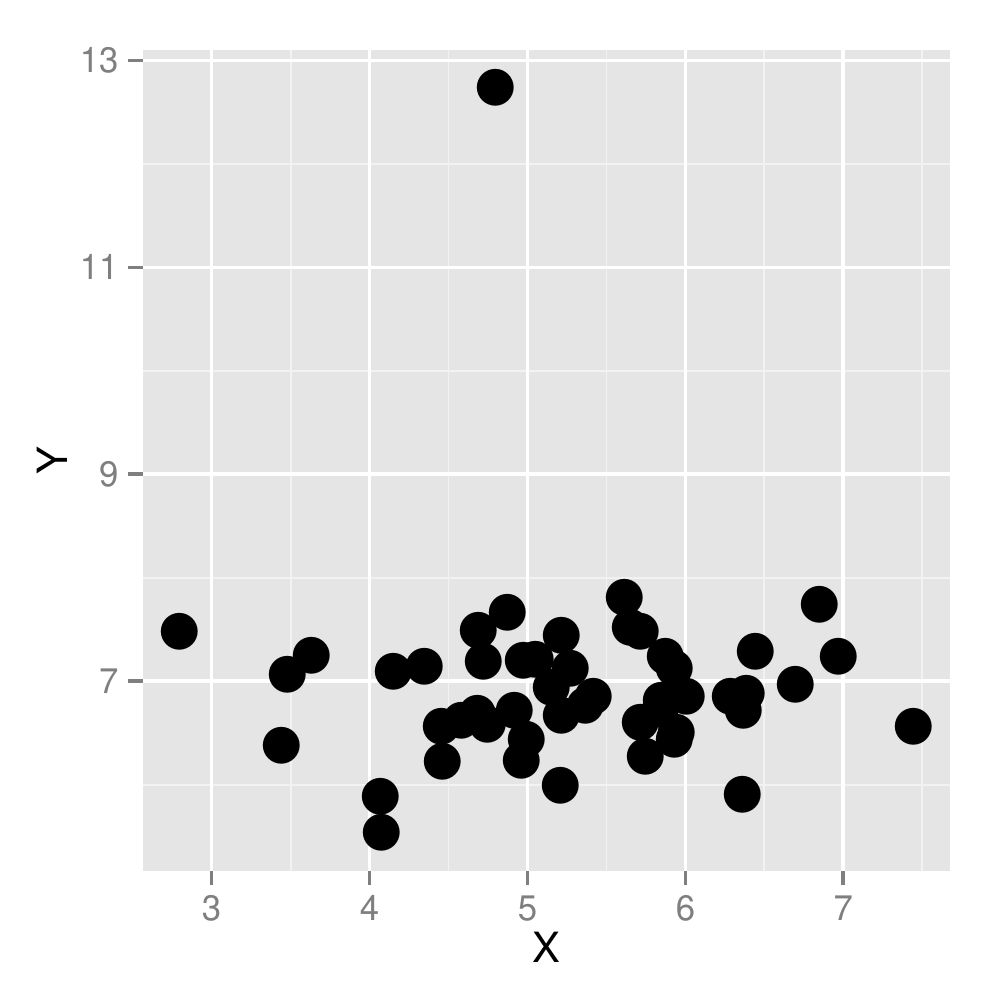}} \end{center} \end{minipage} &&&  \begin{minipage}[h]{1.5cm} \begin{center} \scalebox{0.25}{\includegraphics{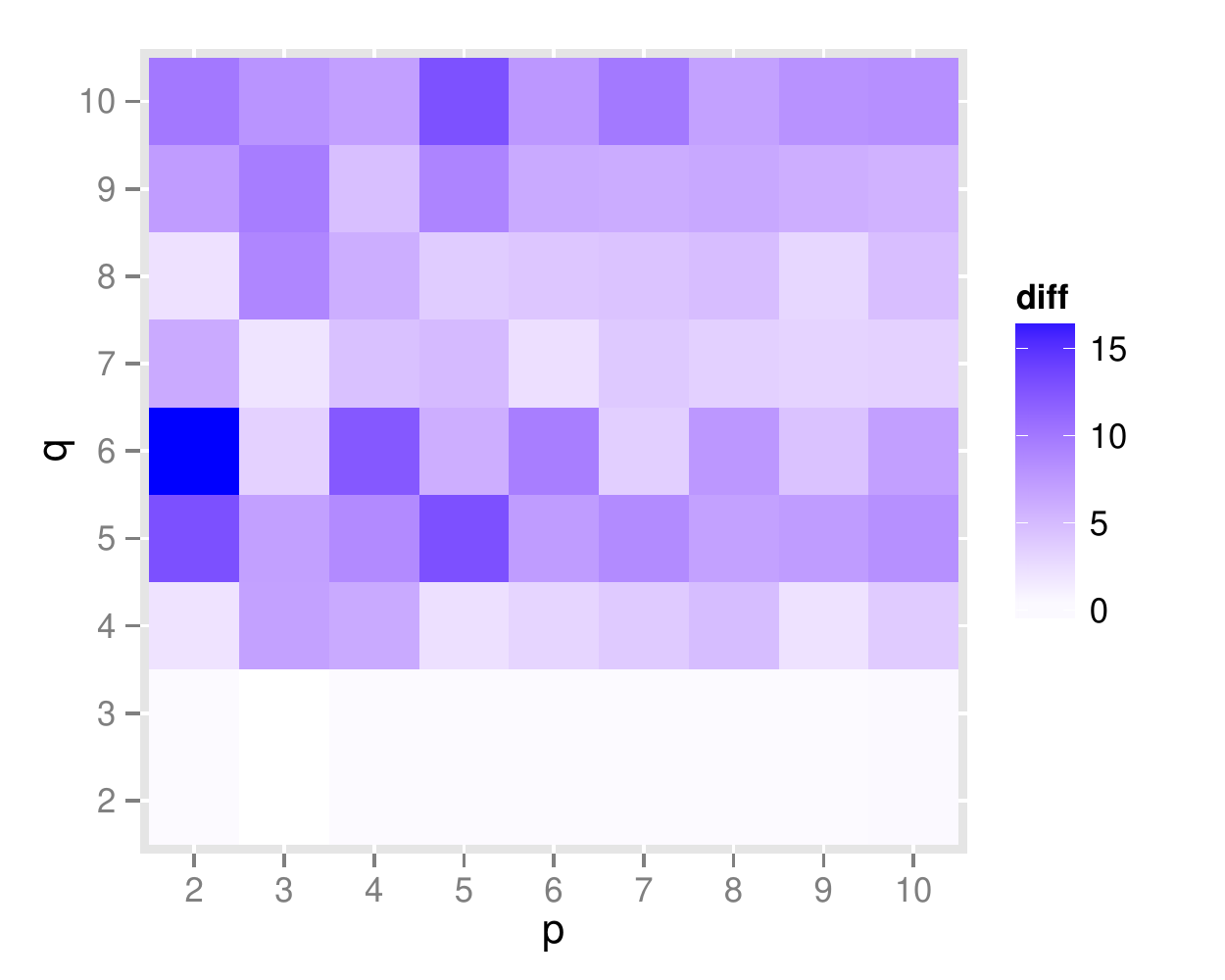}} \end{center} \end{minipage} &&           \hspace{0.8cm} (2, 6, 16.7 ; - 0.4)\\
 \hline
 Same values with one outlier & \begin{minipage}[h]{1.5cm} \begin{center} \scalebox{0.25}{\includegraphics{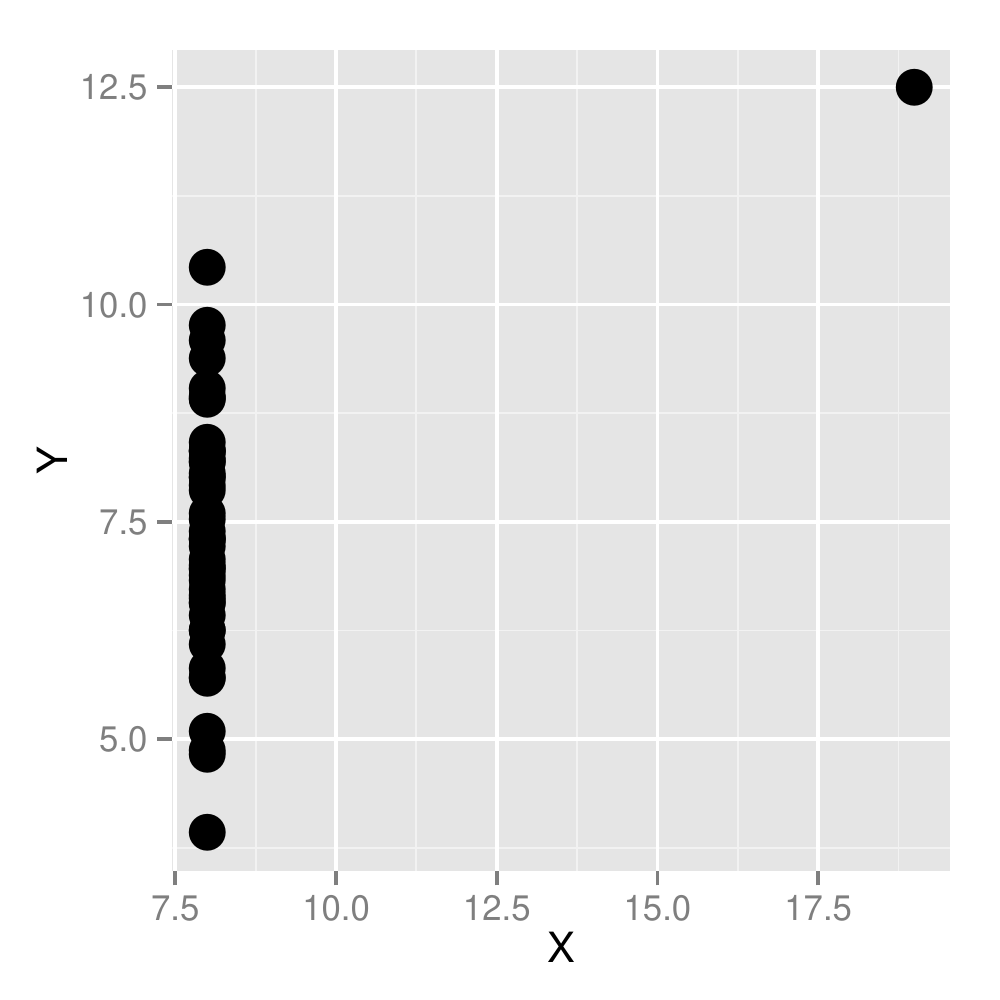}} \end{center} \end{minipage} && Simulation from a \emph{Poi(9)} distribution &  \begin{minipage}[h]{1.5cm} \begin{center} \scalebox{0.25}{\includegraphics{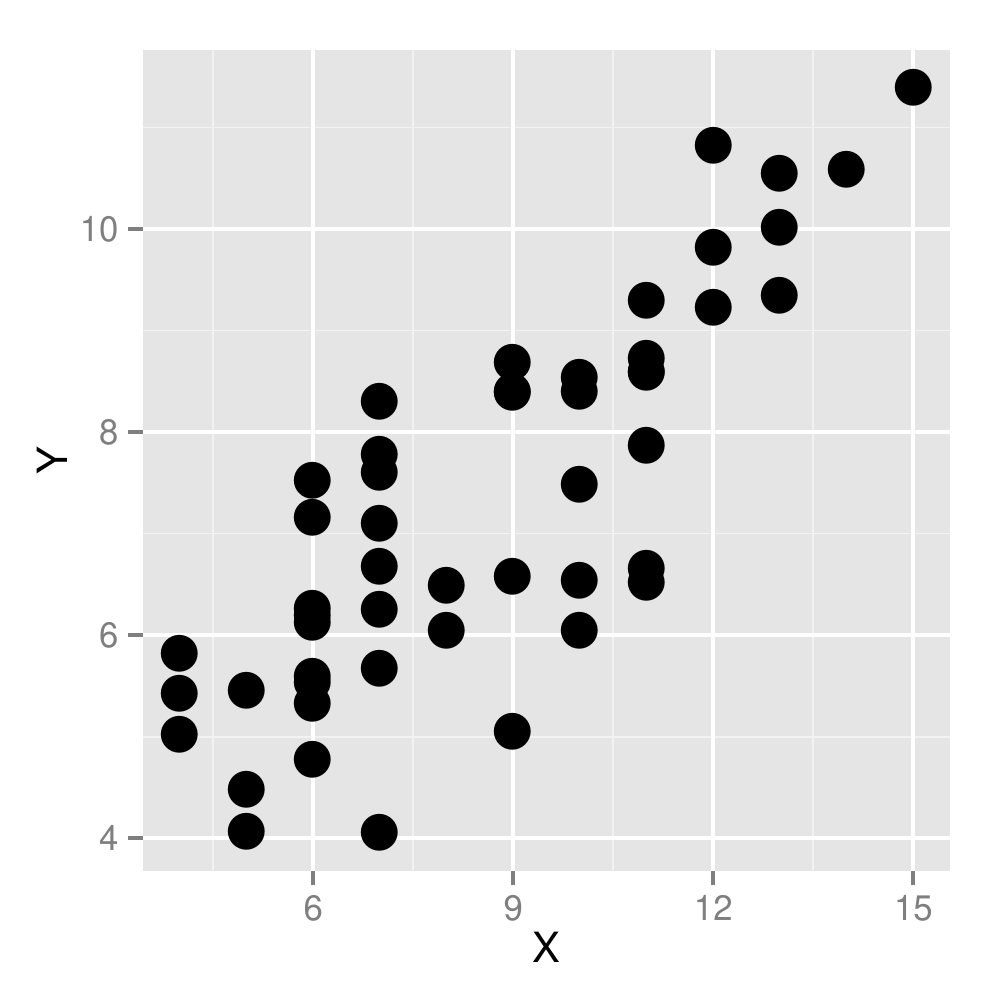}} \end{center} \end{minipage} &&&  \begin{minipage}[h]{1.5cm} \begin{center} \scalebox{0.25}{\includegraphics{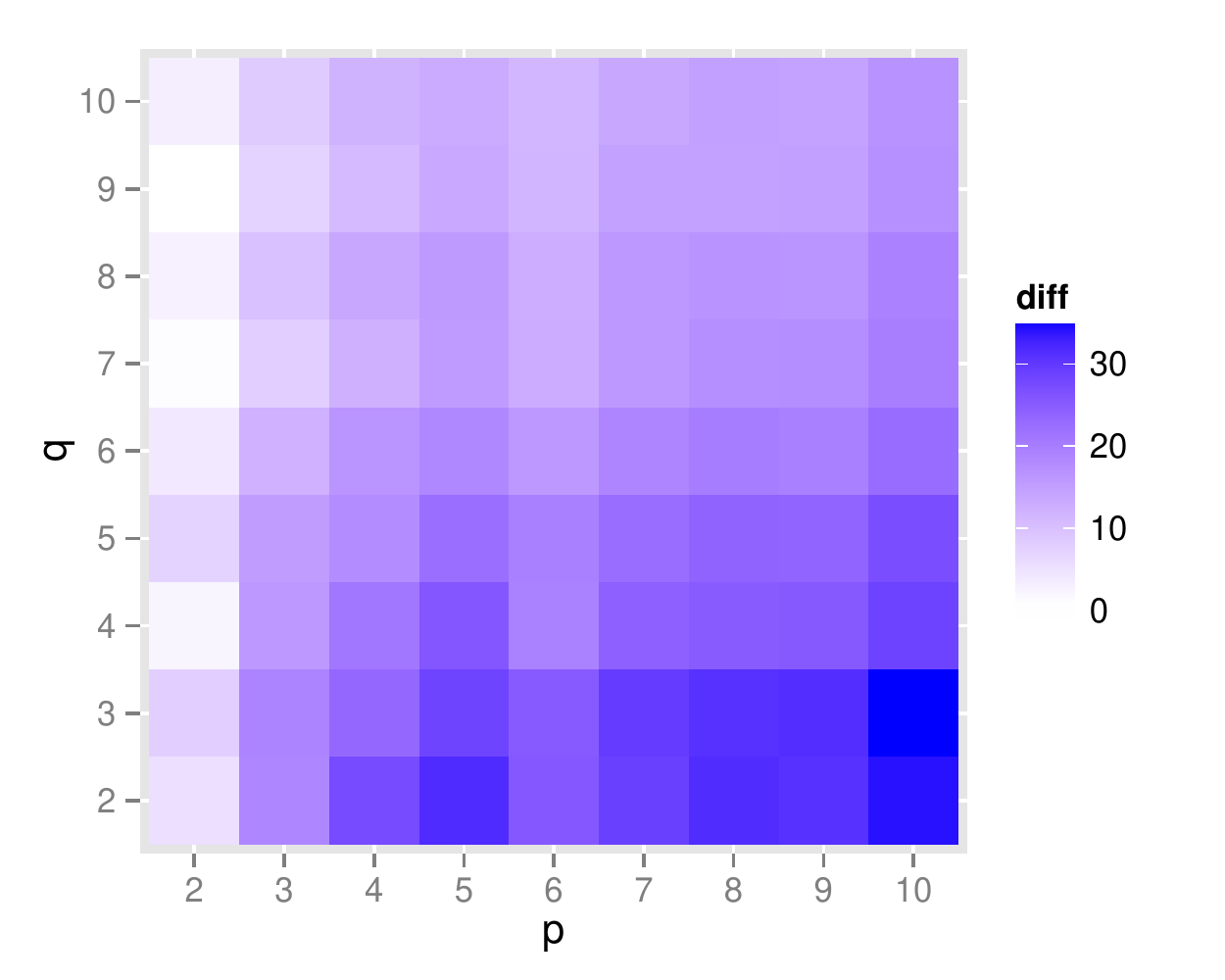}} \end{center} \end{minipage} &&           \hspace{0.8cm}(10, 3, 34.3 ; - 0.1)\\
 \hline
    Clusters & \begin{minipage}[h]{1.5cm} \begin{center} \scalebox{0.25}{\includegraphics{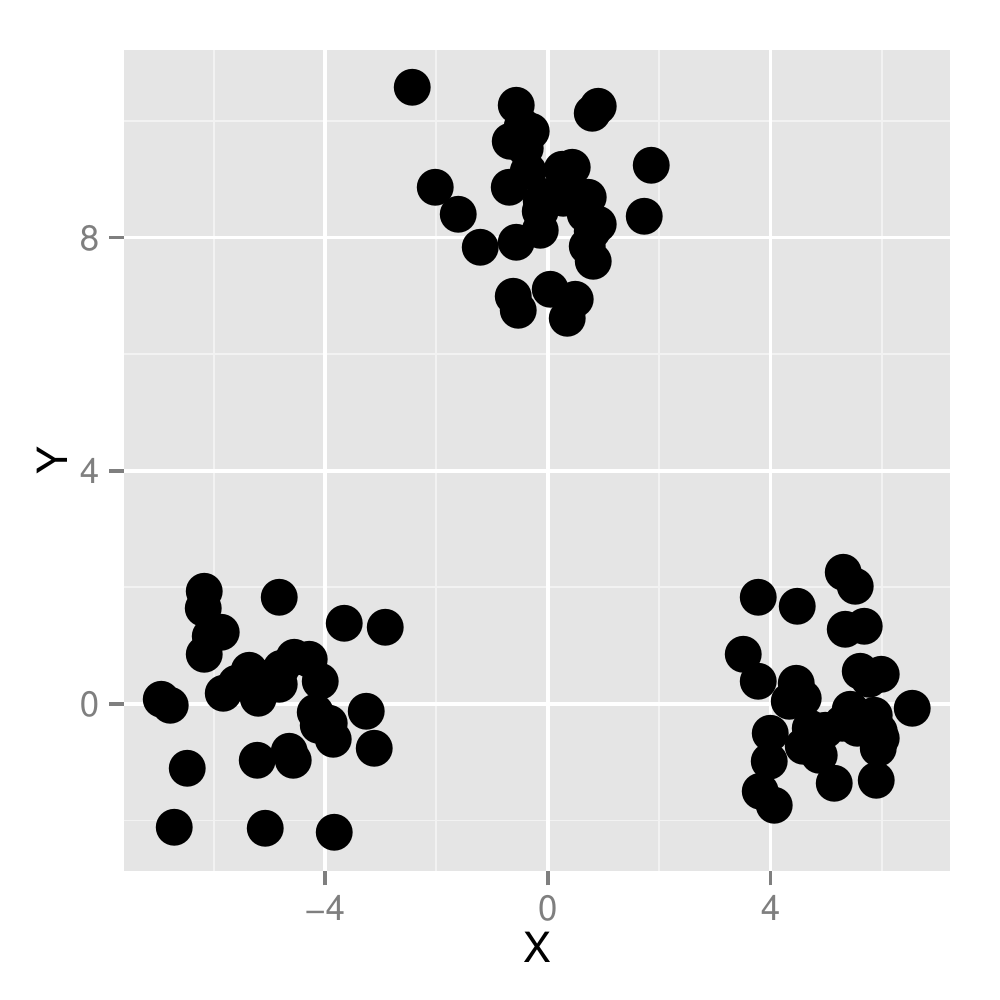}} \end{center} \end{minipage} && Permutation &  \begin{minipage}[h]{1.5cm} \begin{center} \scalebox{0.25}{\includegraphics{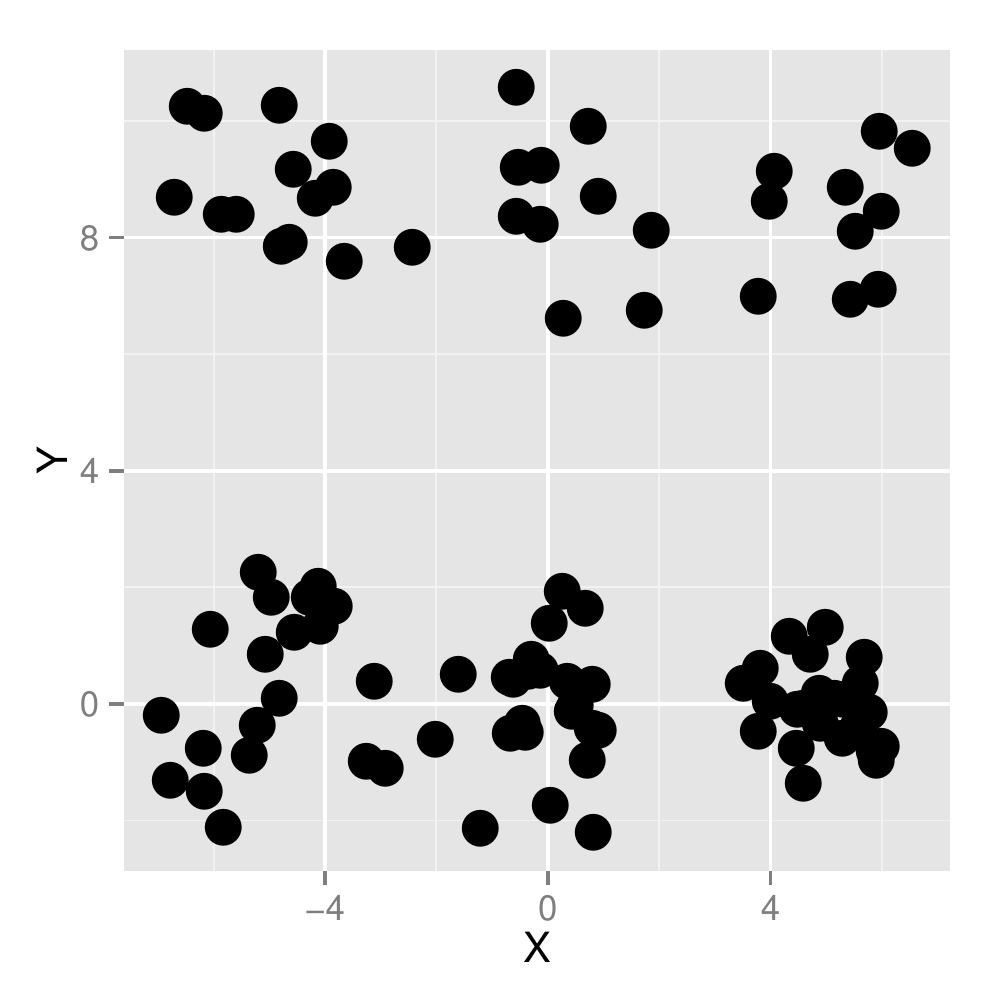}} \end{center} \end{minipage} &&&  \begin{minipage}[h]{1.5cm} \begin{center} \scalebox{0.25}{\includegraphics{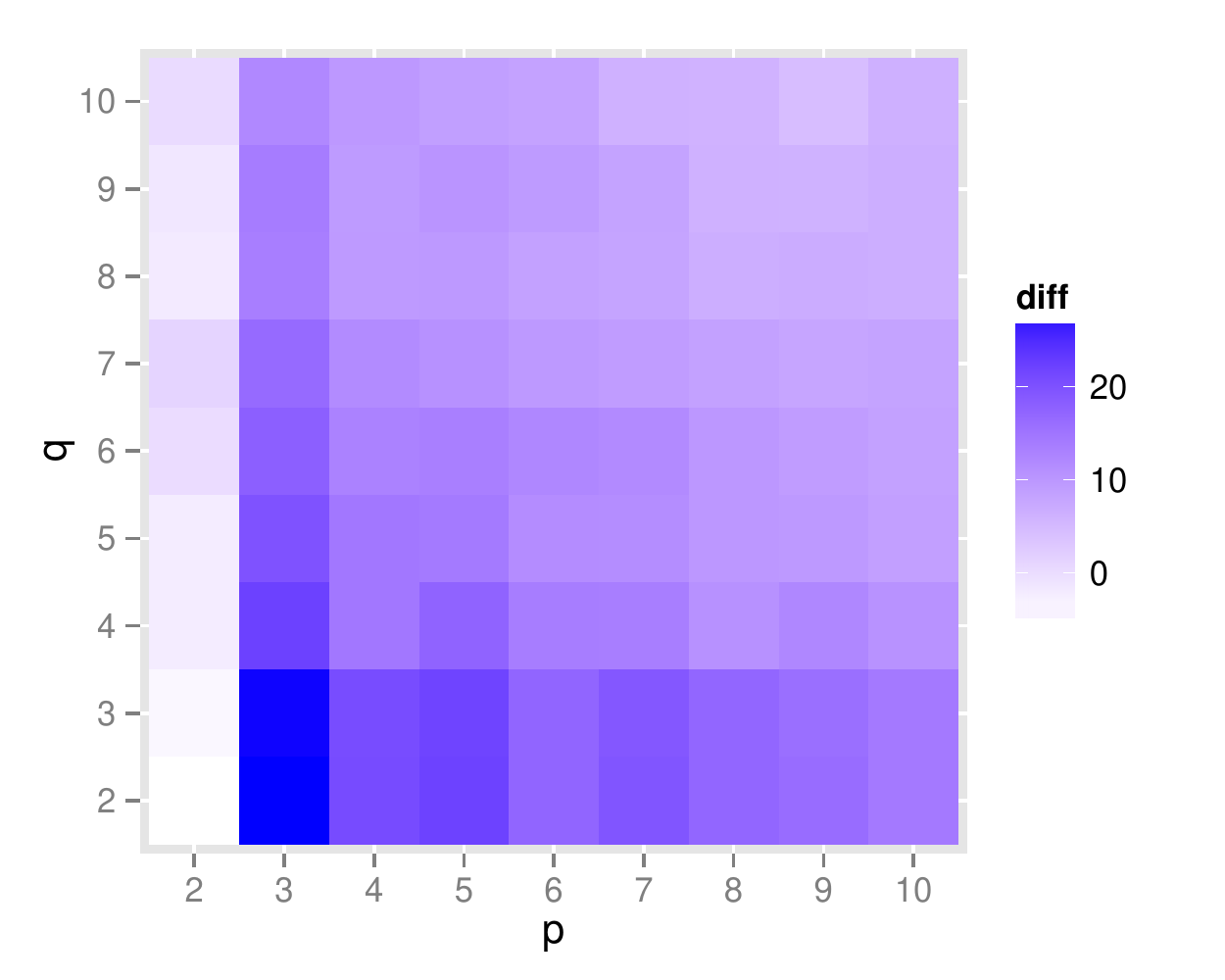}} \end{center} \end{minipage} &&           \hspace{0.8cm} (3, 2, 27.6 ; - 5.7)\\
 \hline
\end{tabular}
\label{tbl:bin1}
\end{table*}

\begin{table*}[hbtp]
\caption{Preferable number of bins for different types of observed data to calculate the binned distance.}
\centering 
\begin{tabular}{p{1.5cm}l c  p{2cm} c cc l c p{4cm}} 
\hline
 Type of Data & Observed Plot && Null Generating Mechanism & A typical null plot && & Difference && (x-bin, y-bin, Max; Min) \\ 
 \hline
Categorical & \begin{minipage}[h]{1.5cm} \begin{center} \scalebox{0.25}{\includegraphics{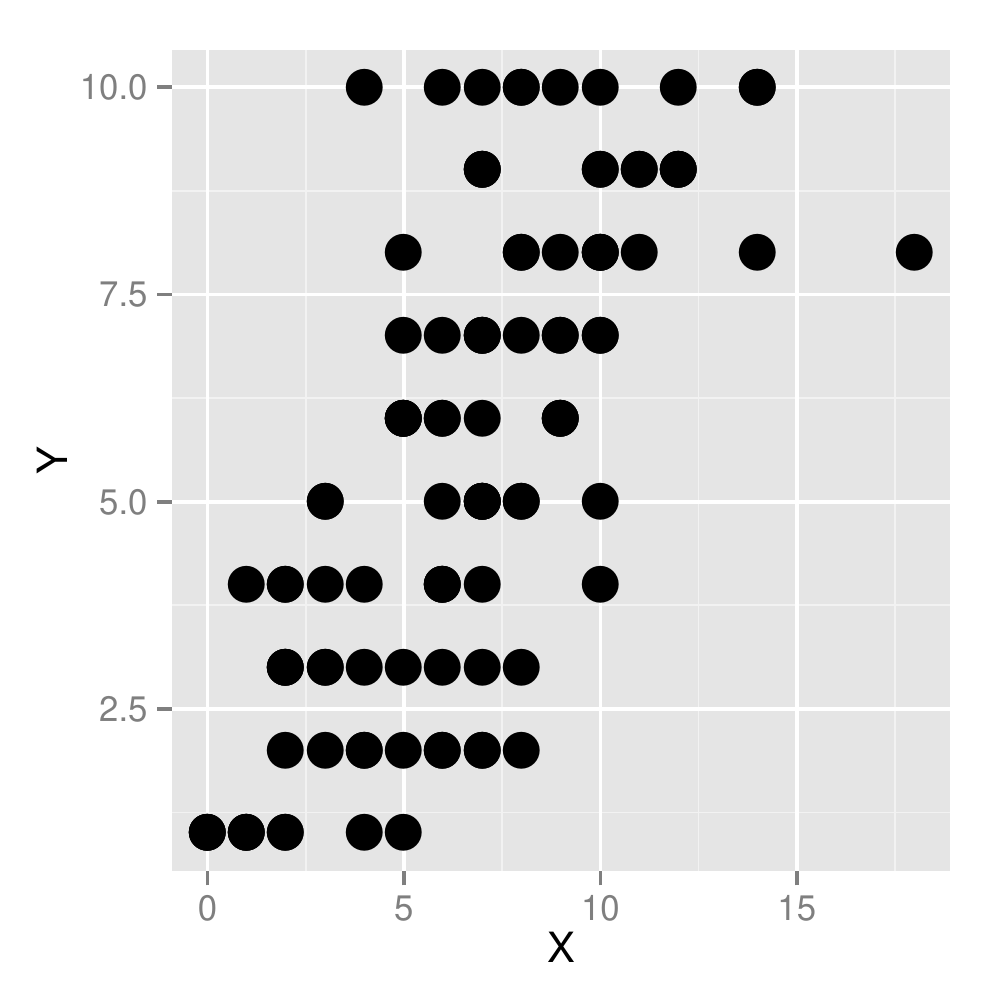}} \end{center} \end{minipage} && Simulation from a Normal distribution &  \begin{minipage}[h]{1.5cm} \begin{center} \scalebox{0.25}{\includegraphics{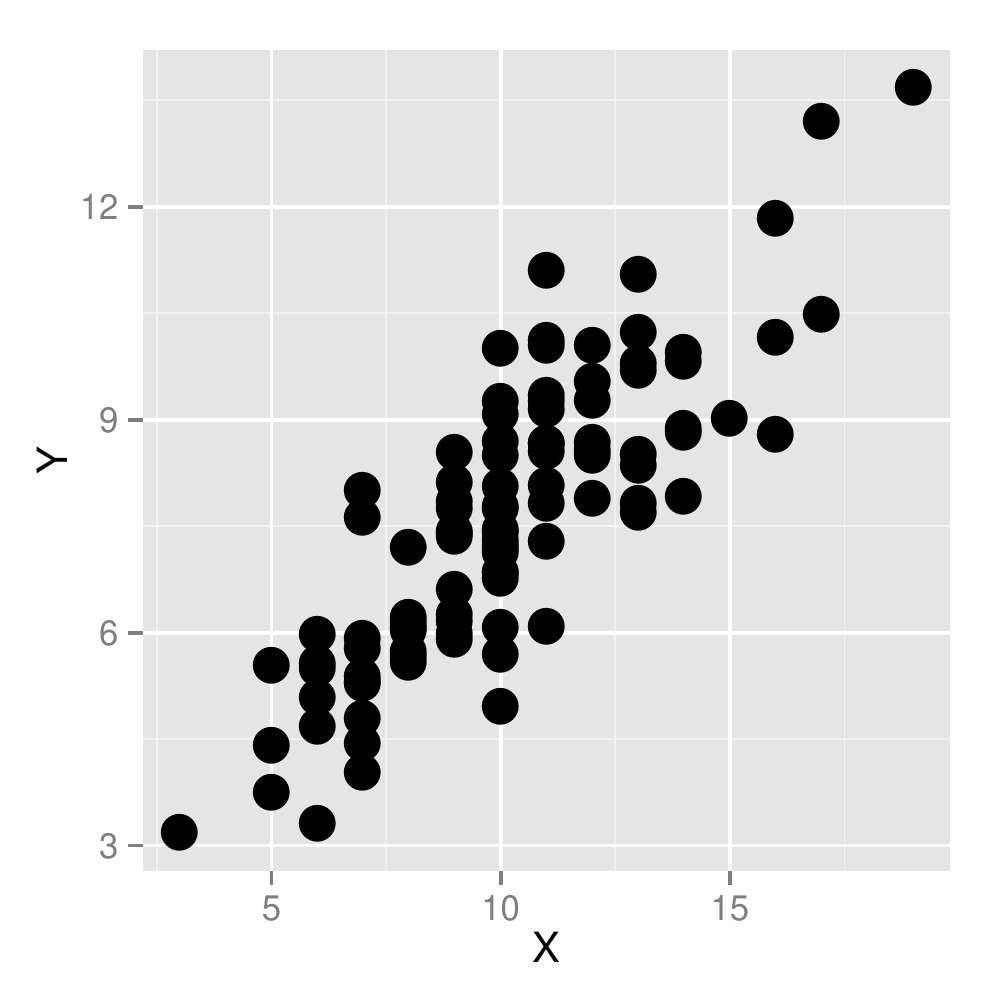}} \end{center} \end{minipage} &&&  \begin{minipage}[h]{1.5cm} \begin{center} \scalebox{0.25}{\includegraphics{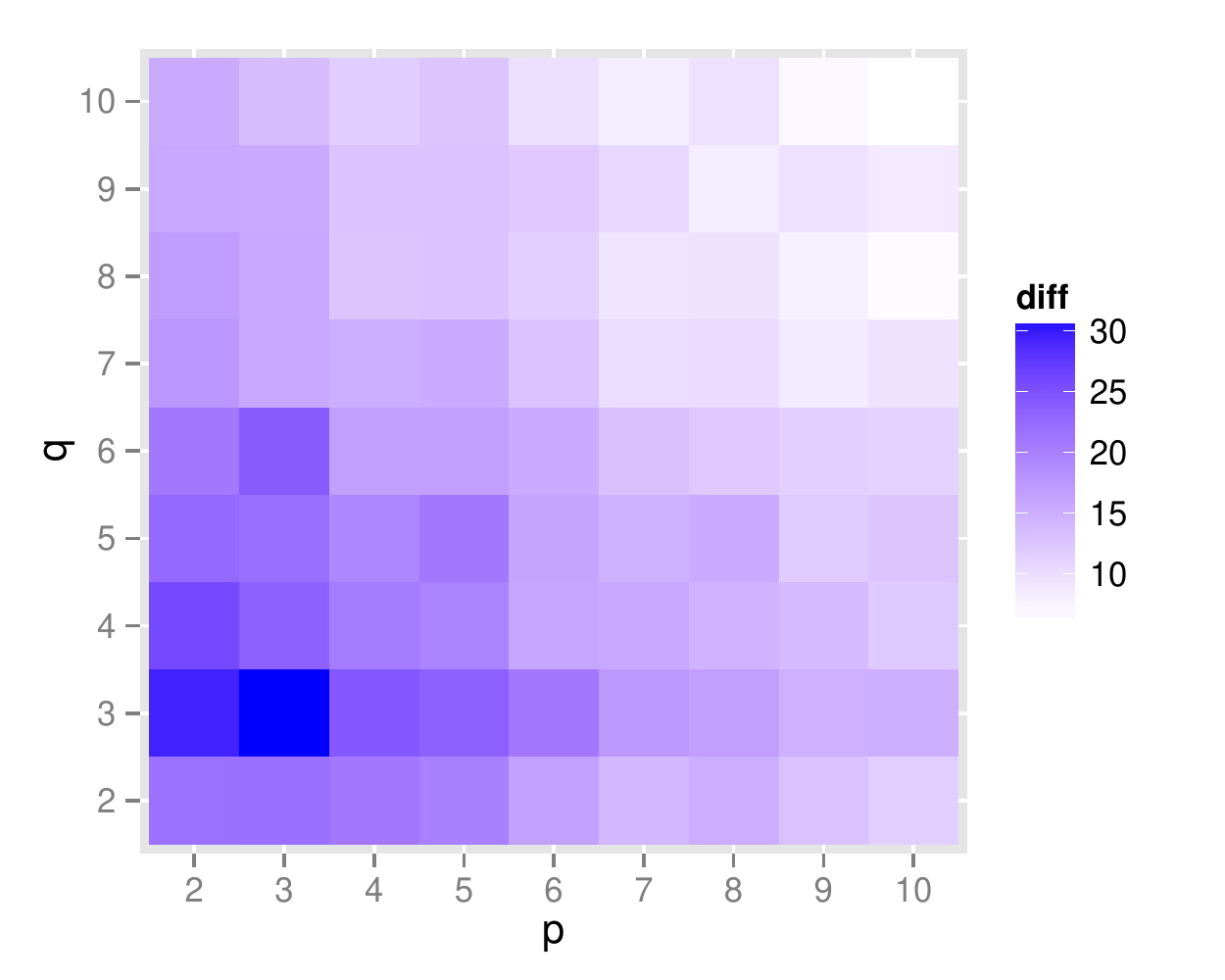}} \end{center} \end{minipage} &&           \hspace{0.8cm} (3, 3, 30.7; 6.2) \\
 \hline
Nonlinear relation with outliers & \begin{minipage}[h]{1.5cm} \begin{center} \scalebox{0.25}{\includegraphics{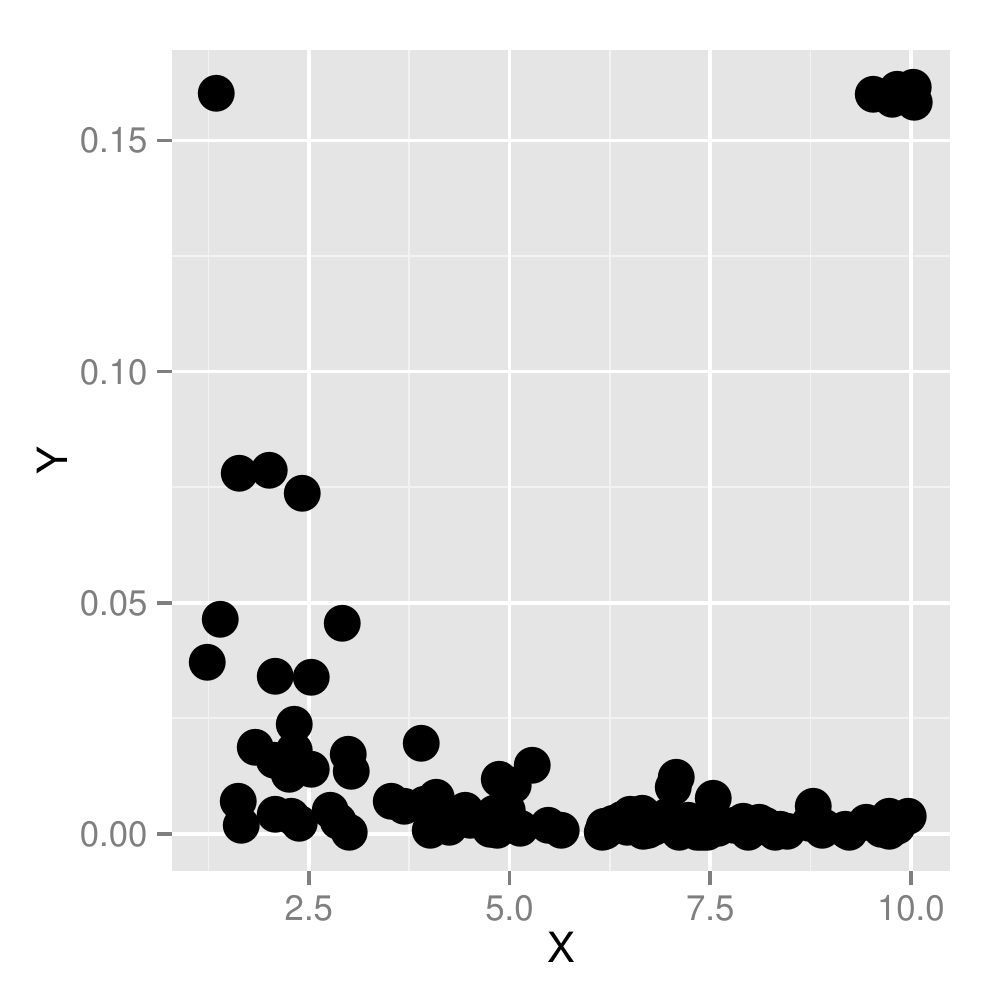}} \end{center} \end{minipage} && Permutation &  \begin{minipage}[h]{1.5cm} \begin{center} \scalebox{0.25}{\includegraphics{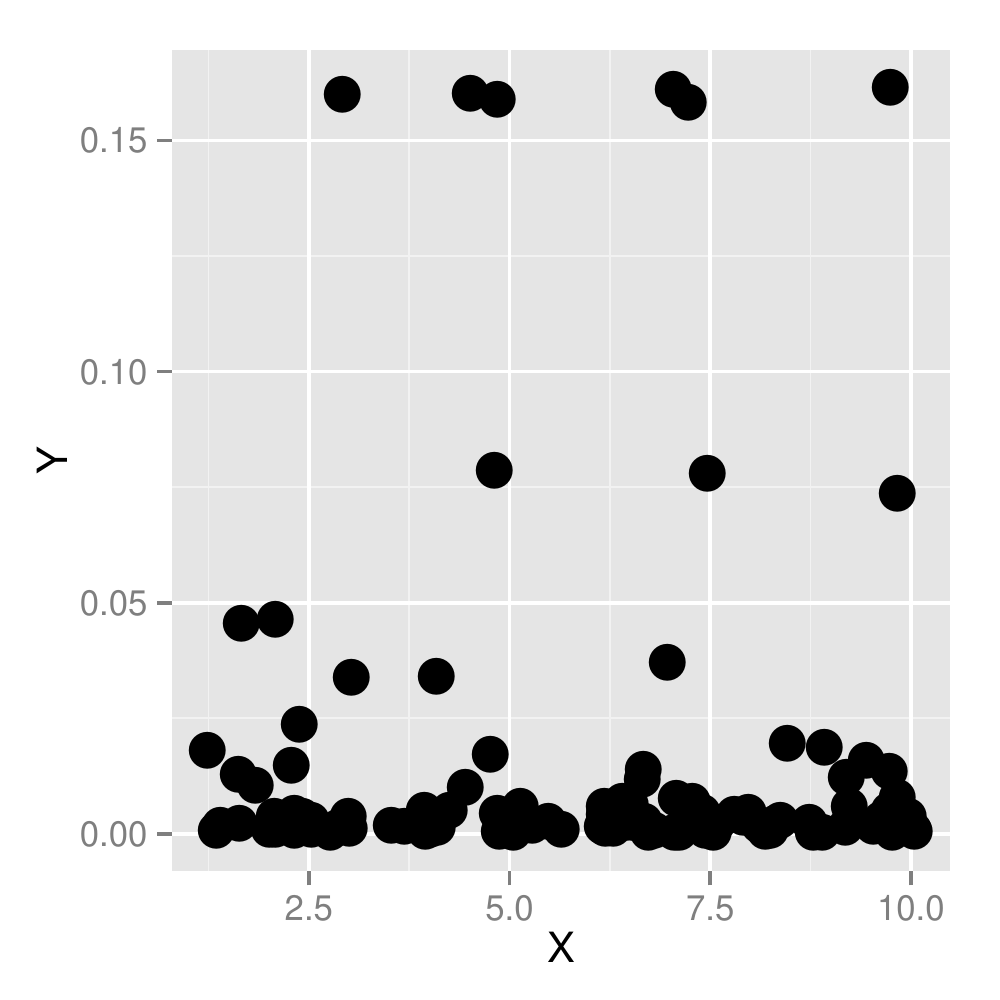}} \end{center} \end{minipage} &&&  \begin{minipage}[h]{1.5cm} \begin{center} \scalebox{0.25}{\includegraphics{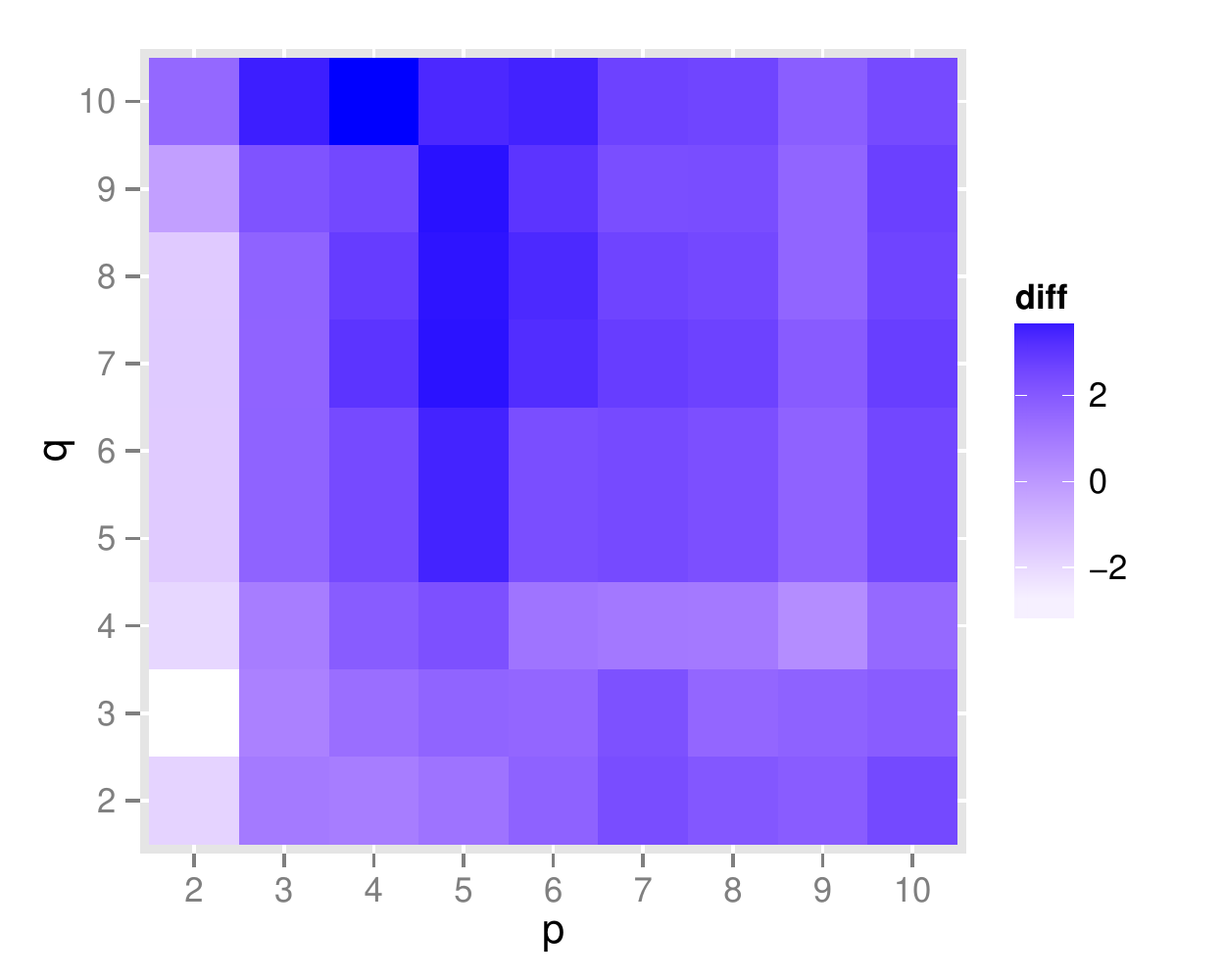}} \end{center} \end{minipage} &&           \hspace{0.8cm} (4, 10, 3.9; -3.4) \\
 \hline
Linear relationship with outlier  & \begin{minipage}[h]{1.5cm} \begin{center} \scalebox{0.25}{\includegraphics{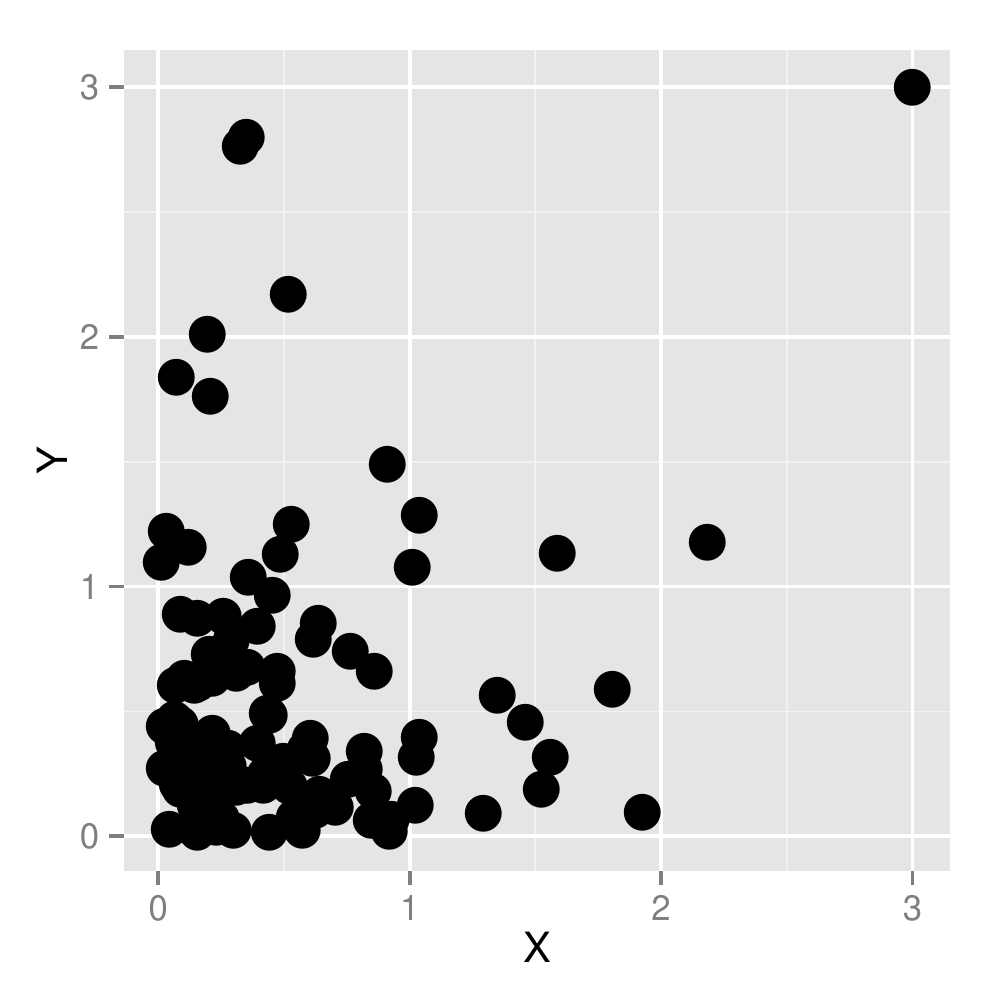}} \end{center} \end{minipage} && Permutation &  \begin{minipage}[h]{1.5cm} \begin{center} \scalebox{0.25}{\includegraphics{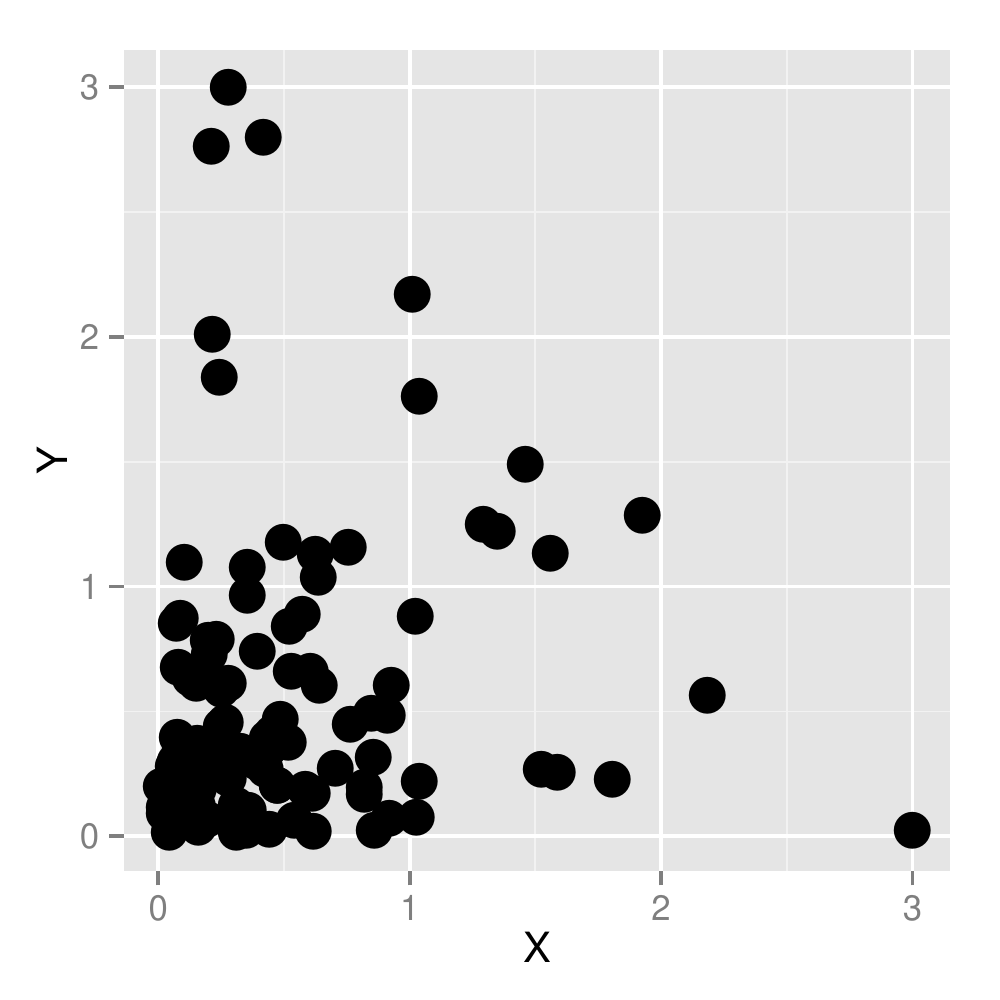}} \end{center} \end{minipage} &&&  \begin{minipage}[h]{1.5cm} \begin{center} \scalebox{0.25}{\includegraphics{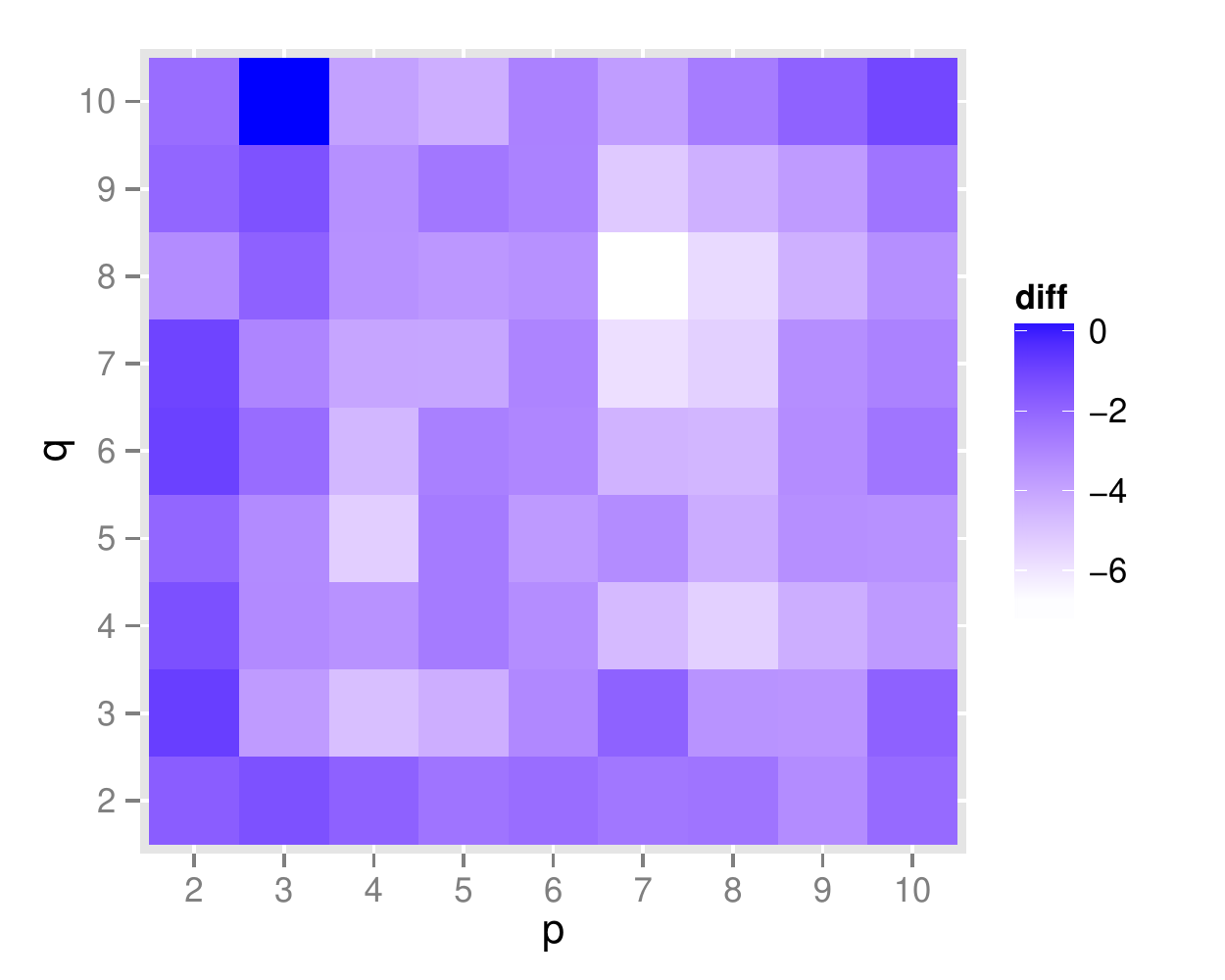}} \end{center} \end{minipage} &&           \hspace{0.8cm} (3, 10, 0.3; -7.1) \\
 \hline
 Residual Plot & \begin{minipage}[h]{1.5cm} \begin{center} \scalebox{0.25}{\includegraphics{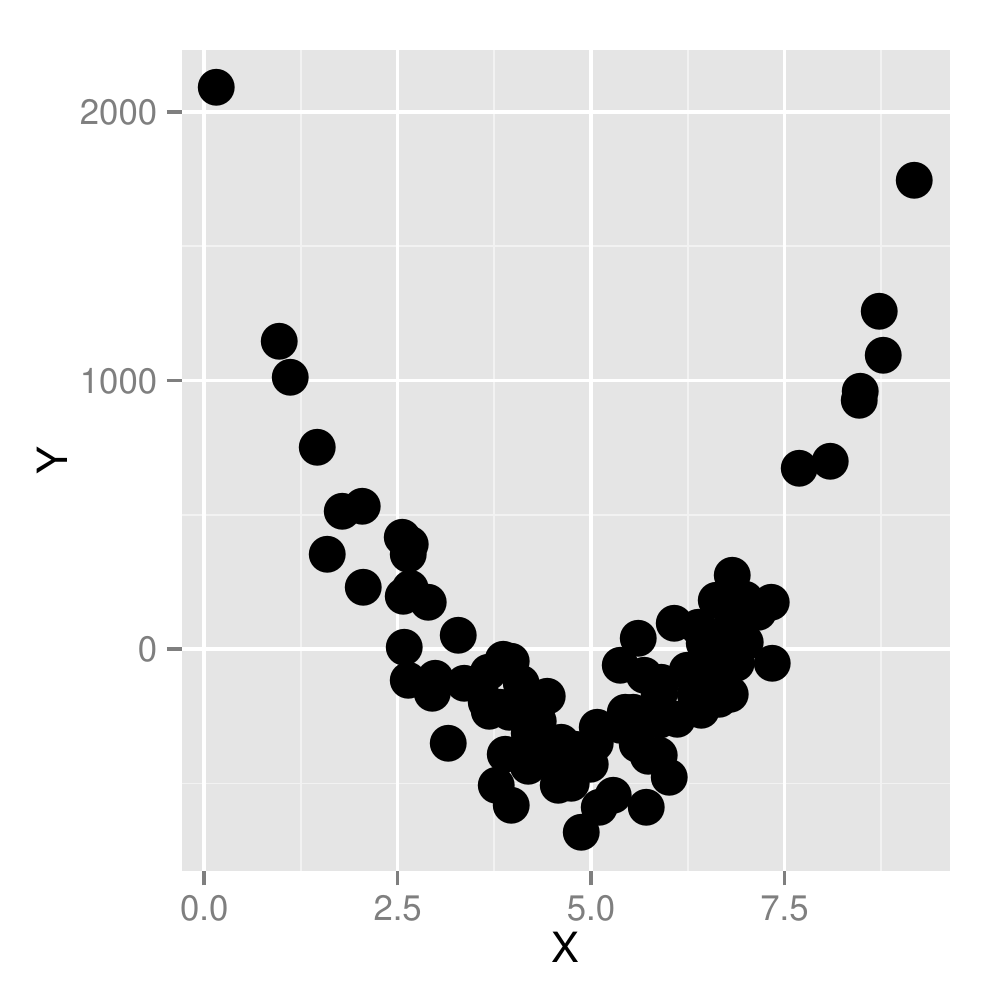}} \end{center} \end{minipage} && Simulation from the null model &  \begin{minipage}[h]{1.5cm} \begin{center} \scalebox{0.25}{\includegraphics{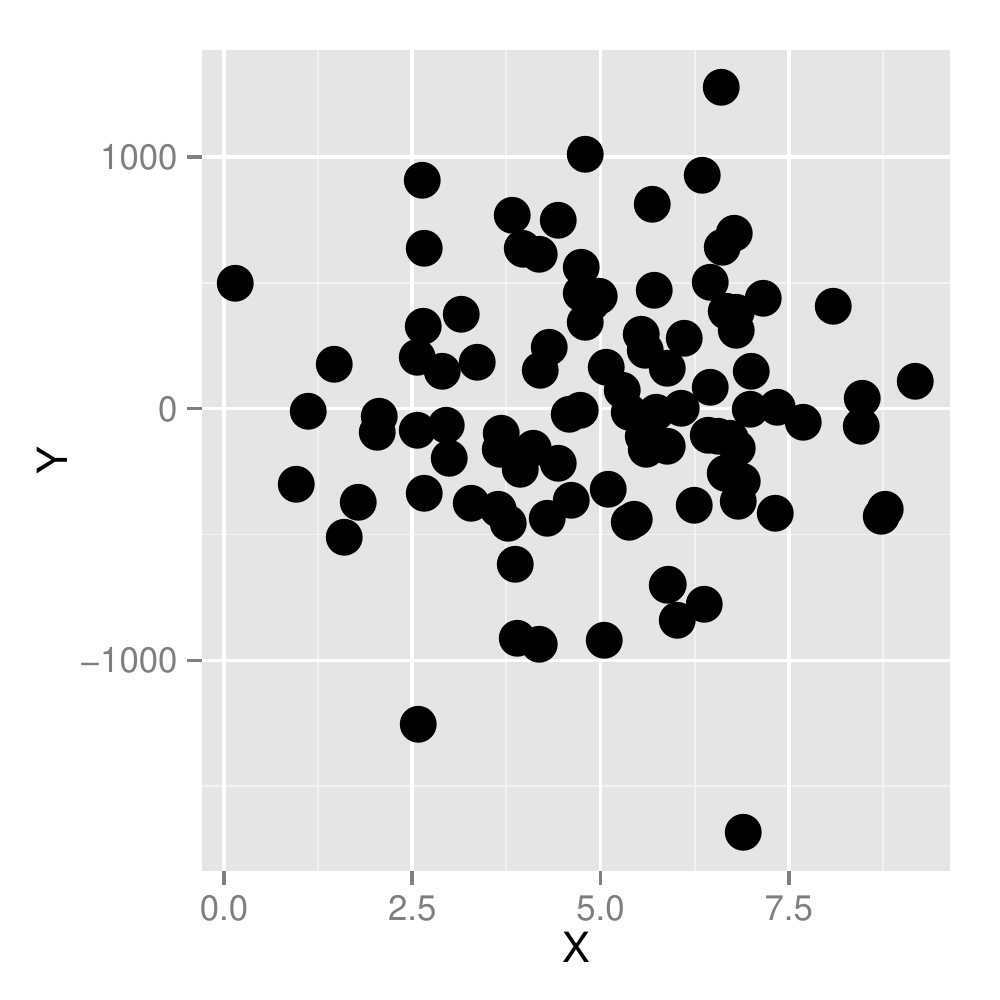}} \end{center} \end{minipage} &&&  \begin{minipage}[h]{1.5cm} \begin{center} \scalebox{0.25}{\includegraphics{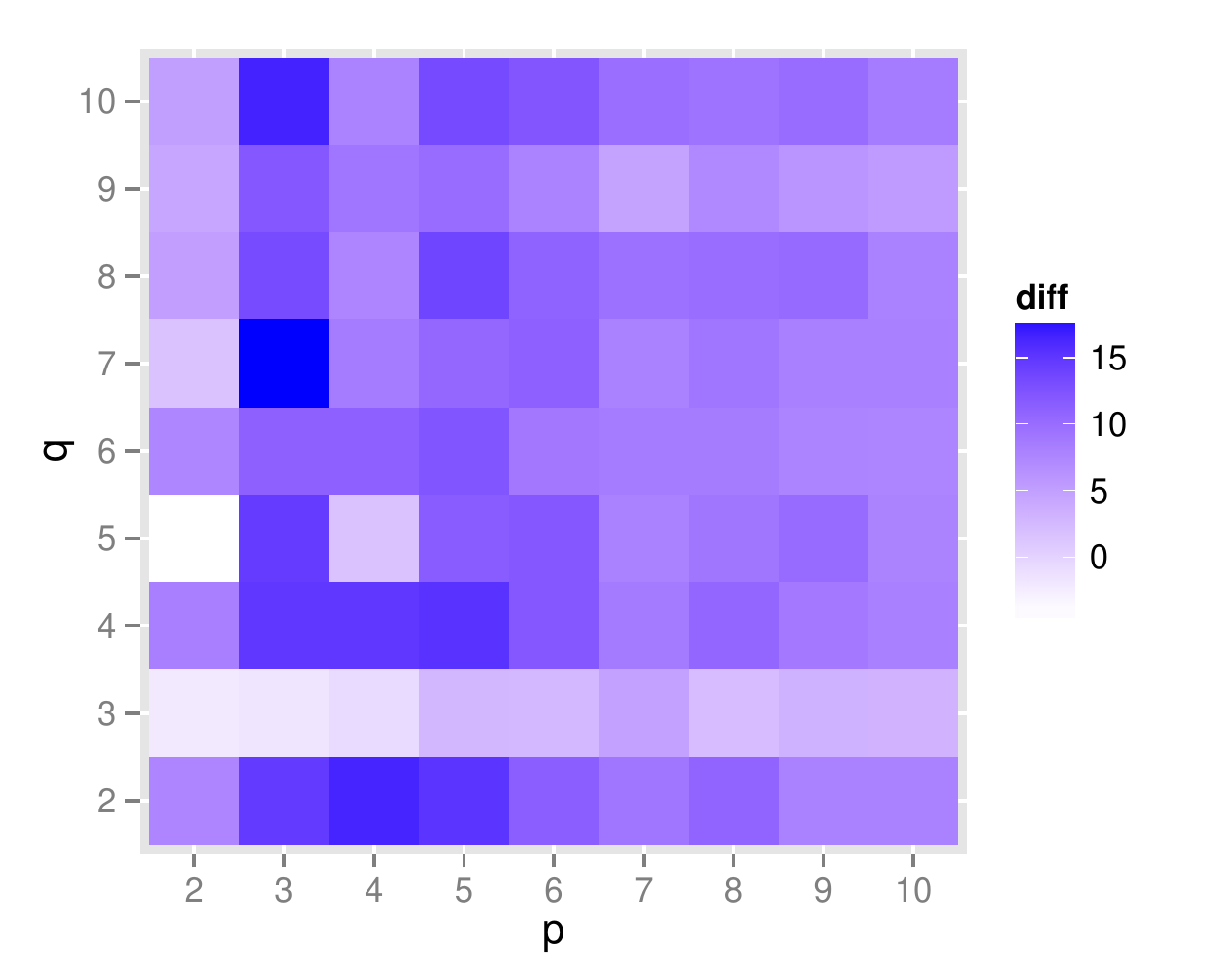}} \end{center} \end{minipage} &&           \hspace{0.8cm} (3, 7, 17.8; -4.5)\\
 \hline
 Residual Plot & \begin{minipage}[h]{1.5cm} \begin{center} \scalebox{0.25}{\includegraphics{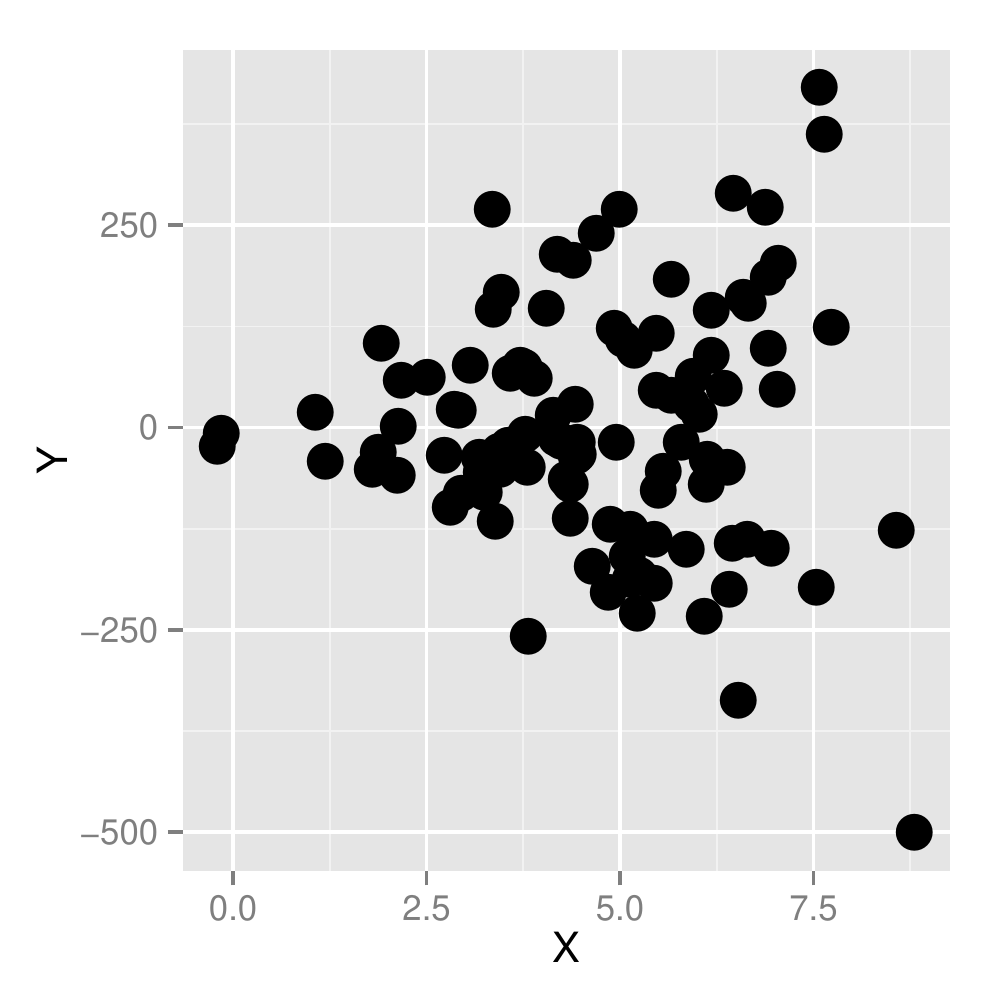}} \end{center} \end{minipage} && Simulation from the null model &  \begin{minipage}[h]{1.5cm} \begin{center} \scalebox{0.25}{\includegraphics{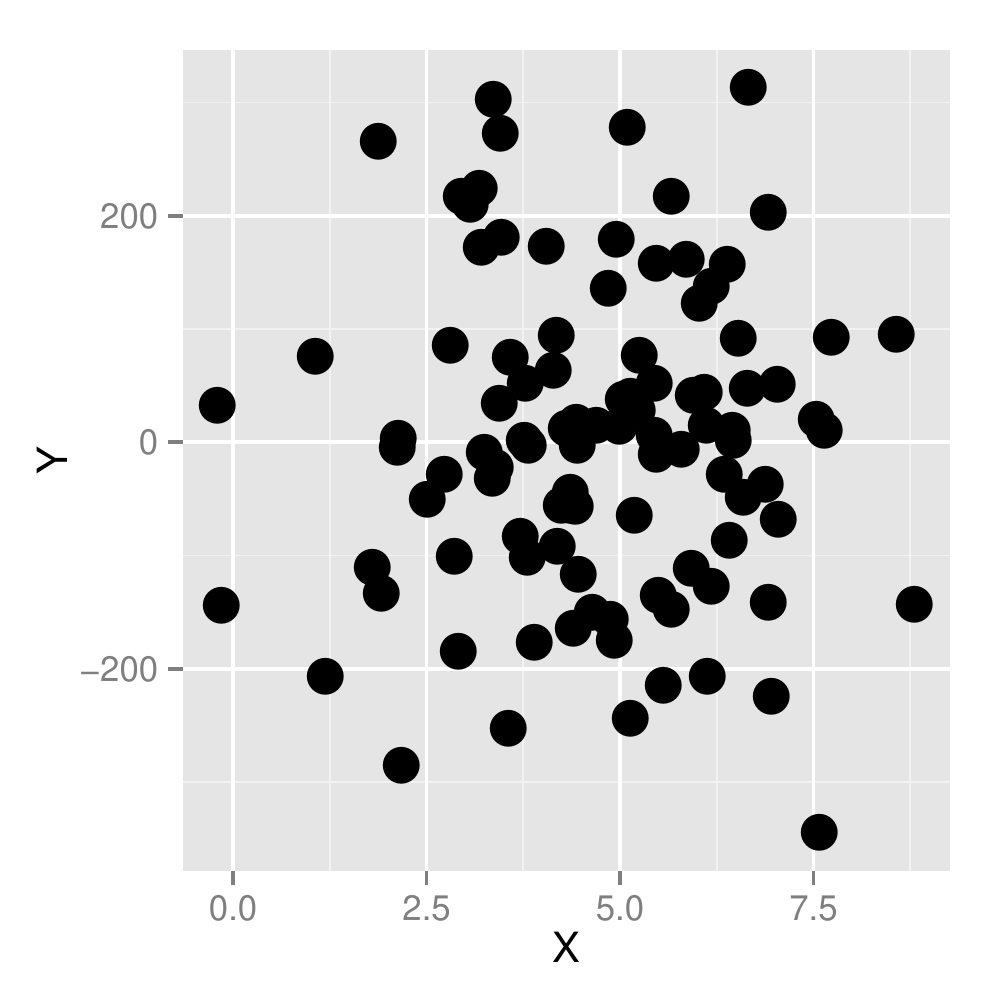}} \end{center} \end{minipage} &&&  \begin{minipage}[h]{1.5cm} \begin{center} \scalebox{0.25}{\includegraphics{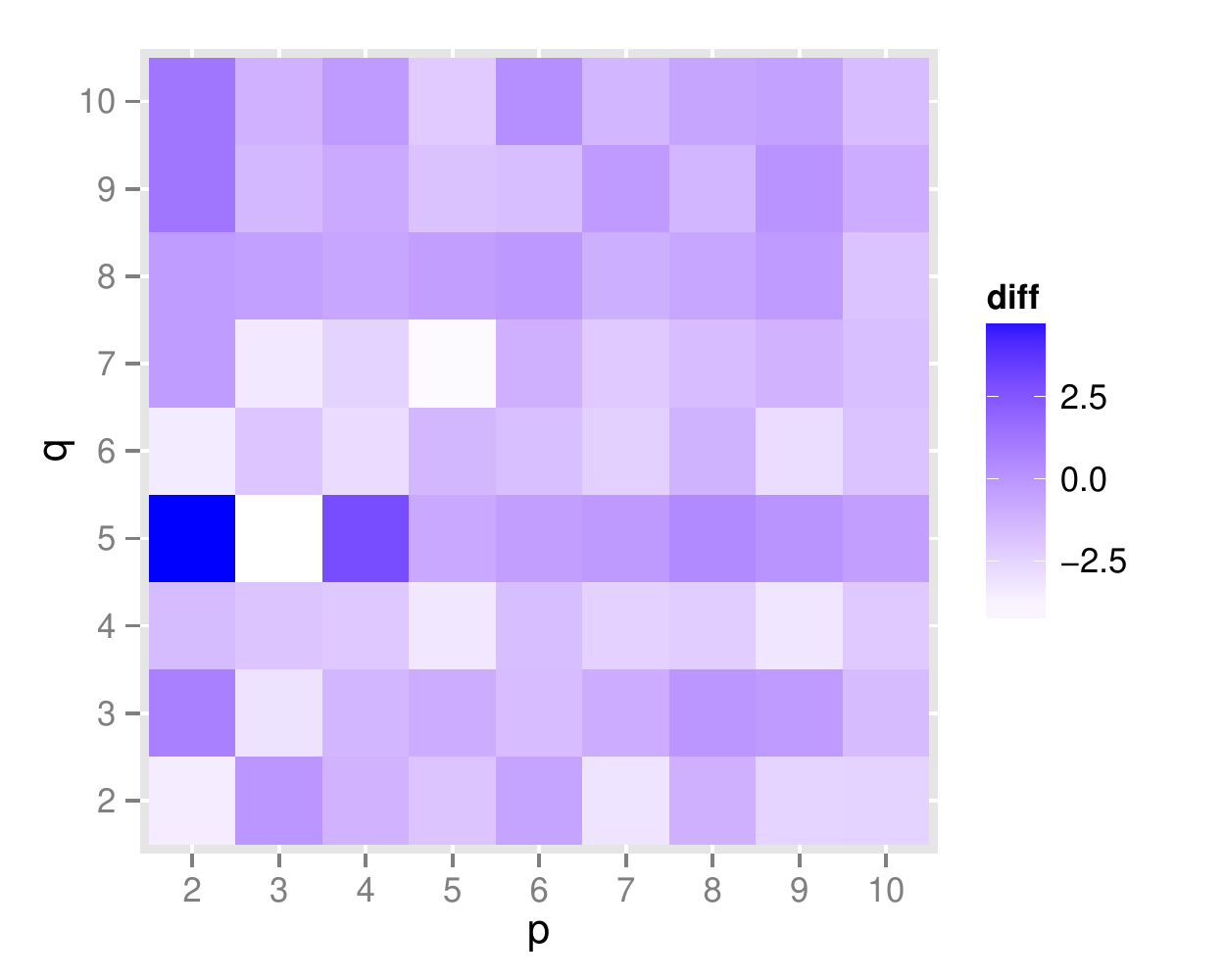}} \end{center} \end{minipage} &&           \hspace{0.8cm} (2, 5, 4.8; -4.4)\\
 \hline
Spiral data & \begin{minipage}[h]{1.5cm} \begin{center} \scalebox{0.25}{\includegraphics{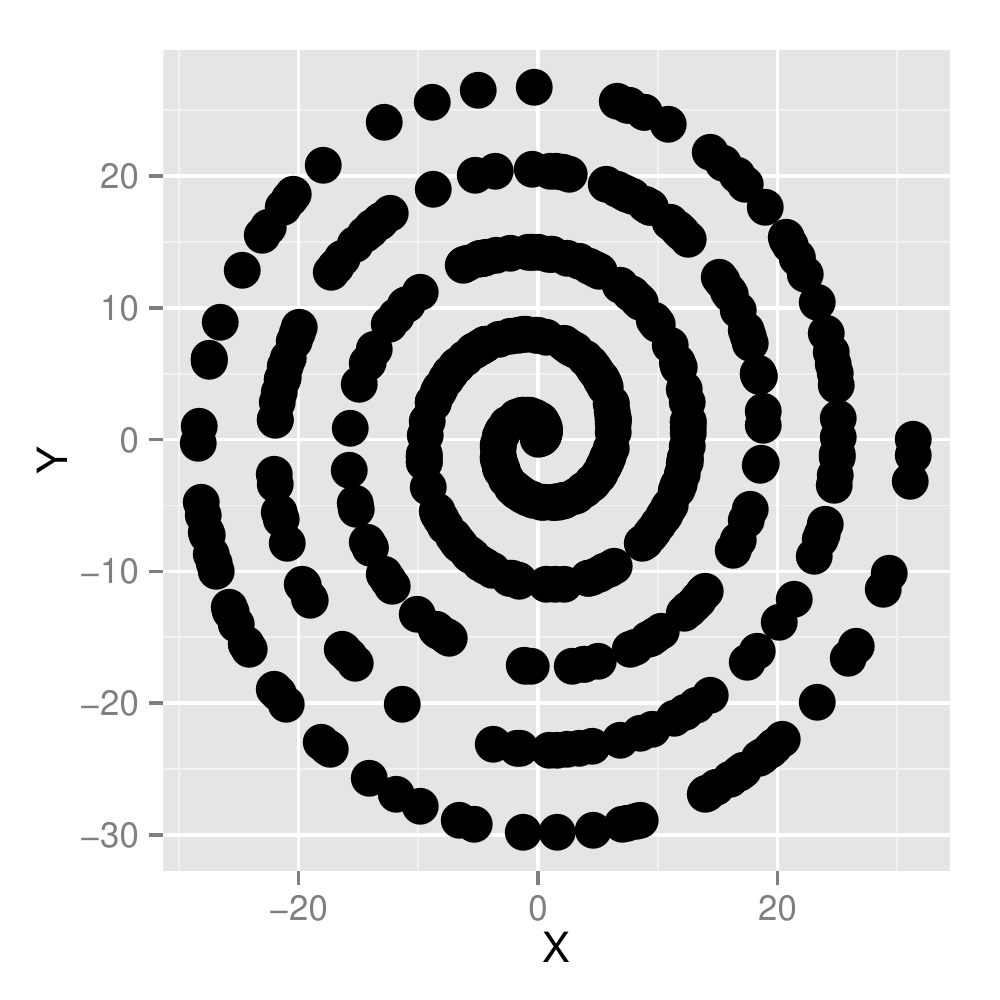}} \end{center} \end{minipage} && Permutation &  \begin{minipage}[h]{1.5cm} \begin{center} \scalebox{0.25}{\includegraphics{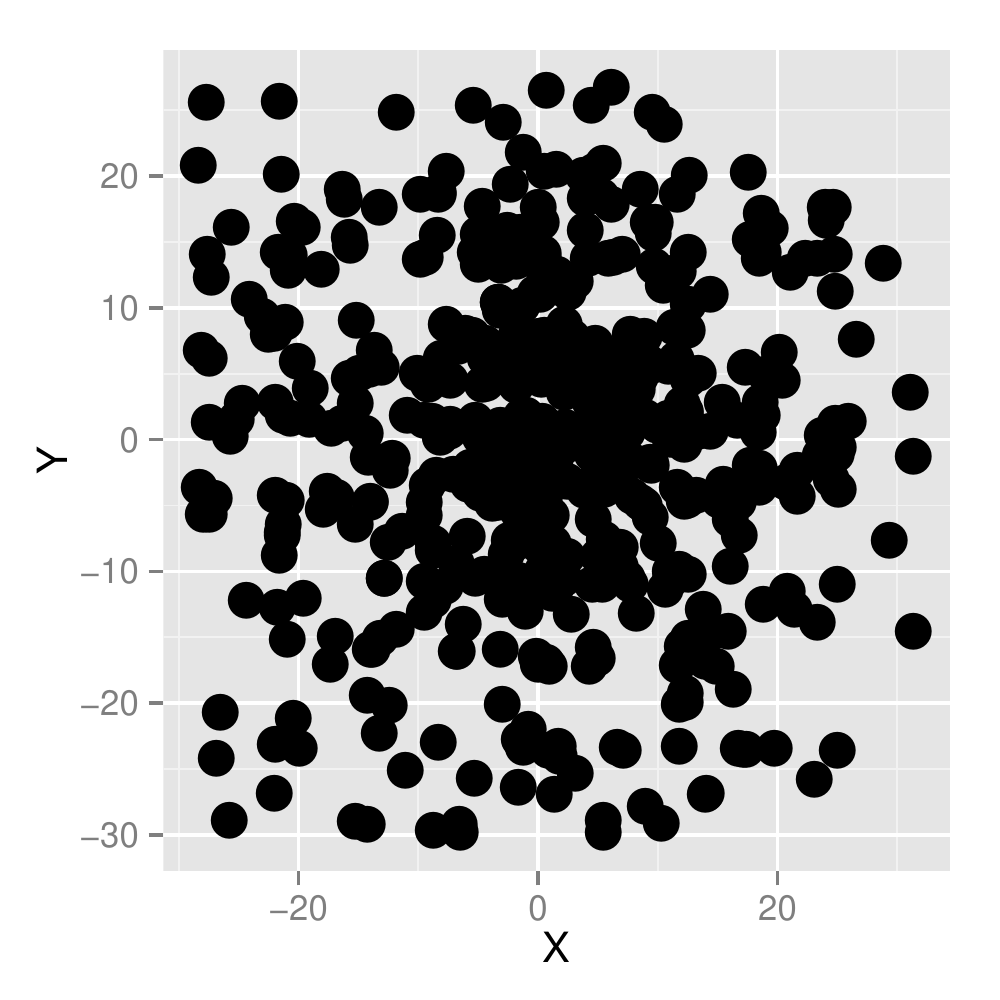}} \end{center} \end{minipage} &&&  \begin{minipage}[h]{1.5cm} \begin{center} \scalebox{0.25}{\includegraphics{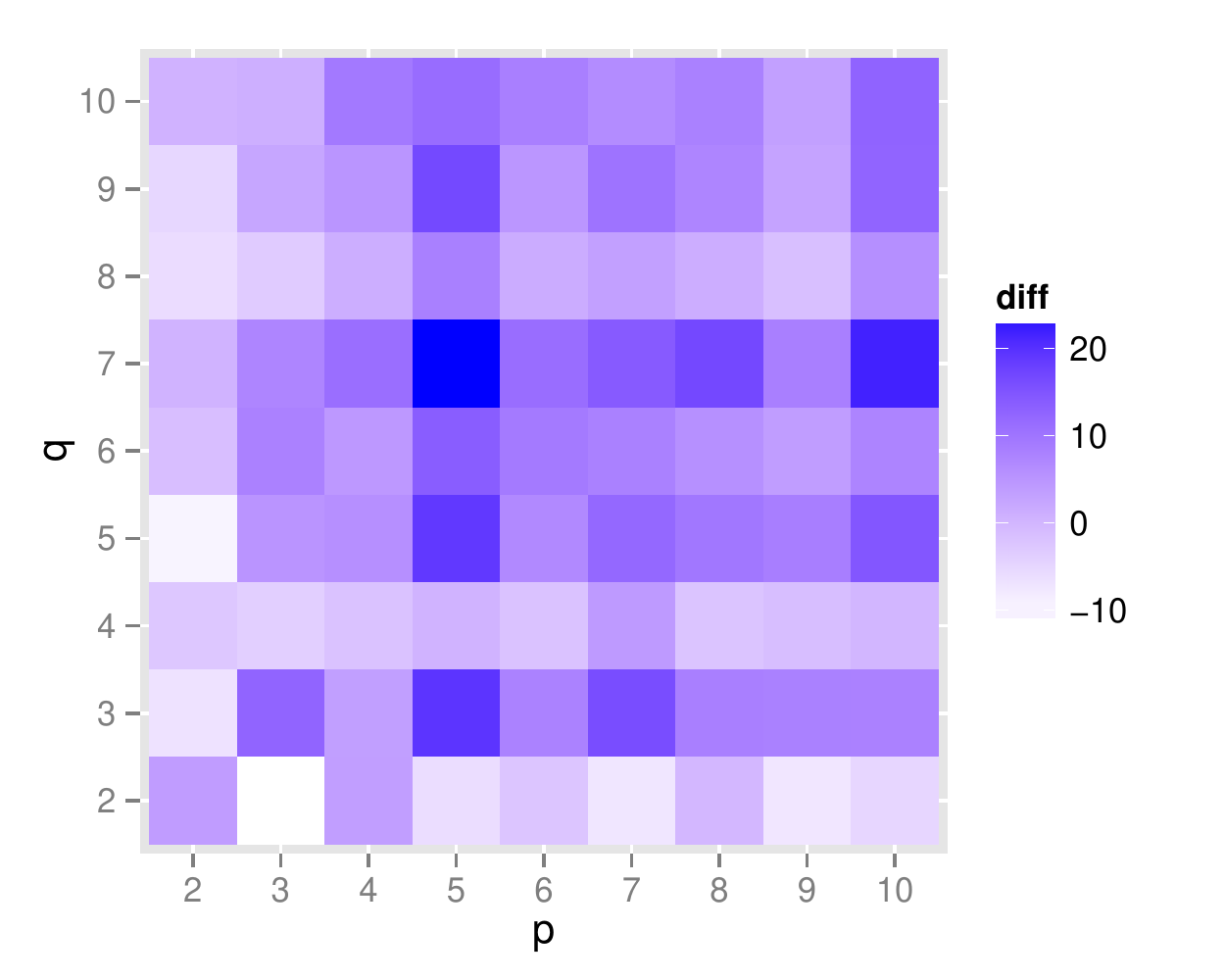}} \end{center} \end{minipage} &&           \hspace{0.8cm} (5, 7, 23.6; -11.9)\\
 \hline
Contaminated data & \begin{minipage}[h]{1.5cm} \begin{center} \scalebox{0.25}{\includegraphics{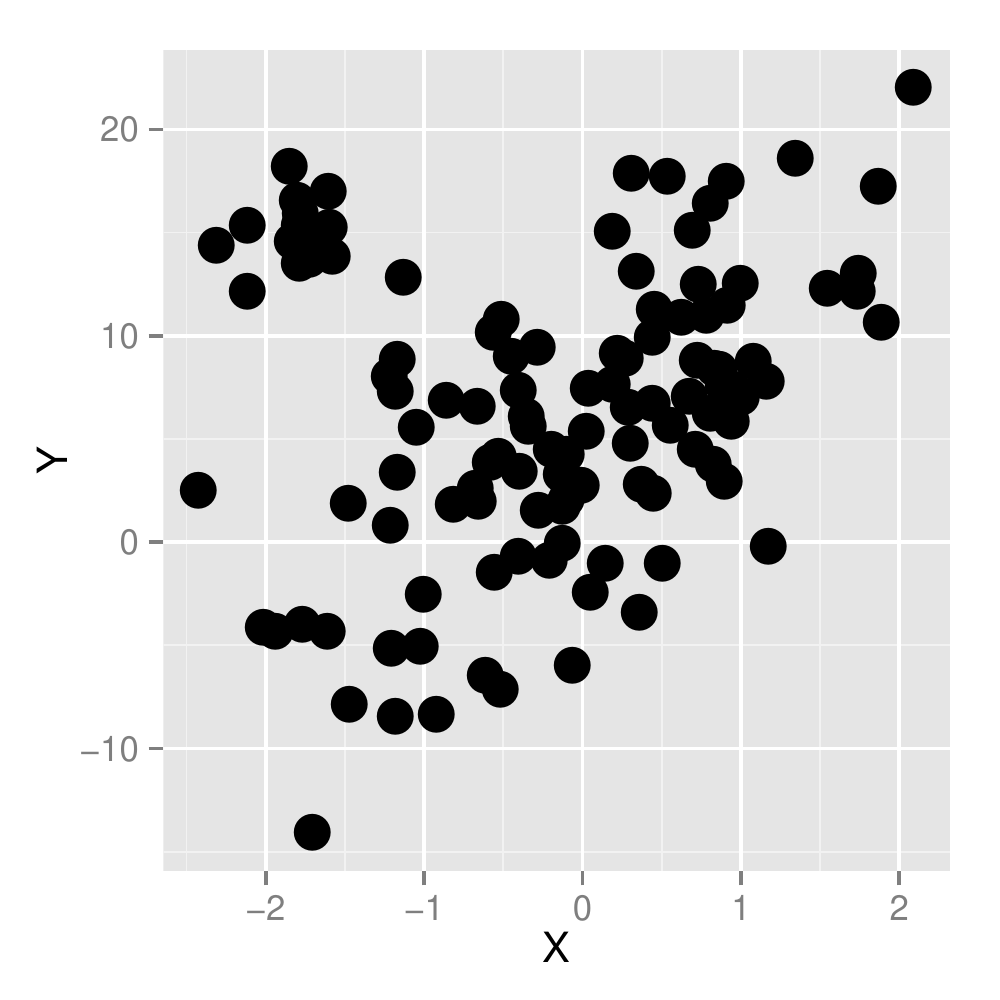}} \end{center} \end{minipage} && Permutation &  \begin{minipage}[h]{1.5cm} \begin{center} \scalebox{0.25}{\includegraphics{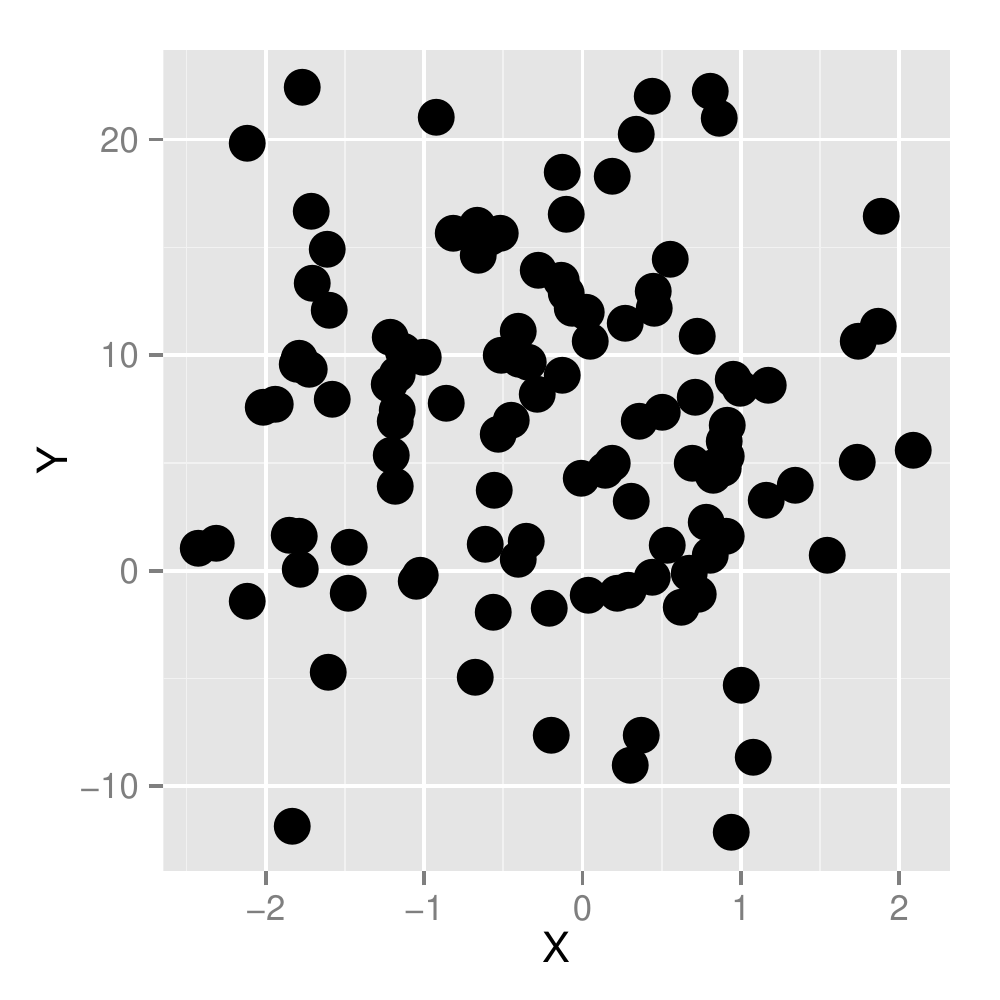}} \end{center} \end{minipage} &&&  \begin{minipage}[h]{1.5cm} \begin{center} \scalebox{0.25}{\includegraphics{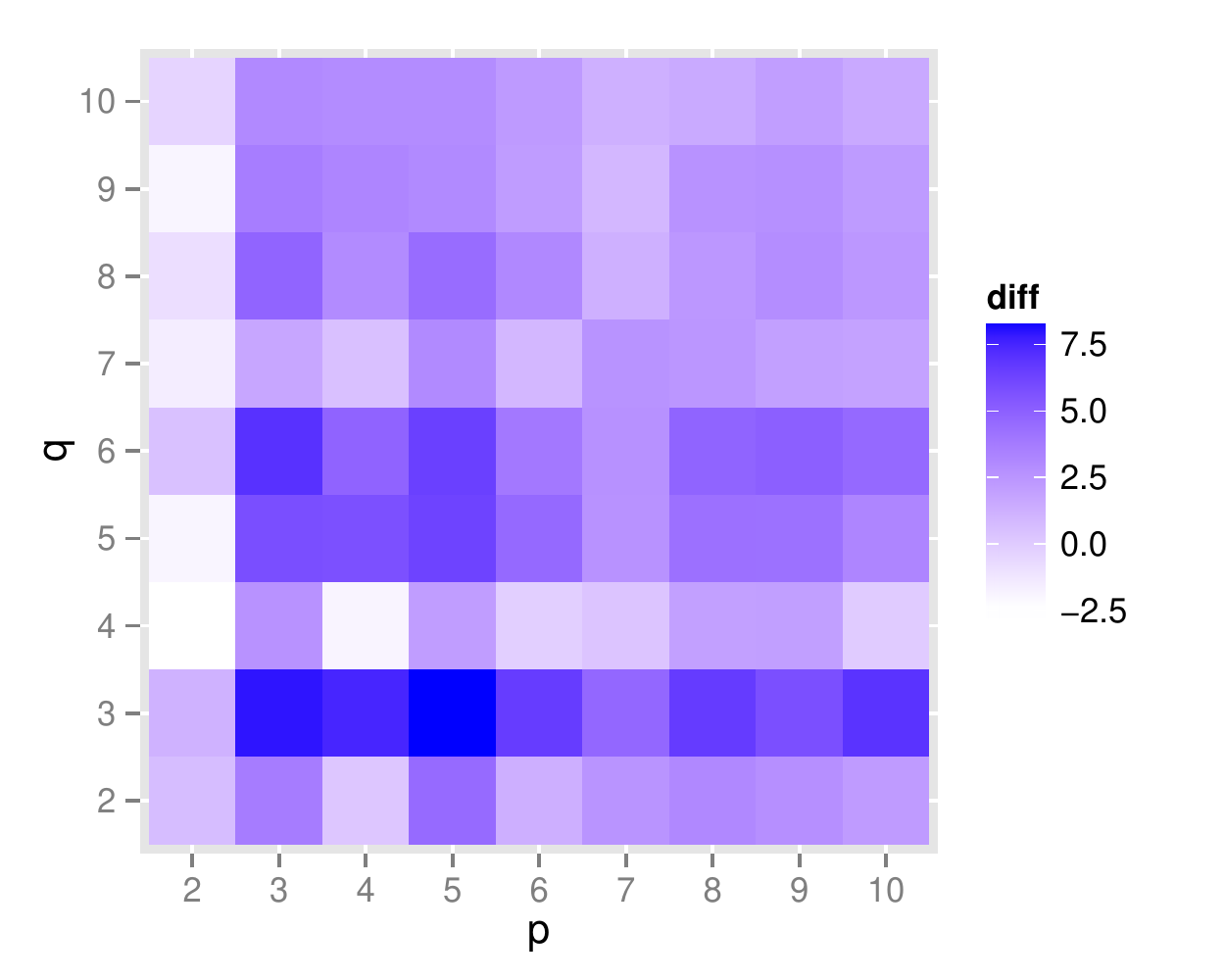}} \end{center} \end{minipage} &&           \hspace{0.8cm} (5, 3, 8.1; -2.5)\\
 \hline
\end{tabular}
\label{tbl:bin2}
\end{table*}

The rationale behind selecting different types of data is to investigate how the optimal number of bins or bin sizes varies with different types of data. The different null generating mechanisms are also selected for the same reason. In Table \ref{tbl:bin1} the first four observed data plots corresponds to the datasets described by Francis Anscombe \citep{anscombe:1972} but with large number of data points. Although the datasets have the same pattern, the datasets do not follow the properties of Anscombe's quartet. The fifth dataset is a data with 3 distinct clusters. In Table \ref{tbl:bin2}, the first dataset shows a categorical data. The second and the third data are non-linear and linear association with the presence of outliers. The fourth and fifth datasets are the residual plots with curved pattern and non-constant variance pattern. The sixth data is a spiral data while the seventh one is a data with contamination. 

The differences, $\delta_{\hbox{lineup}}$, are represented in a tile plot where each tile gives the difference for each combination. The dark blue shows higher values while the white shows lower values. It can be seen that the plots look different for the different datasets. Hence the optimal number of bins varies from data to data. No specific pattern is evident in the plot. But overall it can be seen that for strong linear relationship, small number of bins should be preferred over large number of bins. Also when outlier is present in the data, larger number of bins is preferred at least in one axis.

It is important to mention at this point that Table \ref{tbl:bin1} and Table \ref{tbl:bin2} is not meant to provide any guidelines for the selection of number of bins. The Tables only show that the binned distance is highly affected by the number of bins and the type of data. It is advisable to find the optimal number of bins for a given data before using the binned distance.

\end{document}